\documentclass[acmlarge]{acmart}

\AtBeginDocument{%
  \providecommand\BibTeX{{%
    \normalfont B\kern-0.5em{\scshape i\kern-0.25em b}\kern-0.8em\TeX}}}

\usepackage{graphicx}
\usepackage{ulem}
\usepackage{verbatim}
\usepackage{multirow}

\usepackage[linesnumbered, ruled, vlined]{algorithm2e}
\usepackage{longtable}
\usepackage{array}
\usepackage{subfigure}
\usepackage{stfloats}
\usepackage{balance}
\usepackage{multicol}
\usepackage{color}
\usepackage{bm}
\usepackage{booktabs} 
\usepackage{epsfig}
\usepackage{enumitem}
\usepackage{balance}
\usepackage{amsmath}

\usepackage{amssymb}

\usepackage{arydshln}

\usepackage{cleveref}
\usepackage{graphicx}
\usepackage{textcomp}
\usepackage{xcolor}
\usepackage{subfigure}

\usepackage{tabularx}
\usepackage{bbm}
\usepackage{bbding}

\newcommand{\hide}[1]{} 
\newcommand{\vpara}[1]{\vspace{0.1in}\noindent\textbf{#1}}

\newcommand{\secref}[1]{Section~\ref{#1}} 

\newcommand{\beqn}[1]{\vspace{-0.03in}\begin{eqnarray}#1\end{eqnarray}\vspace{-0.03in}}

\newcommand{\RC}{${\rm PT}$-${\rm GNN}$}
\newcommand{\sRC}{${\rm PT}$-${\rm GNN}$\space}
\newcommand{\nRC}{${\rm MPT}$}
\newcommand{\snRC}{${\rm MPT}$\space}

\setcopyright{acmcopyright}
\copyrightyear{2022}
\acmYear{2022}
\acmDOI{10.1145/1122445.1122456}

\acmJournal{POMACS}
\acmVolume{37}
\acmNumber{4}
\acmArticle{111}
\acmMonth{5}

\begin{document}

\title{A Multi-Strategy based Pre-Training Method for
	Cold-Start Recommendation}

\author{Bowen Hao}
\email{jeremyhao@ruc.edu.cn}
\affiliation{%
  \institution{Renmin University of China}
  \city{Beijing}
  \country{China}
}

\author{Hongzhi Yin}
\authornote{Corresponding Author}
\email{ h.yin1@uq.edu.au}
\affiliation{%
\institution{The University of Queensland}
  \city{Brisbane}
  \country{Australia}
}

\author{Jing Zhang}
\email{zhang-jing@ruc.edu.cn}
\affiliation{%
	\institution{Renmin University of China}
	\city{Beijing}
	\country{China}
}

\author{Cuiping Li}
 \email{ licuiping@ruc.edu.cn}
\affiliation{%
 \institution{Renmin University of China}
 \city{Beijing}
 \country{China}
}

\author{Hong Chen}
\email{chong@ruc.edu.cn}
\affiliation{%
  \institution{Renmin University of China}
  \city{Beijing}
  \country{China}
}

\renewcommand{\shortauthors}{Bowen Hao et al.}

\begin{abstract}
Cold-start issue is a fundamental challenge in Recommender System. 
The recent self-supervised learning (SSL) on Graph Neural Networks (GNNs) model, PT-GNN, pre-trains the GNN model to reconstruct the cold-start embeddings and has 
shown great potential for cold-start recommendation.
However, due to the over-smoothing problem, PT-GNN can only capture up to 3-order relation, which can not provide much useful auxiliary information to depict the target cold-start user or item. Besides, the embedding reconstruction task only considers the intra-correlations within the subgraph of users and items, while ignoring the inter-correlations across different subgraphs.
To solve the above challenges, we propose a multi-strategy based pre-training method for cold-start recommendation (MPT),
which extends PT-GNN from the perspective of  model architecture and  pretext tasks to improve the cold-start recommendation performance\footnote{This paper is the extension of our previously published conference version~\cite{haopretrain21}.}.
Specifically, in terms of the model architecture, in addition to the short-range dependencies of
users and items captured by the GNN encoder, we introduce a
Transformer encoder to capture long-range dependencies. In terms of the pretext task, in addition to considering the intra-correlations of users and items by the embedding reconstruction task,
we add embedding contrastive learning task to capture inter-correlations of users and items. We train the GNN and Transformer encoders on these  pretext tasks under  the meta-learning setting to simulate the real cold-start scenario,  making the model easily
and rapidly being adapted to new cold-start users and items.
Experiments on three public recommendation datasets show
the superiority of the proposed MPT model against the vanilla GNN models, the pre-training GNN model on user/item embedding inference and the recommendation task.
\end{abstract}



\maketitle

\section{Introduction}

As an intelligent system, Recommender System (RS)~\cite{he2017neural,xiangnanhe_lightgcn20,linden2003amazon} has been built successfully in recent years, and aims to alleviate the information overload problem. The most popular algorithm in RS is collaborative filtering, which uses either matrix factorization~\cite{linden2003amazon} or neural collaborative filtering~\cite{he2017neural} to learn the user/item embeddings. However, due to the sparse interactions of the cold-start users/items, it is difficult to learn high-quality embeddings for these cold-start users/items.

To address the cold-start issue, researchers propose to incorporate the side information such as knowledge graphs (KGs)~\cite{wang2019multi,wang2019kgat} or the contents of users/items~\cite{hongzhi_side,YinWWCZ17,ChenYS0GM20} to enhance the representations of users/items.
However, the contents are not always available, and it is not easy to link the items to the entities in KGs due to the incompleteness and ambiguities of the entities. Another research line is to adopt the Graph Neural Networks (GNNs) such as  GraphSAGE~\cite{williamgraphsage17}, NGCF~\cite{wangncgf19} and LightGCN~\cite{xiangnanhe_lightgcn20} to solve the problem.
The basic idea is to incorporate the high-order neighbors to enhance the embeddings of the cold-start users/items.
However, the GNN models for recommendation can not thoroughly solve the cold-start problem, as the embeddings of the cold-start users/items aren't explicitly optimized, and the cold-start neighbors have not been considered in these GNNs.

In our previous work, we propose \RC~\cite{haopretrain21}, which pre-trains the GNN model  to reconstruct the user/item embeddings under the meta-learning setting~\cite{vinyalsmatching16}.
To further reduce the impact from the cold-start neighbors, \sRC incorporates a self-attention based meta aggregator to enhance the aggregation ability of each graph convolution step, and an adaptive neighbor sampler to select proper high-order neighbors according to the feedbacks from the GNN model.

However, \sRC still suffers from the following challenges:
1) In terms of the model architecture, \sRC suffers from lacking the ability to capture long-range dependencies.
Existing researches such as Chen et al.~\cite{CWKDD20} pointed out that GNN can only capture up to 3-hop neighbors due to the overfitting and over-smoothing problems~\cite{xiangnanhe_lightgcn20}. 
However, in the cold-start scenario, both low-order neighbors ($L \leq 2$, where $L$ is the convolution layer) and high-order neighbors ($L> 2$) are important to the target cold-start user/item. On one hand, the low-order neighbors are crucial for representing the target cold-start users/items, as the first-order neighbors directly reflect the user's preference or the item's target audience, and the second-order neighbors directly reflect the signals of the collaborative users/items. On the other hand, due to the extremely sparse interactions of the cold-start users/items, the first- and second-order neighbors are insufficient to represent a user or an item. 
Thus, for the target users/items, it is crucial to capture not only the short-range dependencies from their low-order neighbors, but also the long-range dependencies from their high-order neighbors. 
Nevertheless, due to the limitation of the network structure, most of the existing GNN models can only capture the short-range dependencies, which inspires the first research question: \textit{how to capture both short- and long-range dependencies of users/items?}
2) In terms of the pretext task, \sRC only considers the intra-correlations within the subgraph of the target user or item, but ignores the inter-correlations between the subgraphs of different users or items.
More concretely, given a target cold-start user or item, \sRC samples its high-order neighbors to form a subgraph, and leverages the subgraph itself to reconstruct the cold-start embedding. Essentially, the embedding reconstruction task is a generative task, which focuses on exploring intra-correlations between nodes in a subgraph. On the contrary, some recent pre-training GNN models (e.g., GCC~\cite{qiu2020gcc}, GraphCL~\cite{YCTSCWNIPS20}) depending on the contrastive learning mechanism, can capture the similarities of nodes between different subgraphs, i.e., the inter-correlations.
Thus, this inspires the second research question: \textit{how to capture the inter-correlations of the target cold-start users/items in addition to the intra-correlations?}

 \begin{figure}[t]
	\centering
	\includegraphics[width= 0.65\textwidth]{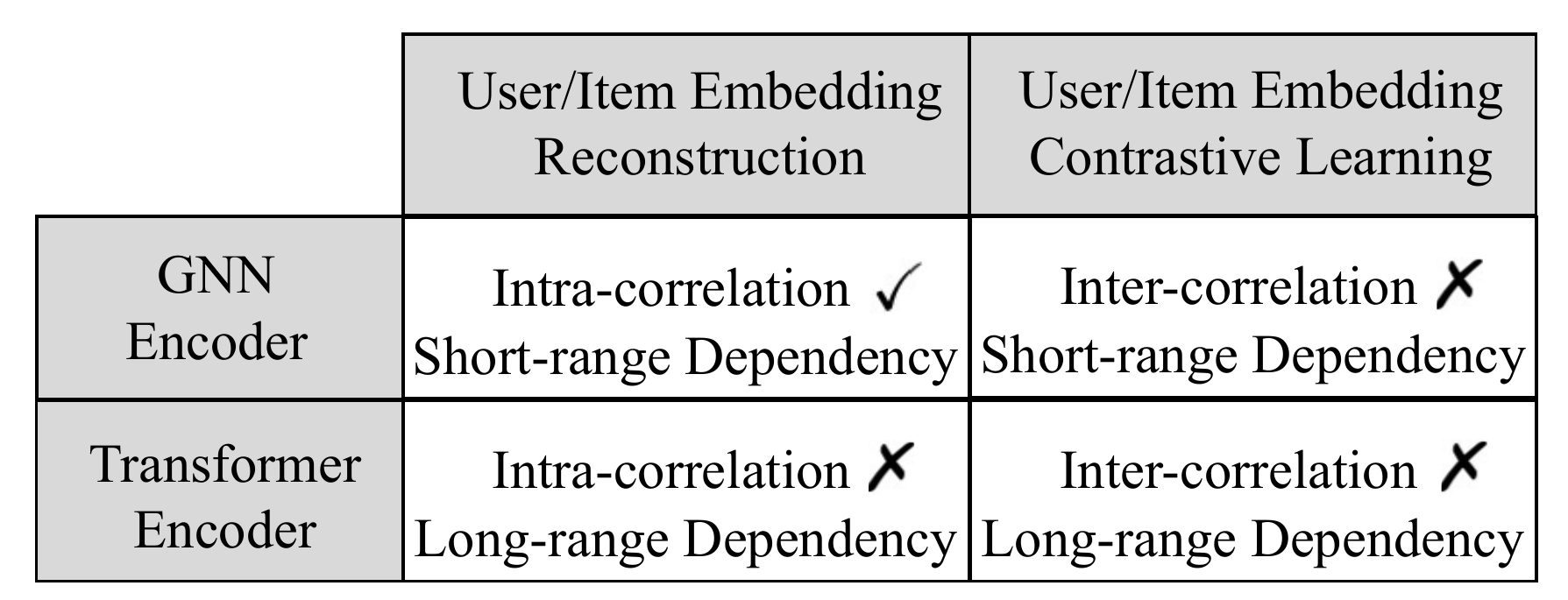}
	\caption{\label{fig:pretext_tasks} The improvement of \snRC on the original \RC. $\checkmark$ denotes the capacity of \sRC and \XSolidBrush denotes the missing capacity of \RC. }
\end{figure}

\vpara{Present work.} 
We propose a \underline{M}ulti-strategy based \underline{P}re-\underline{T}raining method for cold-start recommendation (\nRC), which extends \sRC from the perspective of  model architecture and  pretext tasks to improve the cold-start recommendation performance. The improvements over \sRC are shown in Fig.~\ref{fig:pretext_tasks}. 

First, in terms of the model architecture, in addition to the original GNN encoder which captures the short-range dependencies from the user-item edges in the user-item graph, we introduce a Transformer encoder to capture the long-range dependencies from the user-item paths, which can be extracted by performing random walk~\cite{deepwalk} starting from the target user or item in the user-item graph. The multi-head self-attention mechanism in the Transformer  attends nodes in different positions in a path~\cite{self_attention17}, which can explicitly capture the long-range dependencies between users and items.
Besides, we change the RL-based neighbor sampling strategy in the original GNN encoder into a simple yet effective dynamic sampling strategy, which can reduce the time complexity of the neighbor sampling process.

Second, in terms of the pretext task, in addition to the original embedding reconstruction task which considers the intra-correlations within the subgraph of the target user or item, we add  embedding contrastive learning~\cite{WXYSCVPR18} task to capture the inter-correlations across the subgraphs or paths of different users or items.
Specifically, we first augment a concerned subgraph or path by deleting or replacing the nodes in it. Then we treat the augmented subgraphs or paths of the concerned user or item as its positive counterparts, and those of other users or items as the negative counterparts. By contrastive learning upon the set of the positive and negative instances, we can pull the similar user or item embeddings together while pulling away dissimilar embeddings.

We train the GNN and Transformer encoders by the  reconstruction and  contrastive learning pretext tasks under the meta-learning setting~\cite{vinyalsmatching16} to simulate the cold-start scenario. Specifically, following \RC, we first pick the users/items with sufficient interactions as the target users/items, and learn their ground-truth embeddings on the observed abundant interactions. Then for each target user/item, in order to simulate the cold-start scenario, we mask other neighbors and only maintain $K$ first-order neighbors, based on which we form the user-item subgraph and use random walk~\cite{deepwalk} to generate the paths of the target user or item. Next we  perform graph convolution multiple steps upon the user-item subgraph or self-attention mechanism upon the path to obtain the target user/item embeddings. Finally, we optimize the model parameters with the reconstruction and contrastive losses.

We adopt the pre-training \& fine-tuning paradigm~\cite{qiu2020gcc} to train \nRC. During the pre-training stage, in order to capture the  correlations of users and items from different views (i.e., the short- and long-range dependencies, the intra- and inter-correlations), we assign each pretext task an independent set of initialized embeddings, and train these tasks independently.
During the fine-tuning state, we fuse the pre-trained embeddings from each pretext task and fine-tune the encoders by the downstream recommendation task.  
The contributions can be summarized as:

\begin{itemize}[ leftmargin=10pt ]
	\item We extend  \sRC  from the perspective of  model architecture.  In addition to the short-range dependencies of users and items captured by the GNN encoder, we add a  Transformer encoder to capture long-range dependencies.
	
	\item  We extend  \sRC  from the perspective of  pretext tasks. In addition to considering the intra-correlations of users and items by the embedding reconstruction task, we add embedding contrastive learning task to capture inter-correlations of users and items.

	\item Experiments on both intrinsic embedding evaluation and extrinsic downstream tasks demonstrate the superiority of our proposed \snRC model against the state-of-the-art GNN model and the original proposed \RC.
	
\end{itemize}

\section{ Preliminaries}
\label{sec:preliminaries}
In this section, we first define the problem and then introduce the original proposed \sRC.

\subsection{Notation and Problem Definition}

\vpara{Bipartite Graph.}
\noindent We denote the user-item bipartite graph as $\mathcal{G}=(\mathcal{U},\mathcal{I}, \mathcal{E})$, where $\mathcal{U} = \{  u_1, \cdots, u_{|\mathcal{U}|}  \}$ is the set of users and $\mathcal{I} = \{  i_1, \cdots, i_{|\mathcal{I}|} \}$ is the set of items. $\mathcal{E} \subseteq \mathcal{U} \times \mathcal{I}$ denotes the set of edges that connect the users and items. 
Notation $\mathcal{N}^{l}(u)$ denotes the $l$-order neighbors of user $u$. When ignoring the superscript,  $\mathcal{N}(u)$ denotes the first-order neighbors of $u$. $\mathcal{N}^{l}(i)$ and $\mathcal{N}(i)$ are defined similarly for items.

\vpara{User-Item Path.}
The user-item path is generated by the random walk strategy from the user-item interaction data, and has the same node distribution with the graph $\mathcal{G}$, as Perozzi et al.~\cite{deepwalk} proposed. 
Formally, there exists two types of paths: $\mathcal{P}$ = ${\rm UIUI}$ or $\mathcal{P}$ = ${\rm IUIU}$, where ${\rm U}$ denotes the node type is user and ${\rm I}$ denotes the node type is item.

\hide{Formally, the path $\mathcal{P}$ is in the form of $A_1\stackrel{R_1}{\longrightarrow}A_2\stackrel{R_2}{\longrightarrow}\dots\stackrel{R_l}{\longrightarrow}A_{l+1}$ (abbreviated as $A_1A_2\dots A_{l+1}$), which describes a composite relation $R=R_1\circ R_2\circ \dots\circ R_l$ between node types $A_1$ and $A_{l+1}$, where $\circ$ denotes the composition operator on relations.
In the recommendation scenario, there exists two types of paths: $\mathcal{P}$ = ${\rm UIUI}$ or $\mathcal{P}$ = ${\rm IUIU}$, where ${\rm U}$ denotes the node type is user and ${\rm I}$ denotes the node type is item.
}

\vpara{Problem Definition.}
Let $f:\mathcal{U} \cup \mathcal{I} \rightarrow \mathbf{R}^d$ be the encoding function that maps the users or items to $d$-dimension real-valued vectors.
We denote a user $u$ and an item $i$ with initial embeddings $\textbf{h}_{u}^{0}$ and $\textbf{h}_{i}^{0}$, respectively.
Given the bipartite graph $\mathcal{G}$ or the path $\mathcal{P}$, we aim to pre-train the encoding function $f$ that is able to be applied to the downstream recommendation task to improve its performance.
Note that for simplicity, in the following sections,  we mainly take user embedding as an example to explain the proposed model, as item embedding can be explained in the same way.
\hide{In this paper, we aim to pre-train various GNNs as the encode function $f$.
In the following sections, we mainly take user embedding as an example to explain the proposed model. Item embedding can be explained in the same way. }

\subsection{A Brief Review of PT-GNN }
\label{subsec:pretrain}

The basic idea of \sRC is to leverage the vanilla GNN model as the encoder $f$ to reconstruct the target cold-start  embeddings to explicitly improve their embedding quality. 
To further
reduce the impact from the cold-start neighbors, \sRC incorporates a self-attention based meta aggregator to enhance the aggregation ability of each graph convolution
step, and an adaptive neighbor sampler to select proper
high-order neighbors according to the feedbacks from the
GNN model.

\hide{
\textcolor{red}{It might be more clear to add the subsections in this section. This part is too long for the readers to capture the general idea. You can still begin with an overview and then briefly introduce each part.}
}

\subsubsection{Basic Pre-training GNN Model}
The  basic pre-training GNN model  reconstructs the
cold-start embeddings under the meta-learning setting~\cite{vinyalsmatching16} to simulate the cold-start scenario.
Specifically, for each target user $u$, we mask some neighbors and only maintains at most $K^l$ neighbors ($K$ is empirically selected as a small number, e.g., $K$=3) at the $l$-th layer. Based on which we  perform graph convolution multiple steps to predict the target user embedding. Take GraphSAGE~\cite{williamgraphsage17} as the backbone GNN model as an example, the graph convolution process at the $l$-th layer is:

\beqn{
	\label{eq:basic_graph_convolution}
	\textbf{h}_u^{l} &=&\sigma (\mathbf{W}^{l} \cdot  ( \textbf{h}_{u}^{l-1} \ || \ \textbf{h}^{l}_{ \mathcal{N}(u)} )),
}

\noindent where $\textbf{h}_{u}^{l}$ is the refined user embedding at the $l$-th convolution step, $\textbf{h}_{u}^{l-1}$ is the previous user embedding ($\textbf{h}_{u}^{0}$ is the initialized embedding). $\textbf{h}^{l}_{ \mathcal{N}(u)}$ is the averaged embedding of the neighbors, in which the neighbors are sampled by the random sampling strategy. $\sigma$ is the sigmoid function, $\mathbf{W}^l$ is the parameter matrix, and $||$ is the concatenate operation.

We perform graph convolution $L$-1 steps, aggregate the refined embeddings of the first-order neighbors $\{\textbf{h}_{1}^{L-1}, \cdots, \textbf{h}_{K}^{L-1}\}$  to obtain the smoothed embedding $\textbf{h}^{L}_{  \mathcal{N}(u)}$, and transform it into the target embedding $\textbf{h}_u^{L}$, i.e.,  $\textbf{h}^{L}_{\mathcal{N}(u)} = \text{AGGREGATE}(\{\textbf{h}^{L-1}_i, \forall i \in \mathcal{N}(u)\})$, $\textbf{h}_u^{L} = \sigma ( \mathbf{W}^L \cdot  \textbf{h}_{\mathcal{N}(u)}^{L}  )$.
Then we calculate the cosine similarity between the predicted embedding $\textbf{h}_u^{L}$ and the ground-truth embedding $\textbf{h}_u$ to optimize the parameters of the GNN model:

\beqn{
	\label{eq:cosine_similarity}
	\Theta^{*} &=& \mathop{\arg\max}_{\Theta} \sum_{u} {\rm cos}( \textbf{h}_u^{L}, \textbf{h}_u ),
}

\hide{
	\beqn{
		\label{eq:pre-training-aggregation}
		\textbf{h}^{L}_{\mathcal{N}(u)} &=& \text{AGGREGATE}(\{\textbf{h}^{L-1}_i, \forall i \in \mathcal{N}(u)\}),	\\ \nonumber
		
	}
}

\noindent where the ground-truth embedding $\textbf{h}_u$ is learned by any recommender algorithms (e.g., NCF~\cite{he2017neural}, LightGCN~\cite{xiangnanhe_lightgcn20}\footnote{They are good enough to learn high-quality user/item
	embeddings from the abundant interactions, and we will discuss it in~\secref{sec:ground_truth_embedding}}), $\Theta $ is the model parameters. Similarly, we can obtain the item embedding  $\textbf{h}^L_{i}$ and optimize  model parameters by (Eq. (\ref{eq:cosine_similarity})).

\hide{
\footnote{We choose one classical model NCF~\cite{he2017neural}, as it is good enough to learn high-quality user/item
embeddings from the abundant interactions. We also choose the state-of-the-art model LightGCN~\cite{xiangnanhe_lightgcn20} to explore whether the embedding reconstruction task is sensitive to the ground-truth embeddings.}
}

\subsubsection{Enhanced Pre-training Model: PT-GNN}
\label{subsec:enhanced_ptgnn}
The basic pre-training strategy has two problems. On one hand, it can not explicitly deal with the high-order cold-start neighbors during the graph convolution process. On the other hand, the GNN sampling strategies such as random sampling~\cite{williamgraphsage17} or importance sampling~\cite{chenfastgcn18} strategies may fail
to sample high-order relevant cold-start neighbors due to their
sparse interactions.

To solve the first challenge, we incorporate  a meta aggregator to enhance
the aggregation ability of each graph convolution step. Specifically, the meta aggregator uses self-attention mechanism~\cite{self_attention17} to encode the initial embeddings $\{\textbf{h}_{1}^{0}, \cdots, \textbf{h}_{K}^{0}\}$ of the $K$ first-order neighbors for $u$ as input, and outputs the meta embedding $\tilde{\textbf{h}}_u$ for $u$. The process is:

\beqn{
	\label{eq:meta_learner}
	\nonumber
	\{ \textbf{h}_{1}, \cdots, \textbf{h}_{K} \}  &\leftarrow& {\rm SELF\_ATTENTION} ( \{ \textbf{h}_{1}^{0}, \cdots, \textbf{h}_{K}^{0} \} ), \\
	\tilde{\textbf{h}}_u &=& {\rm AVERAGE} (\{ \textbf{h}_{1}, \cdots, \textbf{h}_{K} \}). 
}

We use the same cosine similarity described in Eq. (\ref{eq:cosine_similarity}) to measure the similarity between the predicted meta embedding $\tilde{\textbf{h}}_u$ and the ground truth embedding $\textbf{h}_u$.

To solve the second challenge, we propose  an adaptive neighbor sampler to select proper high-order
neighbors according to the feedbacks from the GNN model.
The adaptive neighbor sampler is formalized as a hierarchical Markov Sequential Decision Process, which sequentially samples from the low-order neighbors to the high-order neighbors and results in  at most $K^l$ neighbors in the $l$-th layer for each target user $u$. After sampling the neighbors of the former $L$ layers, 
the GNN model accepts the sampled neighbors to produce the reward, which denotes  whether the sampled neighbors are reasonable or not. Through maximizing the expected reward from the GNN model, the adaptive neighbor sampler can sample proper high-order neighbors. 
Once the meta aggregator and the adaptive neighbor sampler are trained, the meta embedding $\tilde{\textbf{h}}_u$ and the averaged sampled neighbor embedding $ \tilde{\textbf{h}}^{l}_{ \mathcal{N}(u)}$ at the $l$-th layer are added into each graph convolution step in Eq. (\ref{eq:basic_graph_convolution}):

\beqn{
	\label{eq:ptgnn_conv}
	\textbf{h}_u^{l} &=&\sigma (\mathbf{W}^{l} \cdot  (\tilde{\textbf{h}}_{u} \ || \  \textbf{h}_{u}^{l-1} \ || \  \tilde{\textbf{h}}^{l}_{ \mathcal{N}(u)} ) ).
}

For the target user $u$, Eq. (\ref{eq:ptgnn_conv}) is repeated $L-1$ steps to obtain the embeddings  $\{\textbf{h}_{1}^{L-1}, \cdots, \textbf{h}_{K}^{L-1}\}$ for its $K$ first-order neighbors, then the predicted embedding $\textbf{h}_{u}^{L}$ is obtained by averaging the first-order embeddings. Finally, Eq. (\ref{eq:cosine_similarity}) is used to optimize the parameters of the pre-training GNN model. The pre-training parameter set $ \Theta = \{\Theta_{gnn}, \Theta_{f}, \Theta_{s} \} $, where $\Theta_{gnn}$ is the parameters of the vanilla GNN, $\Theta_{f}$  is the parameters of the meta aggregator and $\Theta_{s}$ is the parameters of the neighbor sampler.  The item embedding $\textbf{h}^L_{i}$ can be obtained in a similar way.

\subsubsection{Downstream Recommendation Task}
We fine-tune \sRC in the downstream recommendation task.
Specifically, for each target user $u$ and his  neighbors $ \{  \mathcal{N}^{1}(u), \cdots, \mathcal{N}^{L}(u)  \}$ of different order, we first use the pre-trained adaptive neighbor sampler to sample proper high-order neighbors  $\{  \hat{\mathcal{N}}^{1}(u), \hat{\mathcal{N}}^{2}(u) \cdots, \hat{\mathcal{N}}^{L}(u)  \}$, and then use the pre-trained meta aggregator to produce the user embedding $\textbf{h}_u^{L}$. The item embedding $\textbf{h}_i^{L}$ can be obtained in the same way. Then we transform the embeddings to calculate the relevance score between a user and an item, i.e., $y(u, i) = {\sigma ( \mathbf{W} \cdot  \textbf{h}_u^{L}  )}^\mathrm{T}  \sigma ( \mathbf{W} \cdot  \textbf{h}_i^{L}  )$ with parameters	$\Theta_r = \{ \mathbf{W}\}$.  Finally, we adopt the BPR loss~\cite{xiangnanhe_lightgcn20} to optimize $\Theta_r$ and fine-tune $\Theta$: 

\beqn{
		\label{eq:bpr}
		\mathcal{L}_{BPR} = \sum_{ (u,i)\in E, (u,j)\notin E }  - \ln \sigma( y(u, i) - y(u, j) ).
}

However, as shown in Fig.~\ref{fig:pretext_tasks}, due to the limitation of the GNN model,  \sRC can only capture the short-range dependencies of users and items; besides, the embedding reconstruction task only focuses on exploring the intra-correlations within the subgraph of the target user or item.
Therefore, it is necessary to fully capture  both short- and long-range dependencies of users and items, and both intra- and inter-correlations of users or items.

\begin{figure*}[t]
	\centering
	\mbox{ 
		
		\subfigure[\scriptsize  Reconstruction with GNN Encoder
		]{\label{subfig:GNN_Reconstruct}
			\includegraphics[width=0.425\textwidth]{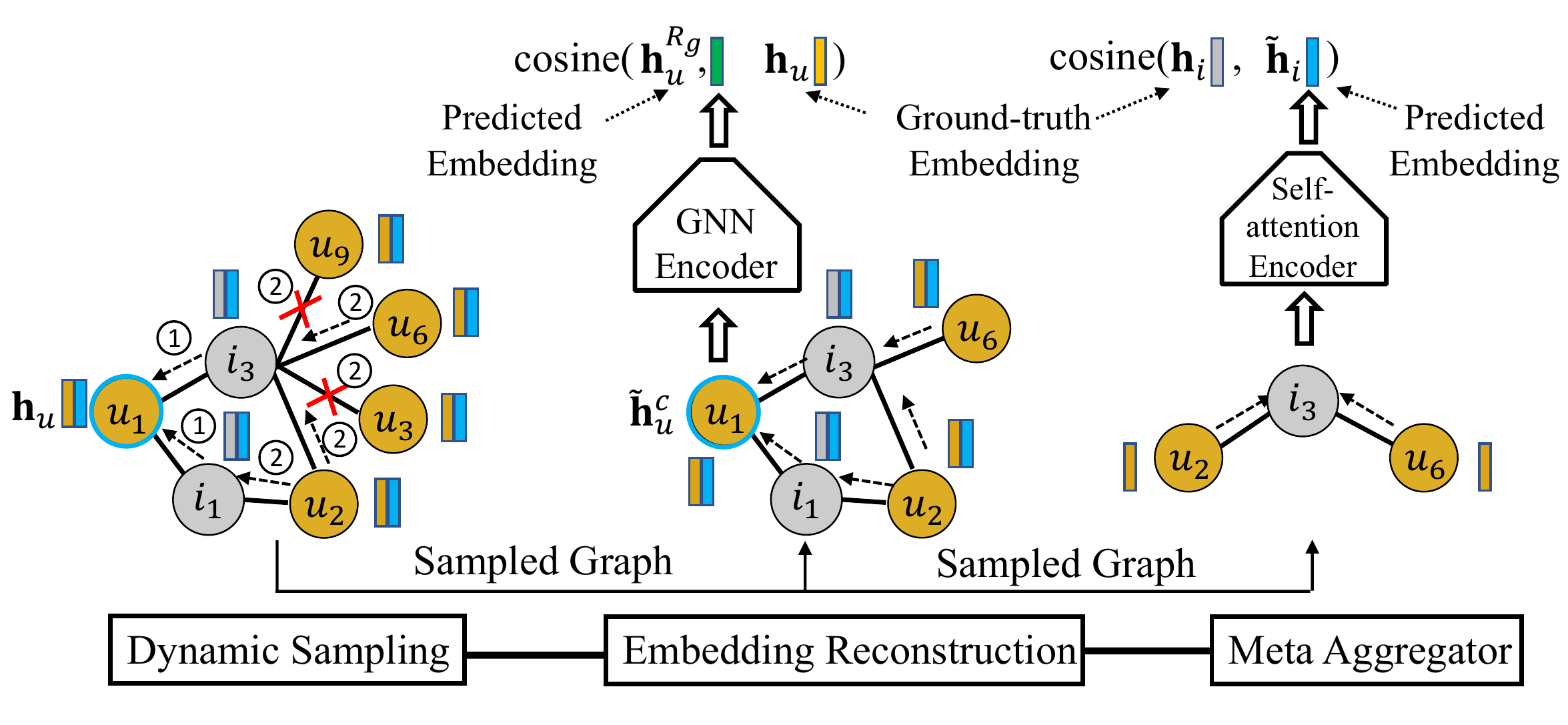}
		}
		
		\hspace{0.9cm}
		
		\subfigure[\scriptsize  Contrastive Learning with GNN Encoder
		]{\label{subfig:GNN_Contrastive}
			\includegraphics[width=0.475\textwidth]{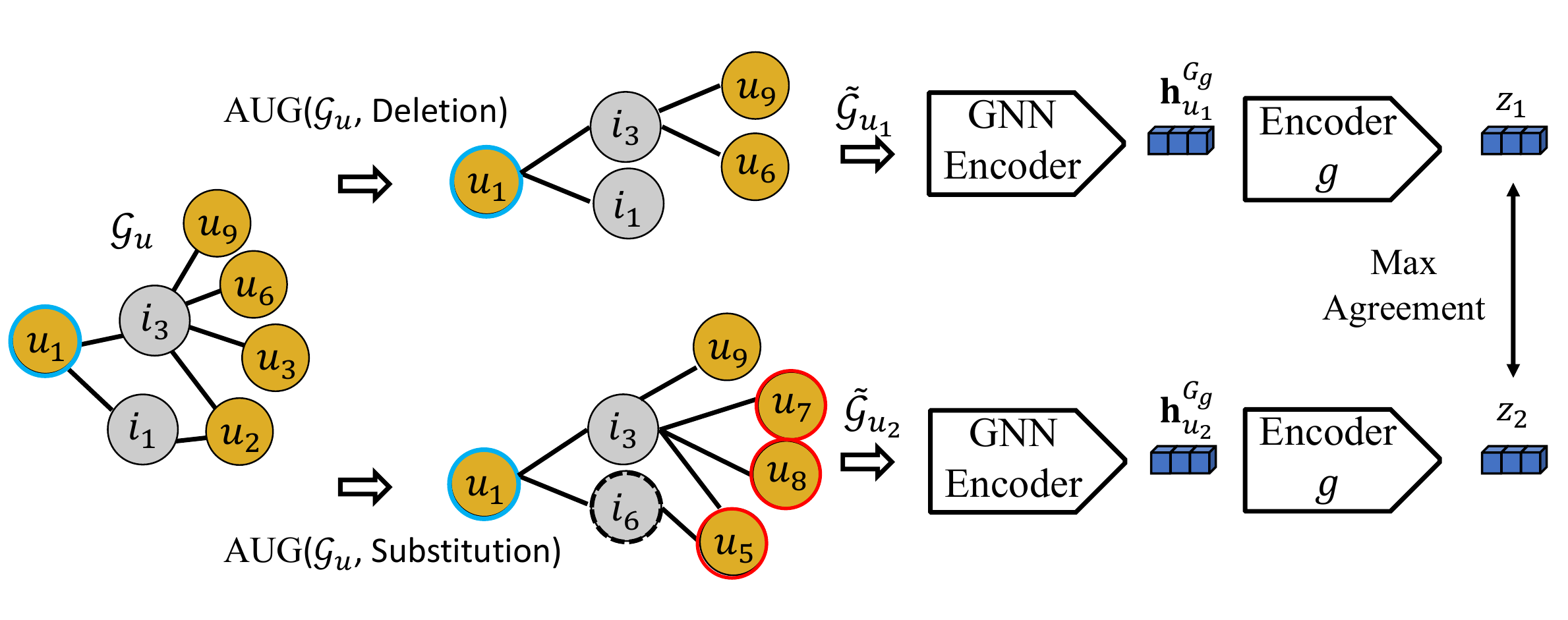}
		}
		
	}
	
	\mbox{ 
		
		\subfigure[\scriptsize  Reconstruction with Transformer Encoder
		]{\label{subfig:MetaPath_Reconstruct}
			\includegraphics[width=0.475\textwidth]{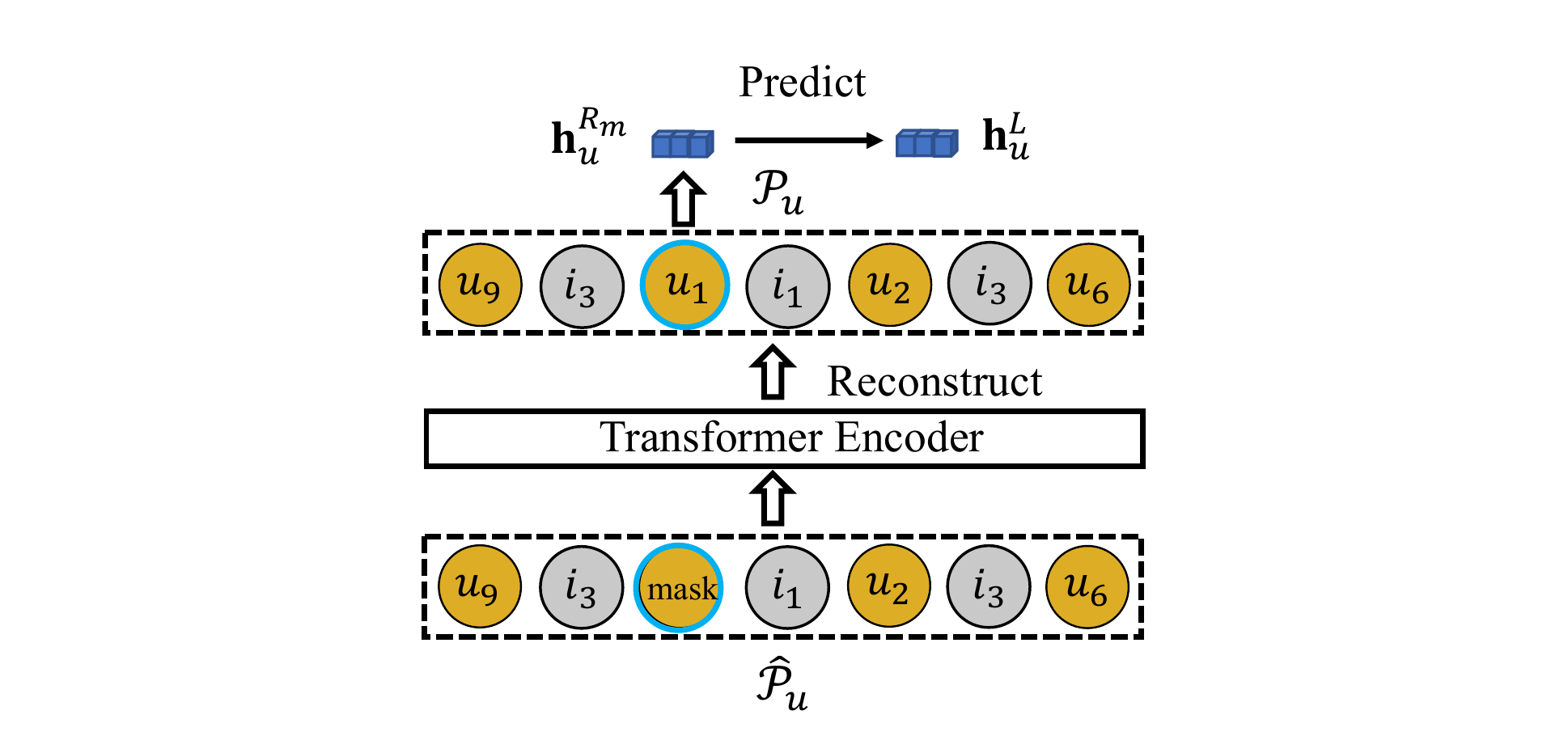}
		}
		
		\hspace{0.2cm}
		
		\subfigure[\scriptsize   Contrastive Learning with Transformer Encoder
		]{\label{subfig:MetaPath_Contrastive}
			\includegraphics[width=0.425\textwidth]{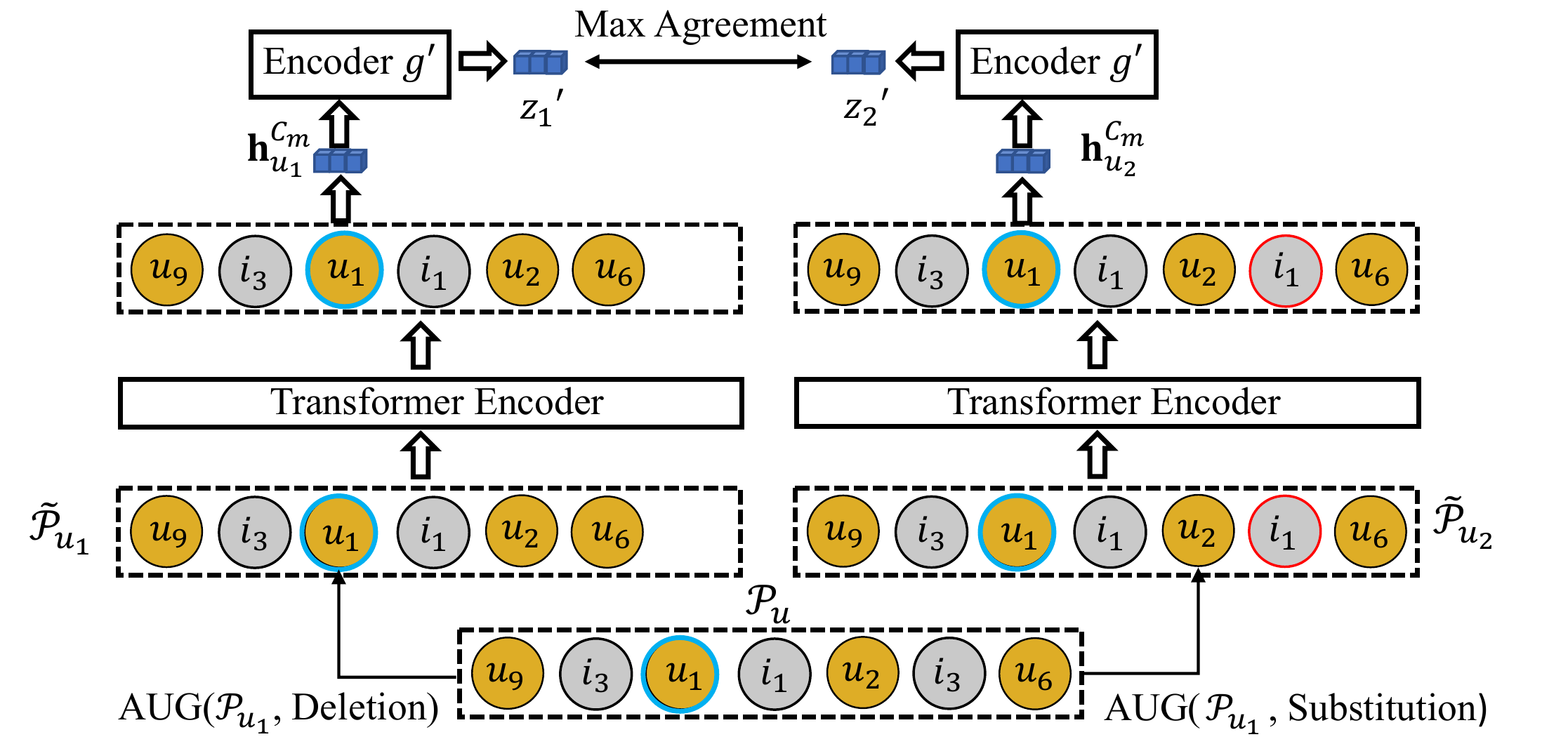}  
		}
	}
	
	\caption{\label{fig:overall} \small{Overview of the proposed pretext tasks.  These pretext tasks are trained independently.} }
\end{figure*}

\section{The Proposed Model MPT }
\label{sec:overall_framework}
We propose a novel  \underline{M}ulti-strategy based \underline{P}re-\underline{T}raining method (\nRC), which extends  \sRC  from the perspective of model architecture and  pretext tasks. 
Specifically, in terms of the model architecture, in addition to using the GNN encoder to capture the short-range dependencies of users and items, we introduce a Transformer encoder to capture long-range dependencies. In terms of the pretext task, in addition to considering the intra-correlations of
users and items by the embedding reconstruction task,
we add embedding contrastive learning task to capture
inter-correlations of users and items. 
Hence, by combining each architecture and each pretext task,  there are four different implementations of pretext tasks, as shown in Fig.~\ref{fig:overall}.  We detail each pretext task implementation  in~\secref{sec:node_level} -~\secref{sec:contrastive_mt}, and then present the overall model training process in~\secref{sec:model_training}.  Finally,  we analyze the time complexity of the pretext task implementation in~\secref{sec:discuss}.

\subsection{ Reconstruction with GNN Encoder}
\label{sec:node_level}
In this pretext task, we primarily follow the original PT-GNN model to reconstruct the user embedding, as proposed in Section~\ref{subsec:pretrain}. We modify PT-GNN in terms of its neighbor sampling strategy to reduce the time complexity.

The adaptive neighbor sampling strategy in \sRC suffers from the  slow convergence issue, as it is essentially a Monte-Carlo based policy gradient strategy (REINFORCE)~\cite{policygradient,williams_policy}, and has the complex action-state trajectories for the entire neighbor sampling process.
To solve this problem,  inspired by the dynamic sampling theory that uses the current model itself to sample instances~\cite{dynamic1,dynamic2}, we propose the dynamic sampling strategy, which samples from the low-order neighbors to the high-order neighbors according to the enhanced embeddings of the target user  and each neighbor. The enhanced embedding for the target user (or each neighbor) is obtained by concatenating the meta embedding produced by the meta aggregator in PT-GNN (Section~\ref{subsec:enhanced_ptgnn}) and the current trained embedding. The meta embedding enables considering  the cold-start characteristics of the neighbors,  and the current trained embedding enables dynamically selecting proper neighbors.
Formally, 
at the $l$-th layer,  the sampling process is:

\beqn{
	\label{eq:dynamic_sampling}
	{a}_{j}^{u} &=& {\rm cos}\ ( \tilde{\textbf{h}}_u \ || \ \textbf{h}_u^l, \tilde{\textbf{h}}_j \ || \ \textbf{h}_{j}^{l} ), \\ \nonumber
	\{ \textbf{h}^{l}_{1}, \cdots, 	\textbf{h}^{l}_{{K^l}} \} & \leftarrow& {\rm S\_TOP}(	{a}_{1}^{u}, \cdots,  	{a}_{j}^{u},   \cdots,	{a}_{| \mathcal{N}^{l}(u) |}^{u} ),  
}

\noindent where $||$ is the concatenate operation, ${a}_{j}^{u} $ is the cosine similarity between the enhanced target user embedding  $\tilde{\textbf{h}}_u \ || \ \textbf{h}_u^l$ and the enhanced $j$-th neighbor embedding $\tilde{\textbf{h}}_j \ || \ \textbf{h}_{j}^{l} $,
$\tilde{\textbf{h}}_u$ and $\tilde{\textbf{h}}_j$ are the meta embeddings produced by the meta aggregator,   $\textbf{h}_u^l$ and  $\textbf{h}_{j}^{l}$ are the current trained user embedding,
$ \{	\textbf{h}^{l}_{1}, \cdots, 	\textbf{h}^{l}_{{K^l}} \} $ is the top $K^l$ relevant neighbors. ${\rm S\_TOP}$ is the operation that selects top neighbor embeddings with top cosine similarity. 
Compared with the adaptive sampling strategy in \RC, the proposed dynamic sampling strategy not only speeds up the model convergence, but also has competitive performance. As it does not need multiple sampling process, and considers the cold-start characteristics of the neighbors during the sampling process.
Notably, the training process of the dynamic sampling strategy is not a fully end-to-end fashion.  In each training epoch, we first select $K^l$ relevant neighbors as the sampled neighbors for each target node according to Eq.~\eqref{eq:dynamic_sampling}, and then we can obtain the adjacency matrix of the user-item graph. Next, we perform  the graph convolution operation to reconstruct the target node embedding to optimize the model parameters. Finally, the updated current node embedding, together with the fixed meta node embedding produced by the meta aggregator are used in the next training epoch, which enables dynamically sampling proper neighbors.

Once the neighbors are sampled, at the $l$-th layer, we use the meta embedding $\tilde{\textbf{h}}_u$, the previous user embedding $\textbf{h}_{u}^{l-1} $ and the  averaged dynamically sampled neighbor embedding $ \tilde{\textbf{h}}^{l}_{ \mathcal{N}(u)}$  to perform graph convolution, as shown in  Eq. (\ref{eq:ptgnn_conv}).
Then we perform graph convolution $L$ steps to obtain the predicted  user embedding $\textbf{h}_{u}^{R_g}$,
and use Eq. (\ref{eq:cosine_similarity}) to optimize the model parameters.  Fig.~\ref{subfig:GNN_Reconstruct} shows this task, and the objective function is as follows:

\beqn{
	\label{eq:objective1}
	\mathcal{L}_{1}: &=&  \mathop{\arg\max}_{\Theta_{1}} \sum_{u} {\rm cos}(\textbf{h}_{u}^{R_g}, \textbf{h}_{u} ).
}

\subsection{ Contrastive Learning with GNN Encoder}
\label{contrastive_gnn}
In this pretext task, we propose to contrast the user embedding produced by the GNN encoder across subgraphs, as shown in Fig.~\ref{subfig:GNN_Contrastive}. 
Similar as \RC, we also train this task under the meta-learning setting to simulate the cold-start scenario. Specifically,  we also select users with 
abundant interactions, randomly select at most $K^l$($1\leq l \leq L$)  neighbors at layer $l$ to form the subgraph $\mathcal{G}_u$. Then we perform contrastive learning (CL) upon $\mathcal{G}_u$ to learn the representations of users. 
In the following part, we first present the four components in the CL framework and then detail the key component, i.e., the data augmentation operation.

\begin{itemize}[ leftmargin=14pt ]
	\item An augmentation module ${\rm AUG(\cdot)}$ augments the user subgraph $\mathcal{G}_u$ by  deleting or replacing the users or items in it, which results in two random augmentations $\tilde{\mathcal{G}}_{u_1}$ = ${\rm AUG}(\mathcal{G}_u, seed_1) $ and $\tilde{\mathcal{G}}_{u_2}$ = ${\rm AUG}(\mathcal{G}_u, seed_2) $, where $seed_1$ and $seed_2$ are two random seeds.
	In this paper, we  evaluate the individual data augmentation operation, i.e., performing either deletion or substitution operation.
	 We	 leave the  compositional data augmentation operation as the future work.
		
	\item A GNN encoder performs the graph convolution $L$ steps using Eq.~\eqref{eq:ptgnn_conv} to generate the representation of the target user for each augmented subgraph.
	 i.e., $\textbf{h}_{u_1}^{C_g} = {\rm GNN}(\tilde{\mathcal{G}}_{u_1})$  and $ \textbf{h}_{u_2}^{C_g} = {\rm GNN}( \tilde{\mathcal{G}}_{u_2})$. 
	
	\item A neural network encoder $g$ maps the encoded augmentations $\textbf{h}_{u_1}^{C_g}$ and $\textbf{h}_{u_2}^{C_g}$ into two vectors $z_1 = g(\textbf{h}_{u_1}^{C_g})$, $z_2 = g(\textbf{h}_{u_2}^{C_g})$. This operation is the same with SimCLR~\cite{simclr}, as its observation that adding a nonlinear projection head can significantly improve the representation quality.
	
	\item A contrastive learning loss module maximizes the agreement between positive augmentation pair $(\tilde{s_1}, \tilde{s_2})$ in the set $\{ \tilde{s}\}$. We construct the set $\{ \tilde{s}\} $ by randomly augmenting twice for all users in a mini-batch $s$ (assuming $s$ is with size $N$), which gets a set $\tilde{s}$ with size $2N$. The two variants from the same original user form the positive pair, while all the other instances from the same mini-batch are regarded as negative samples for them. Then the contrastive loss for a positive pair is defined as:
	
	\beqn{
		\label{eq:contrastive}
		l_c(m,n) = - \log \frac{ \exp ( {\rm cos
			}(z_m, z_n)/ \tau  ) }{\sum_{k=1}^{2N} \mathbbm{1}_{k \neq m} \exp( {\rm cos}(z_m, z_n) / \tau )  },
	}
	
	\noindent where $ \mathbbm{1}_{k \neq m}  $ is the indicator function to judge whether $k \neq m$, $\tau$ is a temperature parameter, and ${\rm cos}(z_m, z_n) = {z_m}^\mathrm{T}z_n/(||z_m||_2||z_n||_2)$  denotes the cosine similarity of two vector $z_m$ and $z_n$.  The overall contrastive loss $\mathcal{L}_2$ defined in a mini-batch is:
	
	\beqn{
		\label{eq:objective2}
		\mathcal{L}_{2}:=  \min_{\Theta_2}   \sum_{m=1}^{2N}\sum_{n=1}^{2N} \mathbbm{1}_{m=n} \ l_c(m,n),
	}
	
	\noindent where $\mathbbm{1}_{m=n}$ is a indicator function returns 1 when $m$ and $n$ is a positive pair, returns 0 otherwise, $\Theta_2$ is the parameters of the CL framework.
	
\end{itemize}

The key method in CL is the data augmentation strategy. In recommender system, since each neighbor may play an essential role in expressing the user profile, it remains unknown whether data augmentation would benefit the representation learning and what kind of data augmentation could be useful. To answer these questions, we explore and test two basic augmentations, \textbf{deletion} and \textbf{substitution}. We believe there exist more potential augmentations, which we will leave for future exploration.

\begin{itemize}[ leftmargin=14pt ]
	\item \textbf{Deletion.}  At each layer $l$, we random select $a$\% users or items and delete them. If a parent user or item is deleted, its child users or items are all deleted. 
	
	\item \textbf{Substitution.} At each layer $l$, we  randomly select  $b$\%  users or items. For each user $u$ or item $i$ in the selected list, we randomly replace $u$ or $i$ with  one of its parent's interacted first-order neighbors.
	Note that in order to illustrate the two data augmentation strategies, we present the deletion and substitution operations in Fig.~\ref{subfig:GNN_Contrastive}, but in practice we perform individual data augmentation strategy, i.e., performing either deletion or substitution operation, and analyze whether the substitution or deletion ratio used for the target user (or target item) can affect the recommendation performance in~\secref{sub:hyper_parameters}.

\end{itemize}

Once the CL framework is trained, we leave out other parts and only maintain the parameters of the trained GNN encoder. When a new cold-start user $u$ comes in, 
same as Section~\ref{sec:node_level}, the GNN encoder performs graph convolution $L$ steps to obtain the enhanced embedding $\textbf{h}_{u}^{C_g}$.

\subsection{ Reconstruction  with Transformer Encoder}
\label{sec:reconstruction_mt}
In this pretext task, we propose using Transformer encoder to reconstruct the embeddings of target users in the user-item path, as shown in Fig.~\ref{subfig:MetaPath_Reconstruct}.
This pretext task is also trained under the meta-learning setting. Specifically, similar to \RC, we  first choose abundant users and use  any recommender algorithms to learn their ground-truth embeddings. Then for each user, we sample  $K^l$($1\leq l \leq L$) neighbors, based on which, we use random walk~\cite{deepwalk} to obtain the user-item path $\mathcal{P}$ = ${\rm IUIU}$ (or $\mathcal{P}$ = ${\rm UIUI}$) with path length $T$. Finally, we  mask the target user in the input path with special tokens “[mask]”, and use Transformer encoder to predict the target user embedding.

Formally, given an original path $\mathcal{P}$=$[x_1,\cdots,x_T]$, where each node $x_i$ in $\mathcal{P}$ represents the initial user embedding $\textbf{h}_{u}^{0}$ or the initial item embedding $\textbf{h}_{i}^{0}$. We first construct a corrupted path $\hat{\mathcal{P}}$ by replacing the target user token
in $\mathcal{P}$ to a special symbol "[mask]" (suppose the target user is at the $t$-th position in $\hat{\mathcal{P}}$). Then we use Transformer encoder to map the input path $\hat{\mathcal{P}}$ into a sequence of hidden vectors ${\rm Tr}(\hat{\mathcal{P}}) = [  {{\rm Tr}(\hat{\mathcal{P}})}_1, \cdots,  {{\rm Tr}(\hat{\mathcal{P}})}_T  ]$. 
Finally, we fetch the $t$-th embedding in ${\rm Tr}(\hat{\mathcal{P}})$ to predict the ground-truth embedding, and use cosine similarity  to optimize the parameters $\Theta_3$ of the Transformer encoder. For simplicity, we use notation $\textbf{h}_{u}^{R_p}$ to represent the $t$-th predicted embedding, i.e., $ {{\rm Tr}(\hat{\mathcal{P}})}_t $  =  $\textbf{h}_{u}^{R_p}$. The objective function is:

\beqn{
	\label{eq:objective3}
	\mathcal{L}_{3}: &=&  \mathop{\arg\max}_{\Theta_{3}} \sum_{u} {\rm cos}(\textbf{h}_{u}^{R_p}, \textbf{h}_{u} ).
}

Once the Transformer encoder is trained, we can use it to predict the cold-start embedding. When a new cold-start user $u$ with his interacted neighbors comes in, we first use random walk to generate the path set $\{\mathcal{P}_1, \cdots, \mathcal{P}_t, \cdots,\mathcal{P}_T  \}$, where in the $t$-th path $\mathcal{P}_t$, $u$ is in the $t$-th position. Then we replace $u$ with the "[mask]" signal to generate the corrupted path $\hat{\mathcal{P}}_t$. Next we feed all the corrupted paths  $\{\hat{\mathcal{P}_1}, \cdots, \hat{\mathcal{P}_T}  \}$ into the Transformer encoder, obtain the predicted user embeddings $\{  \textbf{h}_{u_1}^{R_p}, \cdots, \textbf{h}_{u_T}^{R_p} \}$. Finally, we average these predicted embeddings to obtain the final user embedding $\textbf{h}_{u}^{R_p}$.

\subsection{ Contrastive Learning with Transformer Encoder}
\label{sec:contrastive_mt}
In this task, we propose to contrast the user embedding produced by the Transformer encoder across different paths, as shown in Fig.~\ref{subfig:MetaPath_Contrastive}. 
We train this task under the meta-learning setting. Same as Section~\ref{sec:reconstruction_mt}, we choose abundant users, sample  $K^l$($1\leq l \leq L$) order neighbors, and  use random walk to obtain the path $\mathcal{P}_u$. Then we perform the CL framework to learn the cold-start user embedding:

\begin{itemize}[ leftmargin=14pt ]
	\item An augmentation module ${\rm AUG(\cdot)}$ augments the path $\mathcal{P}_u$=$[x_1,\cdots,x_T]$ by randomly deleting or replacing the users or items in it, i.e.,  $\tilde{\mathcal{P}}_{u_1}$ = ${\rm AUG}(\mathcal{P}_u, seed_1)$ and $\tilde{\mathcal{P}}_{u_2}$ = ${\rm AUG}(T_u, seed_2)$.
	Similar to Section~\ref{contrastive_gnn},  we evaluate the individual  data augmentation operation.

	\item  A Transformer encoder accepts  $\tilde{\mathcal{P}}_{u_1}$, $\tilde{\mathcal{P}}_{u_2}$ as input, and encodes the target user from two augmented paths into  latent vectors, i.e., $\textbf{h}_{u_1}^{C_p}$ = ${\rm Tr}(\tilde{\mathcal{P}}_{u_1})$  and $\textbf{h}_{u_2}^{C_p}$ = ${\rm Tr}( \tilde{\mathcal{P}}_{u_2})$. 
	
	\item A  neural network  encoder $g'$ that maps the encoded augmentations $\textbf{h}_{u_1}^{C_p} $ and $\textbf{h}_{u_2}^{C_p} $ into two vectors $z_{1}' $ = $ g'(\textbf{h}_{u_1}^{C_p}   )$, $z_{2}' $ = $g'(\textbf{h}_{u_2}^{C_p}  )$. 
	
	\item A contrastive learning loss module maximizes the agreement between positive augmentation pair $(\tilde{s_1}, \tilde{s_2})$ in the set $\{ \tilde{s}\}$. 
	Same as Section~\ref{contrastive_gnn}, we also construct the set $\{ \tilde{s}\} $ by randomly augmenting twice for all users in a mini-batch $s$ to get a set $\tilde{s}$ with size $2N$, use the two variants from the same original user as  positive pair,  use all the other instances from the same mini-batch  as negative samples,
	and use Eq.~\eqref{eq:contrastive} as the contrastive loss $l_c(m,n)$ for a positive pair. Similar to Eq.~\eqref{eq:objective2}, the overall contrastive loss $\mathcal{L}_4$ defined in a mini-batch is:

	\beqn{
	\label{eq:objective4}
	\mathcal{L}_{4}:=  \min_{\Theta_4}   \sum_{m=1}^{2N}\sum_{n=1}^{2N}\mathbbm{1}_{m=n} \ l_c(m,n),
}

\noindent where  $\Theta_4$ is the parameters of the CL framework.

\end{itemize}

\noindent The data augmentation strategies are as follows:

\begin{itemize}[ leftmargin=14pt ]
	\item \textbf{Deletion.}  For each path $\mathcal{P}_u$, we random select $a$\% users and items and delete them. 
	
	\item \textbf{Substitution.} For each path $\mathcal{P}_u$, we  randomly select  $b$\%  users and items. 
	For each user $u$ or item $i$ in the selected list, we randomly replace $u$ or $i$ with  one of its parent's interacted first-order neighbors. 
	Note that in order to illustrate the two data augmentation strategies, we present the deletion and substitution operations in Fig.~\ref{subfig:MetaPath_Contrastive}, but in practice we perform individual data augmentation strategy, i.e., performing either deletion or substitution operation, and analyze whether the substitution or deletion ratio used for the target user (or target item) can affect the recommendation performance in~\secref{sub:hyper_parameters}.
	Note that we perform individual data augmentation strategy, i.e., performing either deletion or substitution operation, and analyze whether the substitution or deletion ratio used for the target user (or target item) can affect the recommendation performance in~\secref{sub:hyper_parameters}.

\end{itemize}
Once the CL framework is trained, we leave out other parts and only maintain the parameters of the Transformer encoder. When a new cold-start user $u$ with his interacted neighbors comes in, same as Section~\ref{sec:reconstruction_mt}, we generate the path set $\{\mathcal{P}_1,\cdots,\mathcal{P}_T  \}$, use Transformer encoder to obtain the encoded embeddings  $\{  \textbf{h}_{u_1}^{C_p}, \cdots, \textbf{h}_{u_T}^{C_p} \}$, and average these embeddings to obtain the 
final embedding $\textbf{h}_{u}^{C_p}$.

\subsection{Model Pre-training \&  Fine-tuning Process}
\label{sec:model_training}
We adopt the pre-training and fine-tuning paradigm~\cite{qiu2020gcc} to train the GNN and Transformer encoders.

During the pre-training stage, we independently train each pretext task using the objective functions Eq.~\eqref{eq:objective1}, Eq.~\eqref{eq:objective2}, Eq.~\eqref{eq:objective3} and Eq.~\eqref{eq:objective4} to optimize the parameters $\{ \Theta_{1}, \Theta_{2}, \Theta_{3}, \Theta_{4} \}$. We assign each pretext task an independent set of initialized user and item embeddings, and do not share embeddings for these pretext tasks.
Therefore,  we can train these pretext tasks in a fully  parallel way.

During the fine-tuning process, we initialize the GNN and Transformer encoders with the trained parameters, and fine-tune them via the downstream recommendation task. Specifically, for each target cold-start user and his interacted neighbors $\{  \mathcal{N}^{1}(u), \cdots, \mathcal{N}^{L}(u)   \}$ of each order, we first use the trained GNN and Transformer encoders corresponding to each pretext task to generate the user embedding $\textbf{h}_{u}^{R_g}$, $\textbf{h}_{u}^{C_g}$, $\textbf{h}_{u}^{R_p}$ and  $\textbf{h}_{u}^{C_p}$. Then we concatenate the generated embeddings and transform them into the final user embedding:
	
\beqn{
	\label{eq:fuse}
\textbf{h}_{u}^{*} = \mathbf{W} \cdot (\textbf{h}_{u}^{R_g} || \textbf{h}_{u}^{C_g} || \textbf{h}_{u}^{R_p} || \textbf{h}_{u}^{C_p} ),
}

\noindent where $\Theta_r = \{ \mathbf{W}\}$ is the parameter matrix. We generate the final item embedding $\textbf{h}_{i}^{*}$  in a similar way.
Next we calculate the relevance score as the inner product of user and
item final embeddings, i.e., $y(u, i) = \textbf{h}_{u}^{*^\mathrm{T}} \textbf{h}_{i}^{*}$.
Finally, we use BPR loss defined in Eq.~\eqref{eq:bpr} to optimize $\Theta_r$ and fine-tune $\{ \Theta_{1}, \Theta_{2}, \Theta_{3}, \Theta_{4} \}$.

\subsection{Discussions} 
\label{sec:discuss}
As we can see, the GNN  and Transformer encoders are the main components of the pretext tasks. 
For the GNN encoder, the time complexity  is ${\mathcal O}(T_{GNN} + T_{S} )$, where ${\mathcal O}(T_{GNN})$ represents the time complexity of the layer-wise propagation of the GNN model, and ${\mathcal O}(T_{S})$ represents the time complexity of the sampling strategy.
Since different GNN model has different time complexity ${\mathcal O}(T_{GNN})$, we select classic GNN models and show their ${\mathcal O}(T_{GNN})$.
LightGCN~\cite{xiangnanhe_lightgcn20}: ${\mathcal O}(|\mathcal{R}^{+}| +  T_{S} )$, NGCF~\cite{wangncgf19}: ${\mathcal O}(|\mathcal{R}^{+} | \ d^2 +  T_{S} )$, GCMC~\cite{BRDTKDD2018}:${\mathcal O}(|\mathcal{R}^{+}| \ d^2 +  T_{S} )$, GraphSAGE~\cite{williamgraphsage17}: ${\mathcal O}( |\mathcal{N}^{+}| + T_{S})$, where $| \mathcal{R}^{+}| $ denotes  the number of nonzero entries in the Laplacian matrix,   $|\mathcal{N}^{+}|$ is the number of totally sampled instances and $d$ is the embedding size.
We present the time complexity of dynamic sampling strategy ${\mathcal O}(T_{S})$ = ${\mathcal O}(E*d*|\mathcal{N}^{+}|)$ and the adaptive sampling strategy ${\mathcal O}(T_{S})$ = ${\mathcal O}(E*M*d*|\mathcal{N}^{+}|)$,  where $E$ is the number of convergent epochs, $M$ is the sampling times. Compared with adaptive sampling strategy, the dynamic strategy does not need multiple sampling time $M$, and has fewer convergent epochs $E$, thus has smaller time complexity.
For the Transformer encoder, the time complexity is ${\mathcal O}(T^2*d)$, where $T$ is the length of the user-item path.

\section{Experiments}
\label{sec:experiment}
Following PT-GNN~\cite{haopretrain21}, we conduct intrinsic evaluation task to evaluate the quality of user/item embeddings, and extrinsic task to evaluate the cold-start recommendation performance.
We answer the following research questions:

\begin{itemize}[ leftmargin=10pt ]
\item \textbf{RQ1:} How does \snRC perform embedding inference and cold-start recommendation compared with the state-of-the-art GNN and pre-training GNN models?

\item \textbf{RQ2:} What are the benefits of performing pretext tasks in both intrinsic and extrinsic evaluation? 

\item \textbf{RQ3:} How do different settings influence the effectiveness of the proposed \snRC model?
\end{itemize}

\subsection{Experimental Settings}

\subsubsection{Datasets}

We select on three public datasets MovieLens-1M (Ml-1M)\footnote{https://grouplens.org/datasets/movielens/}~\cite{harper2016movielens}, MOOCs\footnote{http://moocdata.cn/data/course-recommendation}~\cite{zhanghrl19} and Gowalla\footnote{https://snap.stanford.edu/data/loc-gowalla.html}~\cite{gowalla}. Table~\ref{tb:statistics} illustrates the statistics of these datasets. 

\begin{table}[t]
	\newcolumntype{?}{!{\vrule width 1pt}}
	\newcolumntype{C}{>{\centering\arraybackslash}p{4.6em}}
	\caption{
		\label{tb:statistics} Statistics of the Datasets.
		\normalsize
	}
	
	\centering  \small
	\renewcommand\arraystretch{1.3}
	
	\begin{tabular}{@{~}l@{~} @{~}r@{~} @{~}r@{~} @{~}r@{~} @{~}r@{~} }
		\toprule
		\multirow{2}{*}{\vspace{+0.36cm} Dataset}
		&\#Users&\#Items&\#Interactions & \#Sparse Ratio  
		\\
		\midrule
		MovieLens-1M
		&6,040 &3,706&1,000,209& 4.47\% 
		\\
		MOOCs
		& 82,535 & 1,302 & 458,453 	& 0.42\% 
		\\
		Gowalla
		&29,858&40,981&1,027,370& 0.08\%
		\\
		\bottomrule
	\end{tabular}
	
\end{table}

\subsubsection{Comparison Methods}
We select three types of baselines, including the neural collaborative filtering model, the GNN models and the self-supervised graph learning models.

\begin{itemize}[ leftmargin=10pt ]

\item \textbf{NCF~\cite{he2017neural}}: is a neural collaborative filtering model which uses multi-layer perceptron and matrix factorization to learn the representations of users/items.

\item \textbf{PinSage~\cite{pinsage}}:  employs the GraphSAGE~\cite{williamgraphsage17} model, which samples neighbors randomly and aggregates them by the AVERAGE function.

\item \textbf{GCMC~\cite{BRDTKDD2018}}:  employs the standard GCN~\cite{Thomasgcn} model  to learn the embeddings of users and items.

\item \textbf{NGCF~\cite{wangncgf19}}: primarily follows the neural passing based GNN model~\cite{GSSPPMLR17}, but additionally adds second-order interactions during the message passing process.

\item \textbf{LightGCN~\cite{xiangnanhe_lightgcn20}}:  discards the feature transformation and  nonlinear activation functions in NGCF.

\item \textbf{SGL~\cite{WWFHSIGIR2021}}: contrasts the node representation within the graphs from multiple views, where node dropout, edge dropout and random walk are adopted to generate these views. We find edge dropout has the best performance.

\item \textbf{PT-GNN~\cite{haopretrain21}}: takes the embedding reconstruction as the pretext task to explicitly improve the embedding quality.

\end{itemize}

For each vanilla GNN model (e.g., GCMC, NGCF), notation ${ \rm GNN}^{+}$ means we apply GNN in \RC, and notation ${\rm GNN}^{*}$ means we apply GNN in \nRC. Since SGL adopts multi-task learning paradigm for recommendation, for fair comparison, we compare it in the extrinsic recommendation task and use notation ${\rm GNN}$-${\rm SGL}$ to denote it.

\subsubsection{Intrinsic and Extrinsic Settings.} 
Following \RC~\cite{haopretrain21}, we  divide each dataset into the meta-training set $D_T$ and  the meta-test set $D_N$. We train and evaluate the proposed \snRC model in the intrinsic user/item embedding inference task on $D_T$. Once the  the proposed model \snRC is trained, we fine-tune it in the extrinsic  task and evaluate its performance on $D_N$. 
We select the users/items from each dataset with sufficient interactions as the target users/items in $D_T$, as the intrinsic evaluation needs the ground-truth embeddings of users/items inferred from the sufficient interactions. 
In the cold-start user scenario, we divide the users with the number of the direct interacted items (first-order neighbors) more than $n_i$ into $D_T$ and leave the rest users into $D_N$.  We set $n_i$ as 25, 10 and 25 for the dataset Ml-1M, MOOCs and Gowalla, respectively. 
Similarly, in the cold-start item scenario, we divide the items with the number of the direct interacted users (first-order neighbors) more than $n_u$ into $D_T$ and leave the rest items into  $D_N$, where we set $n_u$ as 15, 10 and 15 for Ml-1M, MOOCs and Gowalla, respectively. We set $K$ as 3 and 8 in the intrinsic task, and 8 in the extrinsic task.
By default, we set $d$ as 32, the learning rate as 0.003, $L$ as 4, $T$ as 6, $a$ as 0.2, $b$ as 0.2 and $\tau$ as 0.2. 

\subsubsection{Intrinsic Evaluations: Embedding Inference}
We conduct the intrinsic evaluation task, which aims to infer the  embeddings of  cold-start users and items by the proposed \snRC model. Both the evaluations
on  user embedding inference and  item embedding
inference are performed.

\vpara{Training and Test Settings.} Following \RC~\cite{haopretrain21}, we perform intrinsic evaluation on $D_T$.
Specifically, same as \RC, we train NCF~\cite{he2017neural} to get the ground-truth embeddings for the target users/items in $D_T$. We also explore whether \snRC is sensitive to the ground-truth embedding generated by other models such as LightGCN~\cite{xiangnanhe_lightgcn20}.
We randomly split $D_T$ into the training set $Train_T$ and the test set $Test_T$ with a ratio of 7:3. 
To mimic the cold-start users/items on $Test_T$, we randomly keep $K$ neighbors for each user/item, which results in at most $K^l$ neighbors ($1 \leq l \leq L$) for each target user/item. Thus $Test_T$ is changed into $Test_T'$.

We train NCF  transductively  on the merged dataset $Train_T$ and $Test_T'$. 
We train the vanilla GNN models  by BPR loss~\cite{xiangnanhe_lightgcn20}  on $Train_T$. 
We train \sRC on $Train_T$, where we first perform Eq.~\eqref{eq:ptgnn_conv} $L$-1 steps to obtain the refined first-order neighbors, and then average them to reconstruct the target embedding; finally we use  Eq.~\eqref{eq:cosine_similarity} to measure the quality of the predicted embedding.
We train \snRC on $Train_T$, where we first perform four pretext tasks by  Eq.~\eqref{eq:objective1}, Eq.~\eqref{eq:objective2}, Eq.~\eqref{eq:objective3} and Eq.~\eqref{eq:objective4}, and then use  Eq.~\eqref{eq:fuse} to fuse the generated embeddings; finally we use Eq.~\eqref{eq:cosine_similarity} to measure the quality of the fused embedding.
Note that the embeddings in all the models are randomly initialized.
We use cosine similarity to measure the agreement between the  ground-truth and  predicted embeddings,  due to its popularity as an
indicator for the semantic similarity between embeddings.


\begin{table*}[t]
	\newcolumntype{?}{!{\vrule width 1pt}}
	\newcolumntype{C}{>{\centering\arraybackslash}p{3.2em}}
	\caption{
		\label{tb:intrinsic_overall} \small{Overall performance of user/item embedding inference with cosine similarity. $L$=4. }
		\normalsize
	}
	\centering  \scriptsize
	\renewcommand\arraystretch{1.0}
	\begin{tabular}{@{~}l@{~}?*{1}{CC?}*{1}{CC?}*{1}{CC?}*{1}{CC?}  *{1}{CC?}*{1}{CC} }
		\toprule
		\multirow{2}{*}{\vspace{-0.3cm} Methods }
		&\multicolumn{2}{c?}{Ml-1M (user)}
		&\multicolumn{2}{c?}{MOOCs (user)}
		&\multicolumn{2}{c?}{Gowalla(user)}
		&\multicolumn{2}{c?}{Ml-1M (item)}
		&\multicolumn{2}{c?}{MOOCs (item)}
		&\multicolumn{2}{c}{Gowalla(item)}
		\\
		\cmidrule{2-3} \cmidrule{4-5} \cmidrule{6-7} \cmidrule{8-9} \cmidrule{10-11} \cmidrule{12-13} 
		& {3-shot} & {8-shot} & {3-shot} & {8-shot} &{3-shot} &{8-shot}   &{3-shot}&{8-shot}  &{3-shot}&{8-shot} &{3-shot}&{8-shot} \\
		\midrule 
		${\rm NCF}$
		&0.490 &0.533 
		&0.451 &0.469 
		&0.521 &0.558
		&0.436 &0.496
		&0.482 &0.513
		&0.482 &0.491 		
		\\	
		\midrule
		${\rm PinSage}$
		&0.518 &0.557 
		&0.542 &0.564 
		&0.552 &0.567 
		&0.558 &0.573 
		&0.558 &0.591 
		&0.556 &0.599 
		\\
		${\rm PinSage}^{+}$
	&0.659 &0.689 
	&0.651 &0.669 
	&0.663 &0.692 
	&0.734 &0.745 
	&0.658 &0.668 
	&0.668 &0.676
		\\
		${\rm PinSage}^{*}$
		&\textbf{0.690}&\textbf{ 0.698 }
		&\textbf{0.668}&\textbf{ 0.673 }
		&\textbf{0.687}&\textbf{0.695}
		&\textbf{0.741}&\textbf{ 0.748 }
		&\textbf{0.661}&\textbf{0.669 }
		&\textbf{0.673}&\textbf{0.683 }
		\\
		\midrule
		${\rm GCMC}$
		&0.513 &0.534
		&0.546 &0.569 
		&0.546 &0.562
		&0.559 &0.566
		&0.554 &0.558
		&0.553 &0.557 
		\\
		${\rm GCMC}^{+}$
		&0.600 &0.625
		&0.750 &0.765
		&0.716 &0.734 
		&0.725 &0.733 
		&0.607 &0.615 
		&0.650 &0.667
		\\
		${\rm GCMC}^{*}$
		&\textbf{0.629}&\textbf{ 0.632}
		&\textbf{0.782}&\textbf{ 0.790 }
		&\textbf{0.741}&\textbf{ 0.748} 
		&\textbf{0.744}&\textbf{0.764}
		&\textbf{0.608}&\textbf{ 0.616}
		&\textbf{0.675}&\textbf{ 0.681}
		\\
		\midrule
		${\rm NGCF}$
		&0.534 &0.554
		&0.531 &0.547
		&0.541 &0.557 
		&0.511 &0.516 
		&0.503 &0.509 
		&0.503 &0.506 
		\\
		${\rm NGCF}^{+}$
		&0.591 &0.618 
		&0.541 &0.573 
		&0.552 &0.574 
		&0.584 &0.596 
		&0.549 &0.560 
		&0.576 &0.591 
		\\
		${\rm NGCF}^{*}$
		&\textbf{0.645}&\textbf{ 0.650}
		&\textbf{0.598}&\textbf{0.601 }
		&\textbf{0.582}&\textbf{ 0.589} 
		&\textbf{0.609}&\textbf{ 0.613}
		&\textbf{0.606}&\textbf{ 0.611}
		&\textbf{0.590}&\textbf{ 0.594}
		\\
		\midrule
		${\rm LightGCN}$
		&0.549 &0.569 
		&0.530 &0.534 
		&0.581 &0.592 
		&0.600 &0.631 
		&0.590 &0.616 
		&0.606 &0.622 
		\\
		${\rm LightGCN}^{+}$
		&0.636 &0.644 
		&0.646 &0.654 
		&0.614 &0.617 
		&0.691 &0.701 
		&0.667 &0.676 
		&0.693 &0.701 
		\\
		${\rm LightGCN}^{*}$
		&\textbf{0.641}& \textbf{ 0.647}
		&\textbf{0.652}& \textbf{ 0.657}
		&\textbf{0.695}&\textbf{ 0.700}
		&\textbf{0.699}& \textbf{0.704}
		&\textbf{0.669}&\textbf{ 0.691}
		&\textbf{0.768}&\textbf{ 0.774}
		\\
		
		\bottomrule
	\end{tabular}
	
\end{table*}

\subsubsection{Extrinsic Evaluation: Recommendation}

We apply the pre-training GNN model into the downstream recommendation task and evaluate its  performance. 

\vpara{Training and Testing Settings.}
We consider the scenario of the cold-start users and use the meta-test set $D_N$ to perform recommendation. For each user in $D_N$, we select top $c$\% ($c$\%=0.2 by default) of his interacted items in chronological order into the training set $Train_N$, and leave the rest items into the test set $Test_N$. 
We pre-train our model on $D_T$ and fine-tune it on $Train_N$ according to \secref{sec:model_training}. 

The vanilla GNN and the NCF models are trained by the BPR loss on $D_T$ and $Train_N$. For each user in $Test_N$, we calculate his relevance score to each of the rest 1-$c$\% items. 
We adopt Recall@$\mathcal{K}$ and NDCG@$\mathcal{K}$ as the metrics to evaluate the items ranked by the relevance scores. By default, we set $\mathcal{K}$ as 20 for Ml-1M, MOOCs and Gowalla. 
\hide{In our implementation, the neighbor size $K$ is set as 8,  the embedding size $d$ is set as 256 and the learning rate for the recommendation model is set as 0.003. }
\section{Experimental Results}

\subsection{Performance Comparison (RQ1)}

\subsubsection{Overall  Performance Comparison}

We report the overall performance of  intrinsic embedding inference  and extrinsic recommendation tasks in Table~\ref{tb:intrinsic_overall} and Table~\ref{tb:recommendation}. We find that:

\begin{itemize}[ leftmargin=10pt ]
	\item \snRC (denoted as ${\rm GNN}^{*}$)  is better than NCF and the vanilla GNN model, which indicates the effectiveness of the pre-training strategy. Compared with the pre-training model \sRC(denoted as ${\rm GNN}^{+}$) and SGL (denoted as ${\rm GNN}$-${\rm SGL}$), \snRC also performs better than them. This  indicates the superiority of simultaneously considering the intra- and inter- correlations of nodes as well as the short- and long-range dependencies of nodes.
	
	\item  In Table~\ref{tb:intrinsic_overall}, when $K$ decreases from 8 to 3, the performance of all the baseline methods drops by a large scale, while \snRC drops a little. This indicates \snRC is able to learn high-quality embeddings for users or items that have extremely sparse interactions.
	
	\item In the intrinsic task (Cf. Table~\ref{tb:intrinsic_overall}), we notice that most models perform generally better on the item embedding inference task than the user embedding inference task. The reason is that the users in $D_T$ have more interacted neighbors than the items in $D_T$,
	thus the learned ground-truth user embedding shows much more diversity than the learned ground-truth item embedding, and only use $K^l$ (K is small) neighbors to learn the user embedding is more difficult than learning the item embedding.
	To validate our point of view, we calculate the average number of  interacted neighbors of users and items in $D_T$ among these three datasets. In ML-1M, MOOCs and Gowalla, when the layer size $L = 1$, 
	the users interact with an average of 178.26, 134.90 and 178.27 items, while the items interact with an average of 32.14, 15.75 and 32.14 users. When the layer size $L$ gets larger, the interacted neighbors' number of users and items show the same trend. 
	The results indicate the abundant users have much more interacted neighbors than the abundant items,
	which verifies the above analysis.
	
	\item Aligning Table~\ref{tb:intrinsic_overall} and Table~\ref{tb:recommendation}, we notice  that using  GCMC, LigtGCN as the backbone GNN model can always obtain better performance than using PinSage and NGCF. The possible reason is that compared with PinSage and NGCF, GCMC and LightGCN have discarded the complex feature transformation operation, which will contribute to learn high-quality embeddings and better recommendation performance. This phenomena is consistent with LightGCN's~\cite{xiangnanhe_lightgcn20} findings. 

\end{itemize}

\begin{table}[t]
	\newcolumntype{?}{!{\vrule width 1.2pt}}
	\newcolumntype{C}{>{\centering\arraybackslash}p{6.8em}}
	\caption{
		\label{tb:recommendation} \small{Overall recommendation performance with  sparse rate $c$\%=20\%, $L$=4. }
		\normalsize
	}
	\centering  \scriptsize
	\renewcommand\arraystretch{1.0}
	\begin{tabular}{@{~}l@{~}?*{1}{CC?}*{1}{CC?}*{1}{CC}}
		\toprule
		\multirow{2}{*}{\vspace{-0.3cm} Methods }
		&\multicolumn{2}{c?}{Ml-1M}
		&\multicolumn{2}{c?}{MOOCs}
		&\multicolumn{2}{c}{Gowalla}
		\\
		\cmidrule{2-3} \cmidrule{4-5} \cmidrule{6-7}
		& {Recall} & {NDCG} & {Recall} & {NDCG} &{Recall} &{NDCG} \\
		\midrule 
		${\rm NCF}$
		&0.004&0.009 
		&0.132 & 0.087 
		&0.039 & 0.013
		\\
		\midrule
		${\rm PinSage}$
		&0.006 &0.012
		&0.085 &0.066
		&0.003 &0.011
		\\
		${\rm PinSage}$-${\rm SGL}$
		&0.040 &0.033
		&0.172 &0.086
		&0.011 &0.021
		\\
		${\rm PinSage}^{+}$
		&0.047&\textbf{0.038}
		&0.150&0.089
		&0.008&0.019
		\\
		${\rm PinSage}^{*}$
		&\textbf{0.054}&0.036
		&\textbf{0.182}&\textbf{0.099}
		&\textbf{0.045}&\textbf{0.021}
		\\
		\midrule
		${\rm GCMC}$
		&0.008&0.006 
		&0.232&0.172
		&0.006&0.033
		\\
		${\rm GCMC}$-${\rm SGL}$
		&0.038 &0.037
		&0.239 &0.177
		&0.033 &0.035
		\\
		${\rm GCMC}^{+}$
		&0.044&0.041
		&0.248&0.187
		&0.018&0.032
		\\
		${\rm GCMC}^{*}$
		&\textbf{0.071}&\textbf{0.076}
		&\textbf{0.272}&\textbf{0.212}
		&\textbf{0.065}&\textbf{0.043}
		\\
		\midrule
		${\rm NGCF}$
		&0.052&0.058
		&0.208&0.171
		&0.009&0.013
		\\
		${\rm NGCF}$-${\rm SGL}$
		&0.062 &\textbf{0.079}
		&0.216 &0.177
		&0.028 &0.014
		\\
		${\rm NGCF}^{+}$
		&0.064&0.071
		&0.217&0.181
		&0.037&0.014
		\\
		${\rm NGCF}^{*}$
		&\textbf{0.083}&0.077
		&\textbf{0.241}&\textbf{0.208}
		&\textbf{0.047}&\textbf{0.016}
		\\
		\midrule
		${\rm LightGCN}$
		&0.058&0.064
		&0.217&0.169
		&0.011&0.023
		\\
		${\rm LightGCN}$-${\rm SGL}$
		&0.066 &0.077
		&0.275 &0.178
		&0.034 &0.028
		\\
		${\rm LightGCN}^{+}$
		&0.078& 0.071
		&0.308&0.184
		&0.049&0.031
		\\
		${\rm LightGCN}^{*}$
		&\textbf{0.094}&\textbf{0.101}
		&\textbf{0.346}&\textbf{0.223}
		&\textbf{0.051}&\textbf{0.036}
		\\
		\bottomrule
	\end{tabular}
	
\end{table}

	\begin{figure}[t]
		\centering
		\includegraphics[width= 0.92 \textwidth]{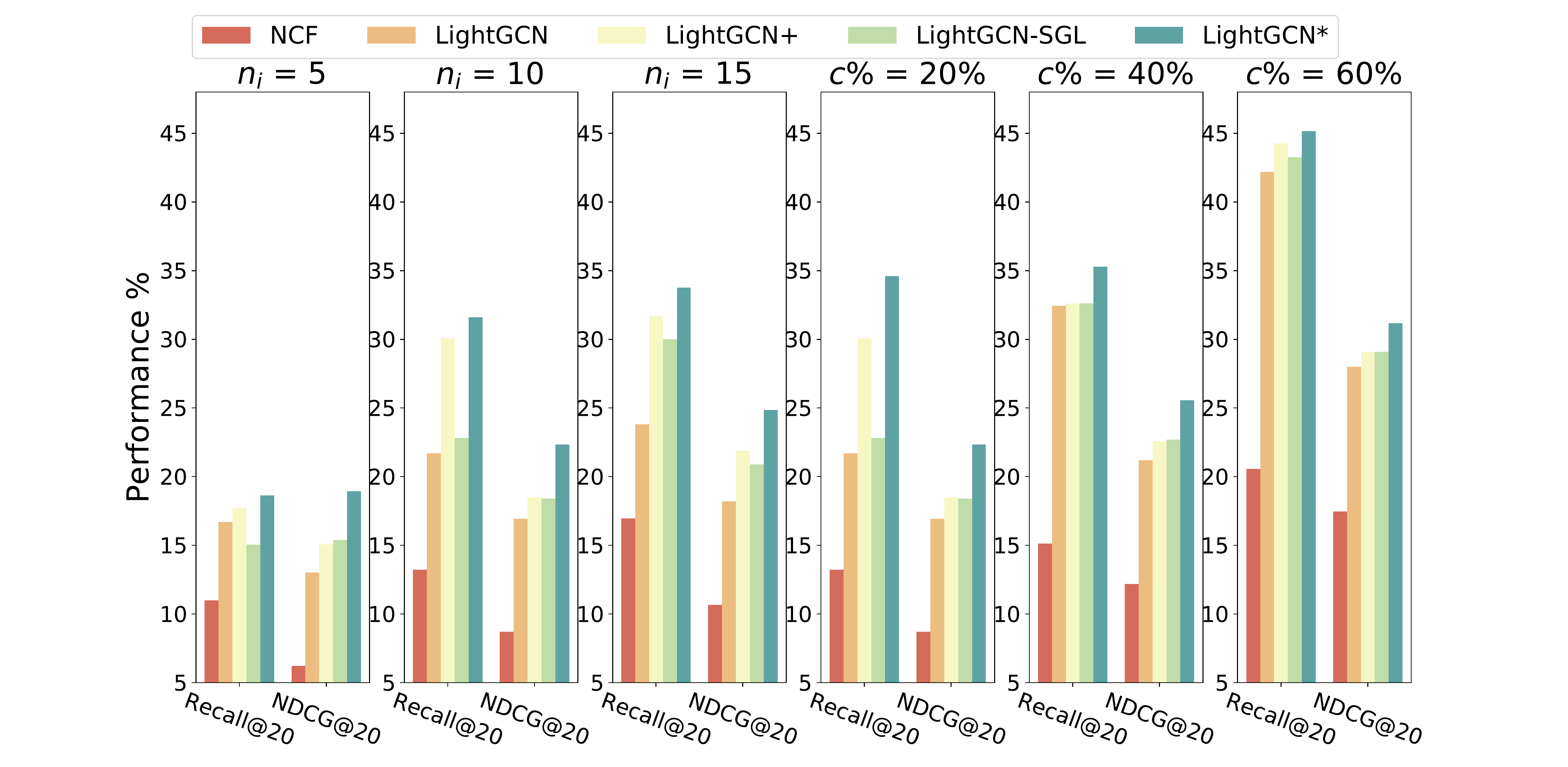}
		\caption{\label{fig:cold_start_recommendation}   \small{Cold-start recommendation under different $n_i$ and $c$\%. } }
	\end{figure}

\subsubsection{Interacted Number and Sparse Rate Analysis}
It is still unclear how does \snRC handle the cold-start users with different interacted items $n_i$ and sparse rate $c$\%. 
To this end, we change the interaction number $n_i$ in the range of $\{5, 10, 15  \}$ and the sparse rate $c$\% in the range of $\{ 20\%,40\%,60\% \}$, select LightGCN as the backbone GNN model  
and  report the cold-start recommendation performance on MOOCs dataset in Fig.~\ref{fig:cold_start_recommendation}. The smaller $n_i$ and $c$\% are, the cold-start users in $D_N$ have fewer interactions. 
We find that: 
\begin{itemize}[ leftmargin=10pt ]
	\item  \snRC (denoted as ${ \rm LightGCN}^{*}$) is consistently superior to all the other baselines, which  justifies the superiority  of \snRC in handling cold-start recommendation with different $n_i$ and $c$\%.

	\item  When $n_i$ decreases from 15 to 5, \snRC has a larger improvement compared with other baselines, which verifies its capability to solve the cold-start users with extremely sparse interactions. 
	
	\item When $c$\% decreases from 60\% to 20\%, \snRC has a larger improvement compared with other baselines, which again verifies the superiority of \snRC in handling cold-start users with different sparse rate.

\end{itemize}

\subsection{Ablation Study (RQ2)}

\begin{figure*}[t]
	\centering

	\mbox{ 
		\subfigure[\scriptsize Cosine similarity  (user) \%
		]{\label{subfig:insic1}
			\includegraphics[width=0.30\textwidth]{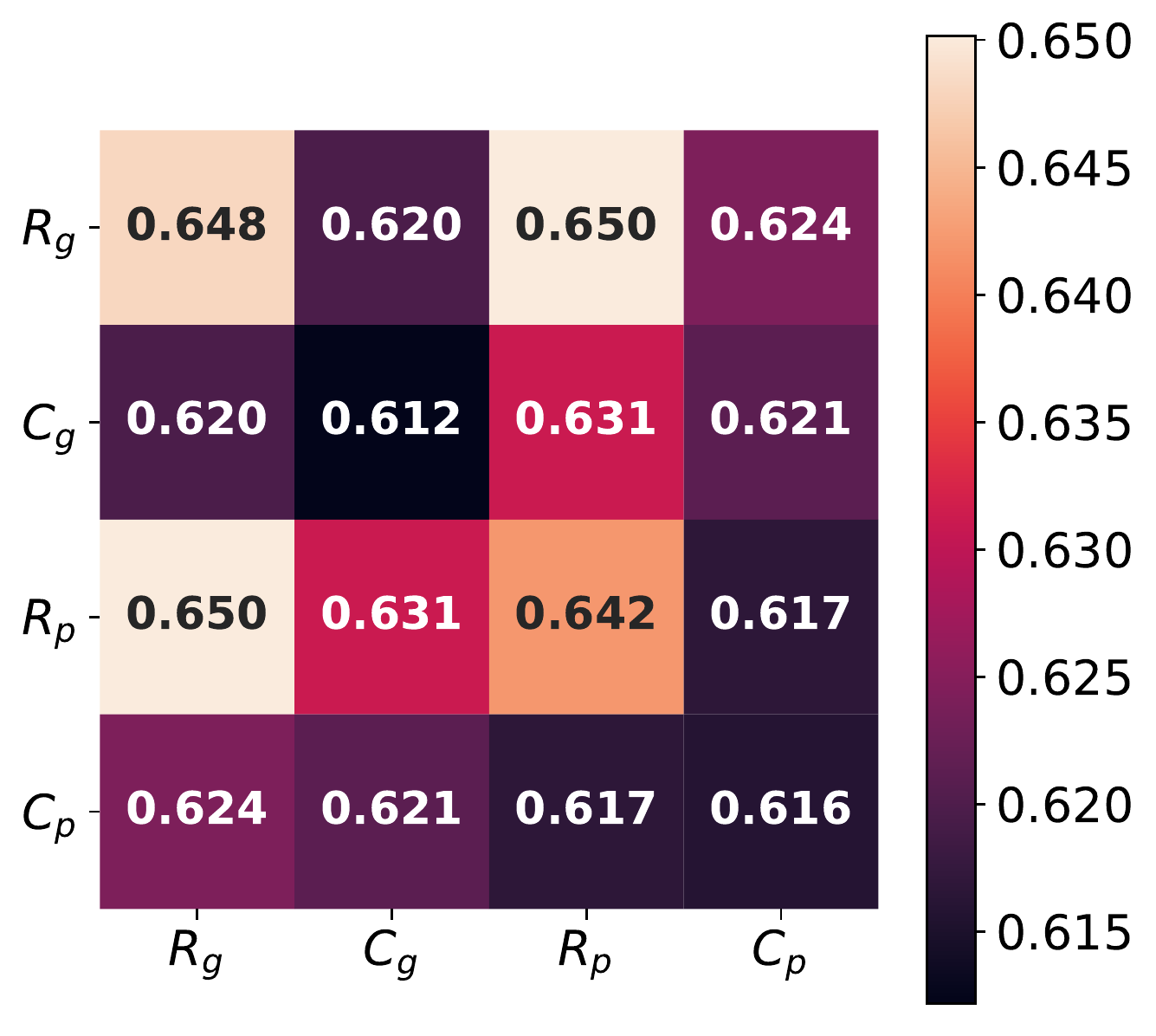}
		}

		\subfigure[\scriptsize Cosine similarity (item)  \%
		]{\label{subfig:insic2}
			\includegraphics[width=0.30\textwidth]{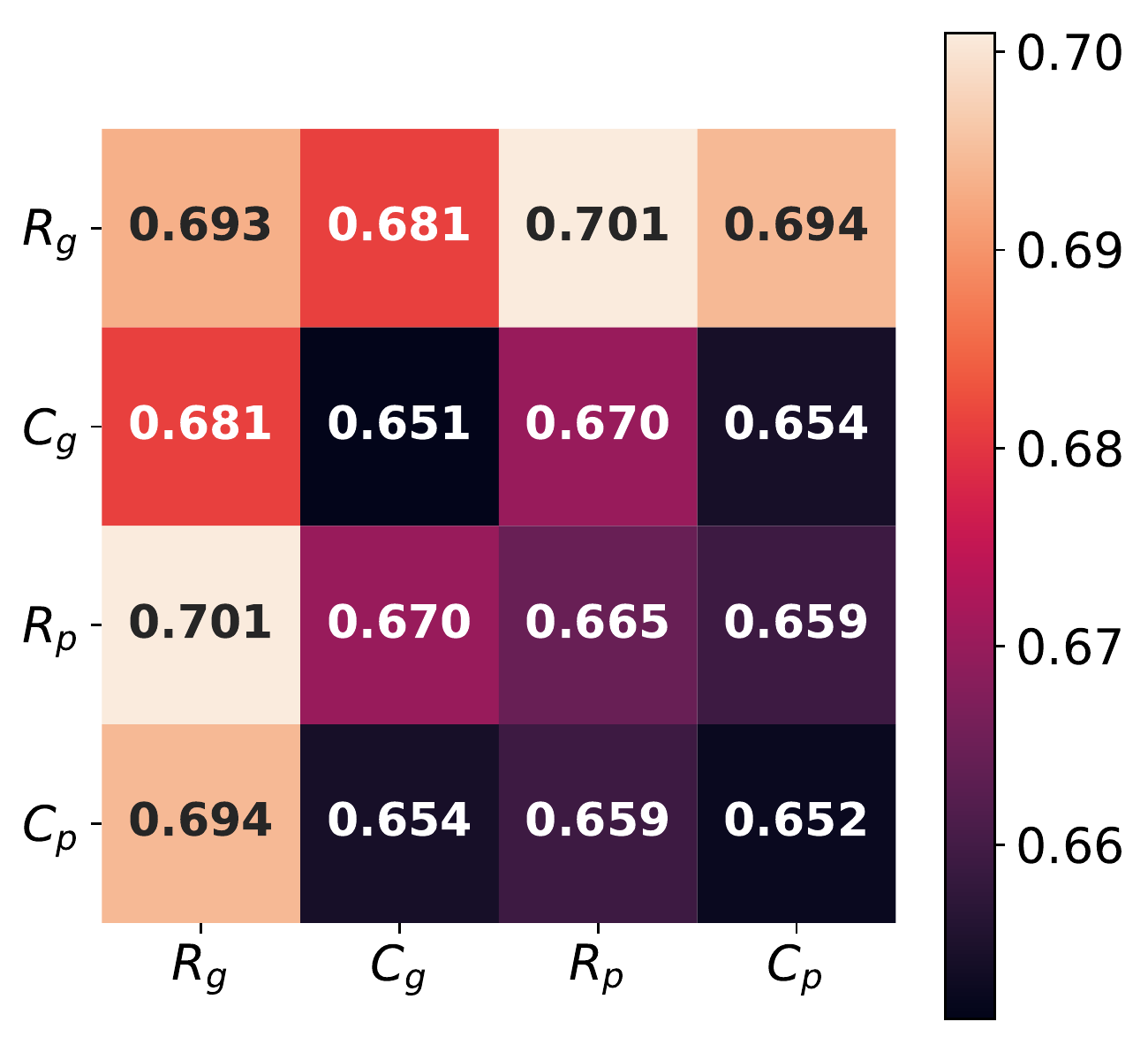}
		}
		
		\subfigure[\scriptsize  Cosine similarity
		]{\label{subfig:insic_pretext}
			\includegraphics[width=0.30\textwidth]{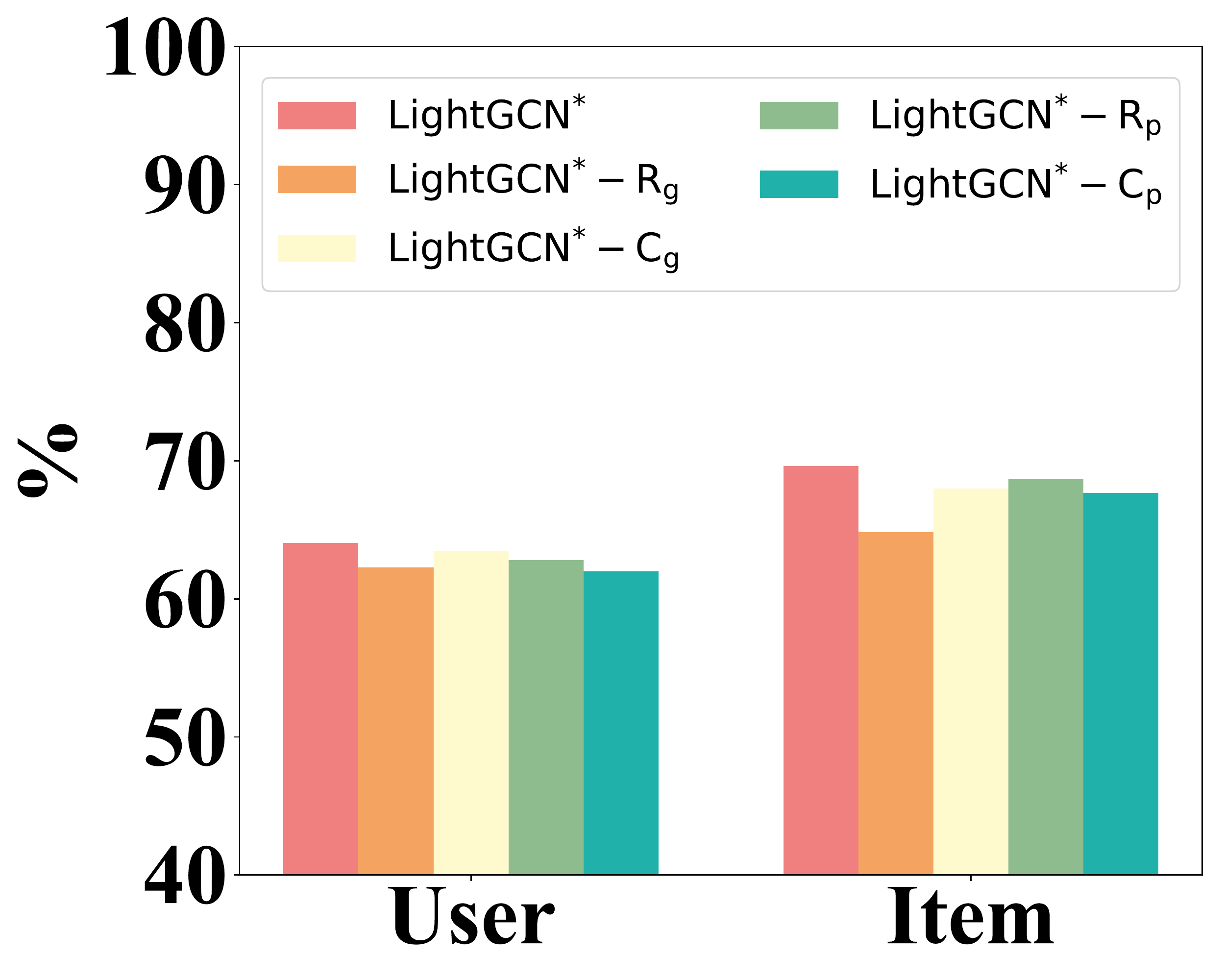}
		}
	}
	
	\mbox{
		
		\subfigure[\scriptsize Recall@20
		]{\label{subfig:rec1}
			\includegraphics[width=0.30\textwidth]{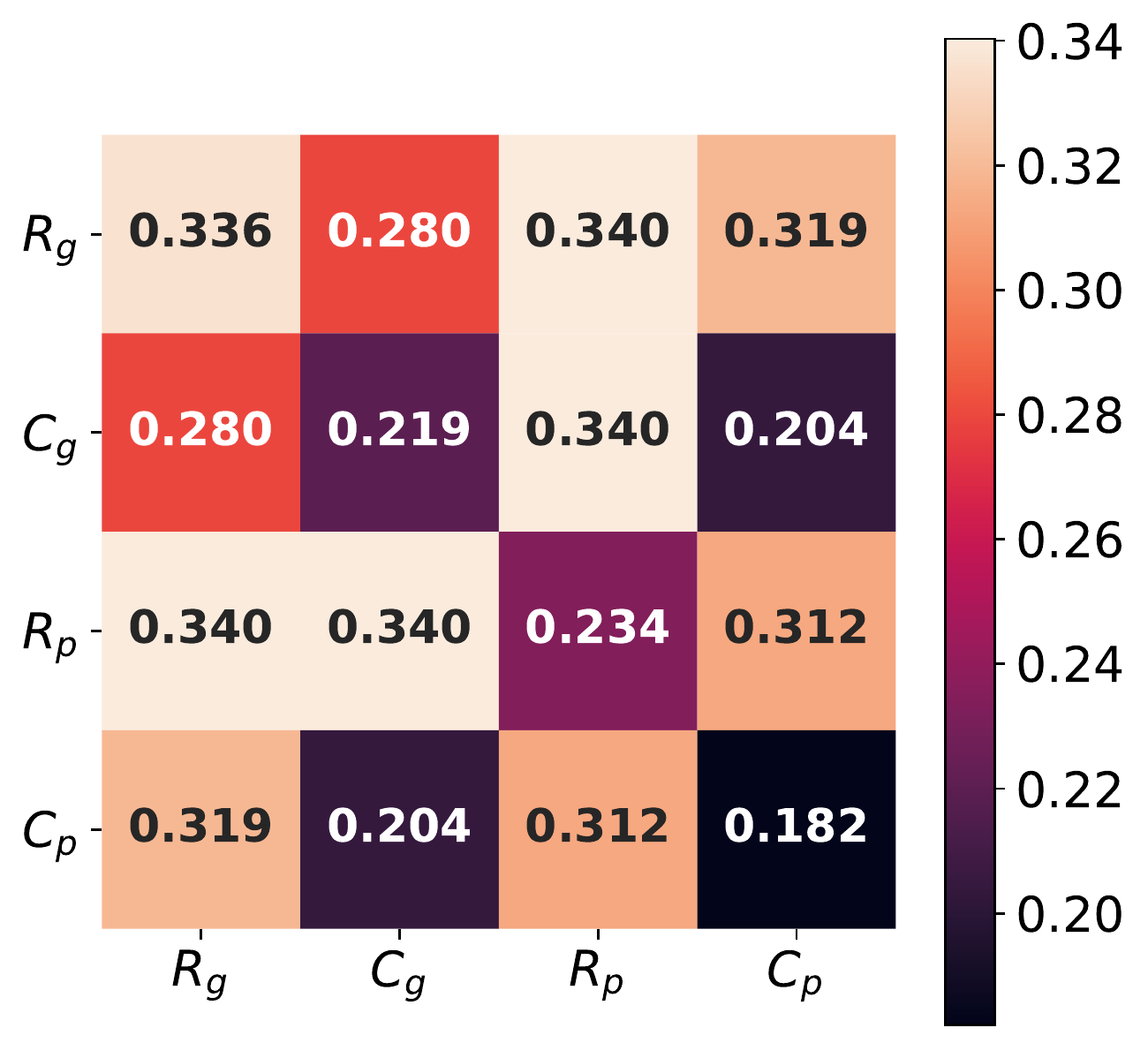}
		}

		\subfigure[\scriptsize  NDCG@20  
		]{\label{subfig:rec2}
			\includegraphics[width=0.30\textwidth]{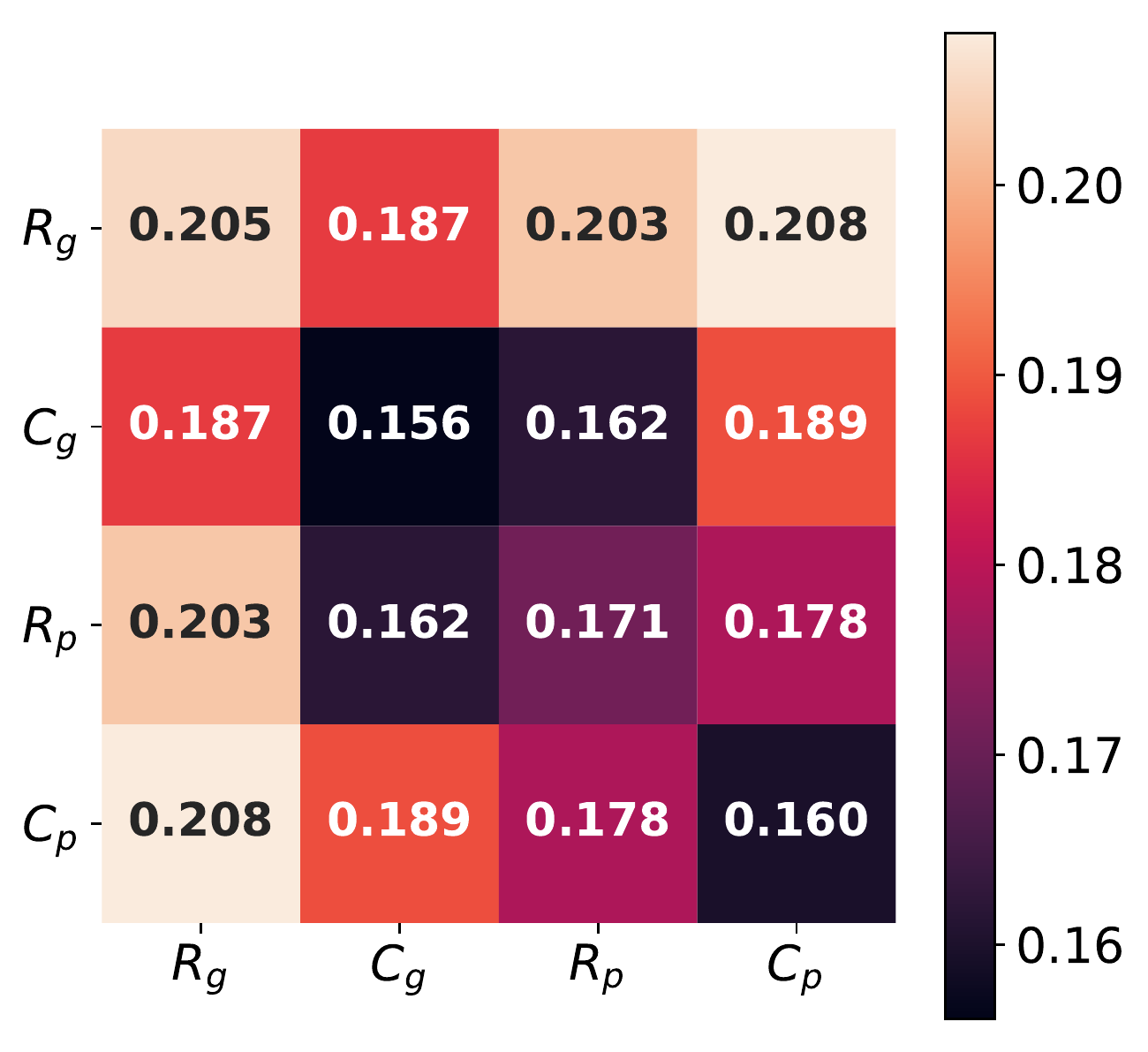}
		}
		
		\subfigure[\scriptsize Recall@20 \& NDCG@20
		]{\label{subfig:rec_pretext}
			\includegraphics[width=0.30\textwidth]{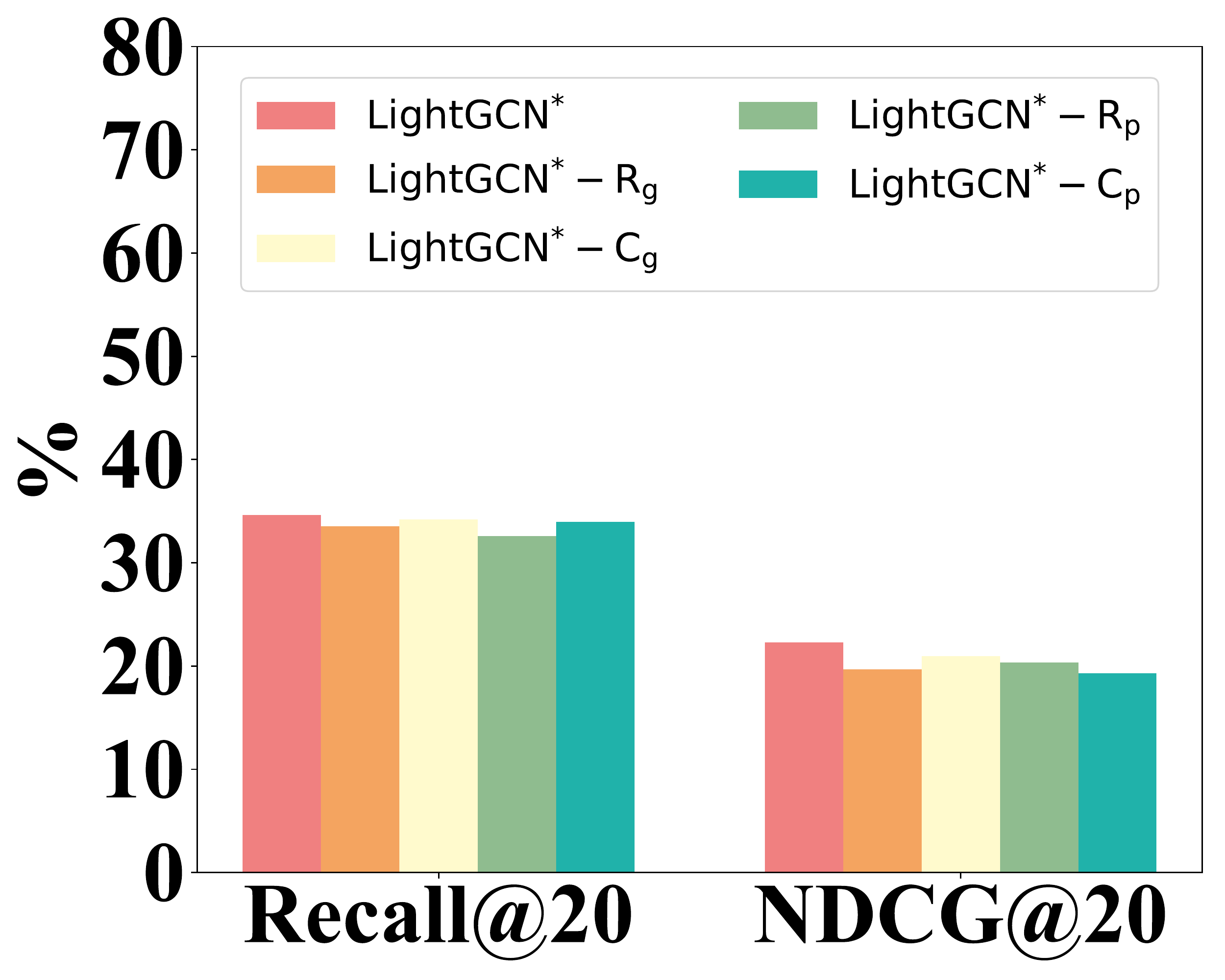}
		}
	}

	\caption{\label{fig:insic_performance} 
		{\small Intrinsic and extrinsic task evaluation under individual or composition of pretext tasks. $K$=3, $L$=4, $T$=6, $c$\%=20\%.  We evaluate the variant models on MOOCs (results on Ml-1M and Gowalla show the same trend which are omitted for space).}}
\end{figure*}

\subsubsection{Impact of Pretext Tasks}

It is still not clear which part of the  pretext tasks is responsible for the good performance in \nRC. To answer this question, we apply LightGCN in \nRC, perform individual or compositional pretext tasks, report the performance of intrinsic and extrinsic task in Fig.~\ref{fig:insic_performance}, where notation $R_g$, $C_g$, $R_p$ and $C_p$ denote the pretext task of reconstruction with GNN (only use the objective function Eq.~\eqref{eq:objective1}), contrastive learning with GNN (only use the objective function Eq.~\eqref{eq:objective2}), reconstruction with Transformer (only use the objective function Eq.~\eqref{eq:objective3}) and contrastive learning with Transformer (only use the objective function Eq.~\eqref{eq:objective4}), respectively;  notation ${\rm LightGCN}^{*}$-$R_g$  means we only discard reconstruction task with GNN encoder and use the other three pretext tasks (use the objective functions Eq.~\eqref{eq:objective2}, Eq.~\eqref{eq:objective3} and Eq.~\eqref{eq:objective4}) as the variant model.
Other variant models  are named in a similar way.
Aligning Table~\ref{tb:intrinsic_overall} with Fig.~\ref{fig:insic_performance},
we find that: 
\begin{itemize}[ leftmargin=10pt ]

\item Combining all the pretext tasks can benefit both  embedding quality and recommendation performance. This indicates simultaneously consider intra- and inter-correlations of users and items can benefit cold-start representation learning and recommendation.
\item Compared with other variant models,
performing the contrastive learning  task with the GNN or Transformer encoder
independently, or performing the contrastive learning task with both the GNN and Transformer encoder leads poor performance in the intrinsic task,
 but has satisfactory performance in the recommendation task. The reason is that contrastive learning does not focus on predicting the target embedding, but  can capture the inter-correlations of users or items, and thus can benefit the recommendation task.

\end{itemize}

\begin{figure}
	\centering

	\mbox{ 
		
		\subfigure[\scriptsize  Recall@20
		]{\label{subfig:ml_dynamic_recall}
			\includegraphics[width=0.30 \textwidth]{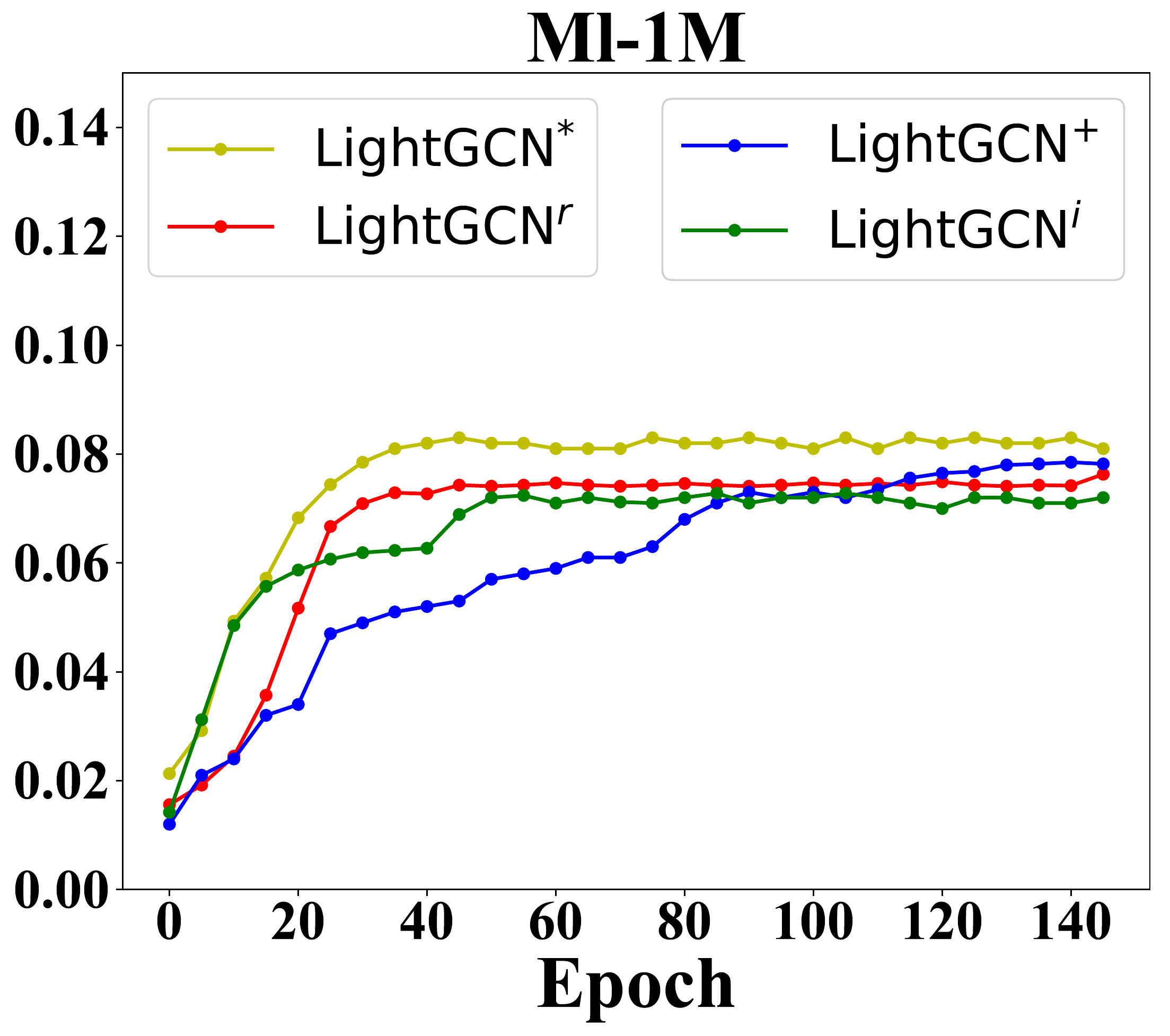}
		}
		
		\subfigure[\scriptsize  NDCG@20
		]{\label{subfig:ml_dynamic_ndcg}
			\includegraphics[width=0.30\textwidth]{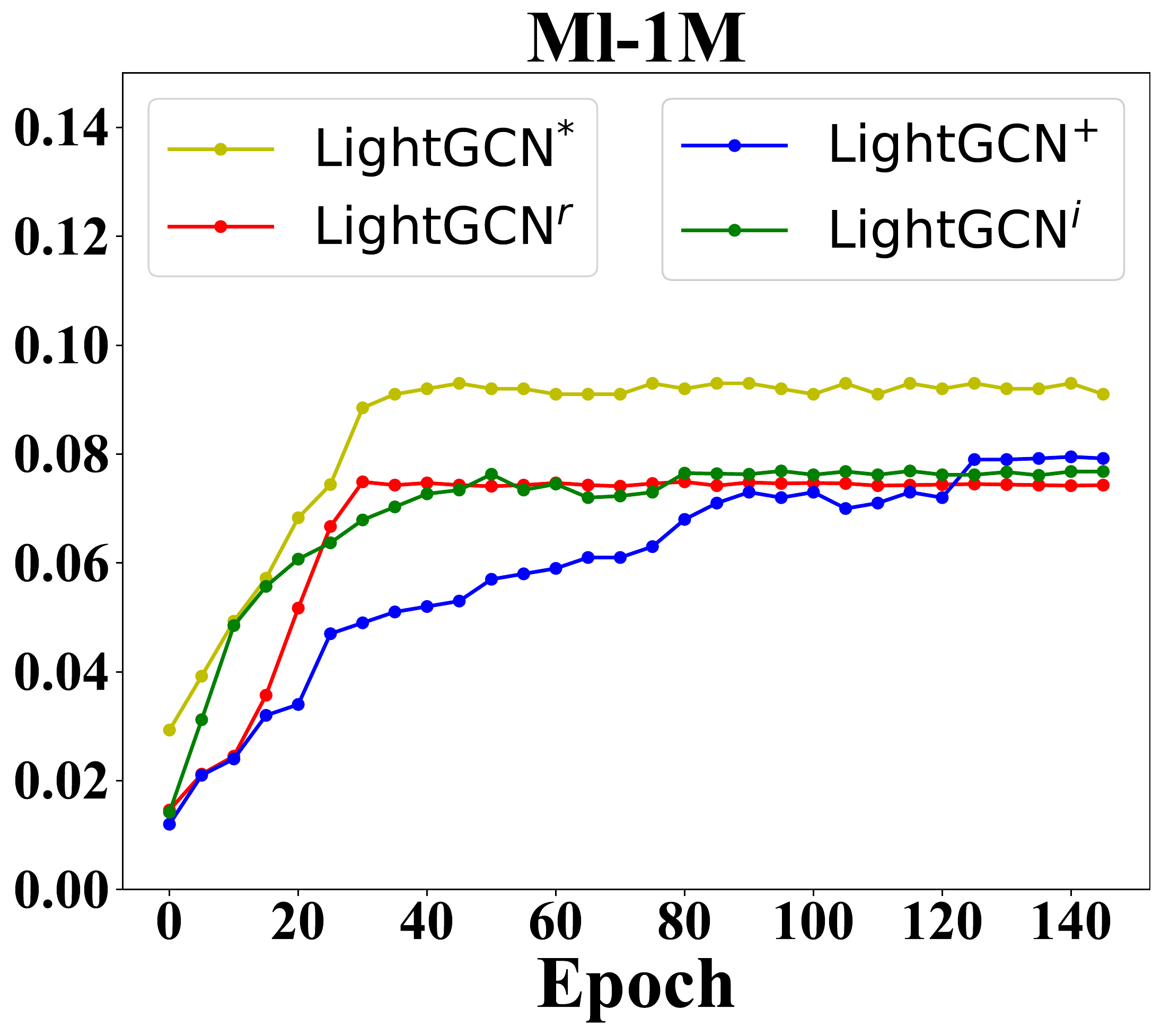}
		}
		
	}
	
	\mbox{ 
		
		\subfigure[\scriptsize  Recall@20
		]{\label{subfig:dynamic_recall}
			\includegraphics[width=0.30 \textwidth]{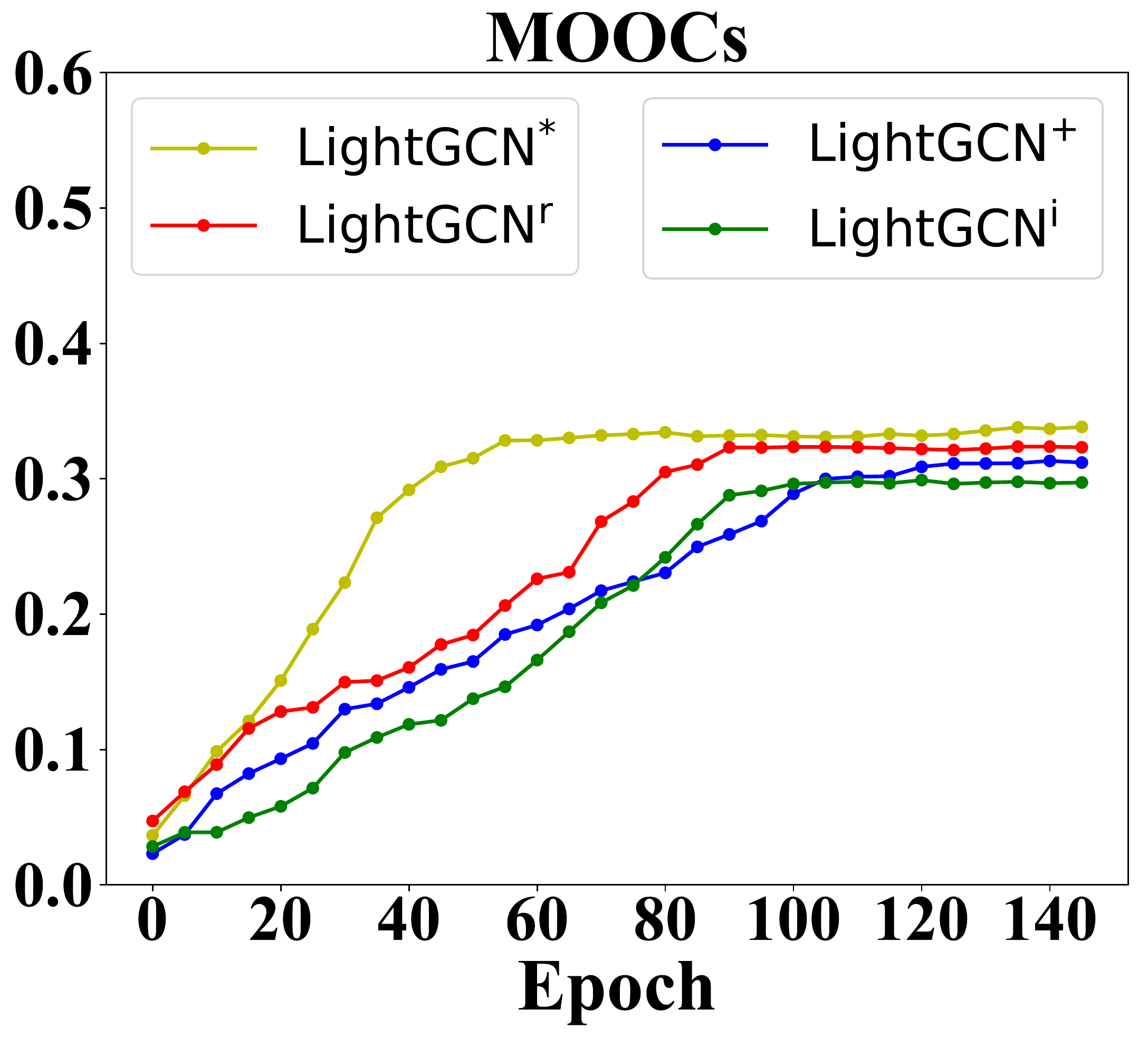}
		}
		
		\subfigure[\scriptsize  NDCG@20
		]{\label{subfig:dynamic_ndcg}
			\includegraphics[width=0.30\textwidth]{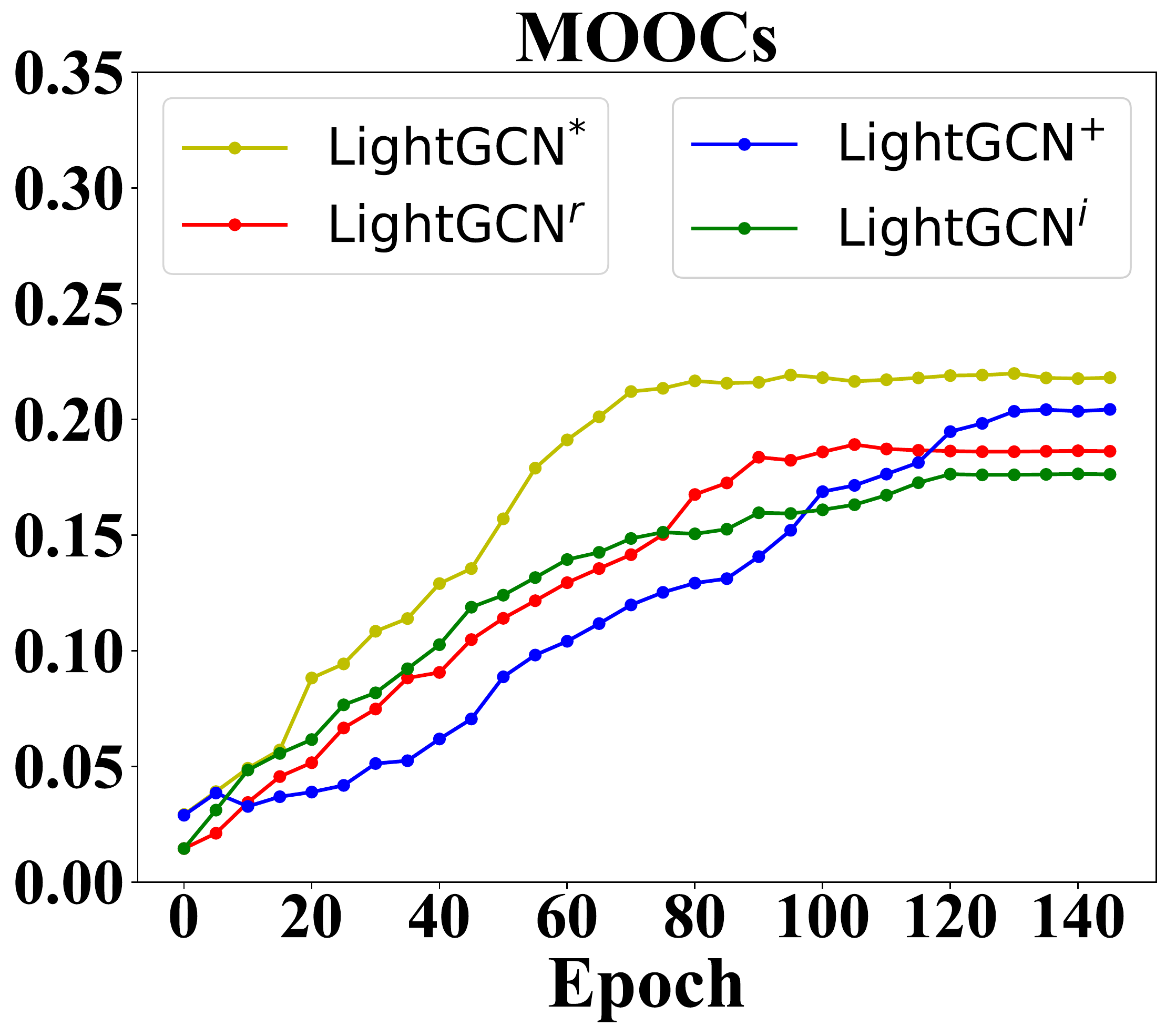}
		}
		
	}

	\caption{\label{fig:dynamic_sampling}   \small{Comparison among different  sampling strategies.}}
\end{figure}

\begin{table*}[t]
	\caption{Analysis of the average sampling/training time in each epoch
	 for different sampling strategies. We run each variant model  ten times, and calculate the average sampling and training time per epoch.} 
	\vspace{-10pt}
	\resizebox{0.90\textwidth}{!}{
		\begin{tabular}{l|cc|ccc}
			\hline
			\multicolumn{1}{c|}{Dataset} & \multicolumn{2}{c|}{Ml-1M} & \multicolumn{2}{c}{MOOCs} \\ \hline
			\multicolumn{1}{c|}{Method}  & Avg. Sampling Time    & Avg. Training Time       & Avg. Sampling Time     & Avg. Training Time     \\ \hline
			${\rm LightGCN}^r$          
			&0.56s &  \textbf{201.02s}    & \textbf{2.78s} & \textbf{237.63s}             \\
			${\rm LightGCN}^i$    & 6.56s               &  204.64s  & 2.86s    & 240.12s    \\ 
			${\rm LightGCN}^+$                                        &134.64s & 205.13s    & 160.37s  &241.75s         \\
			${\rm LightGCN}^*$                   
			&8.93s & 205.76s  &3.20s     & 240.67s &   \\ \hline
	\end{tabular}}
	\label{tb:sampling_strategies_analysis}
	\vspace{-5pt}
\end{table*}

\subsubsection{Impact of the Sampling Strategy}
\label{subsec:dynamic}
We study the effect of the proposed dynamic sampling strategy.
In this paper, we compare four variants of the LightGCN model: ${\rm LightGCN}^{+}$ that adaptively samples neighbors~\cite{haopretrain21},  ${\rm LightGCN}^{i}$ that samples neighbors according to the importance sampling strategy~\cite{chenfastgcn18}, ${\rm LightGCN}^r$ that randomly samples neighbors~\cite{williamgraphsage17} and ${\rm LightGCN}^{*}$ that dynamically samples neighbors. Notably, for fair comparison, for each target node, at the $l$-th layer, the number of the sampled neighbors are at most $K^l$.
We report the average sampling time and the average training time per epoch (the average training time includes the training time of all the four pretext tasks, but does not include the neighbor sampling time) for these sampling strategies on MOOCs and Ml-1M in Table~\ref{tb:sampling_strategies_analysis}, and report the recommendation performance on MOOCs and Ml-1M in  Fig.~\ref{fig:dynamic_sampling}. 
We find that: 
\begin{itemize}[ leftmargin=10pt ]
	
\item The sampling time of the proposed dynamic sampling strategy is in the same magnitude with the importance sampling strategy and the random sampling strategy, which is totally acceptable. While the adaptive sampling strategy takes too much sampling  time, since it adopts 
a Monte-Carlo based policy gradient strategy, and has the complex action-state trajectories
for the entire neighbor sampling process.
Besides, we report the average training time in each epoch and find that the training time of all the variant models is almost the same,
since each variant model performs the same pretext tasks., i.e., the graph convolution operation in the GNN encoder and self-attention operation in the Transformer encoder.  By aligning Table~\ref{tb:sampling_strategies_analysis} and Fig.~\ref{fig:dynamic_sampling}, we find that \snRC has a quick convergence speed.

\item Although the adaptive sampling strategy has competitive performance than the random and importance sampling strategies, it has slow convergence speed due to the complex RL-based sampling process.
\item Compared with the other sampling strategies, the proposed dynamic sampling strategy  not only has the best performance, but also  has quick convergence speed. 
\end{itemize}

\subsection{Study of MPT (RQ3)}

\subsubsection{Effect of Ground-truth Embedding}
\label{sec:ground_truth_embedding}
\begin{figure}
	\centering

	\mbox{ 
		
		\subfigure[\scriptsize  LightGCN
		]{\label{subfig:ground-truth_gnns}
			\includegraphics[width=0.46 \textwidth]{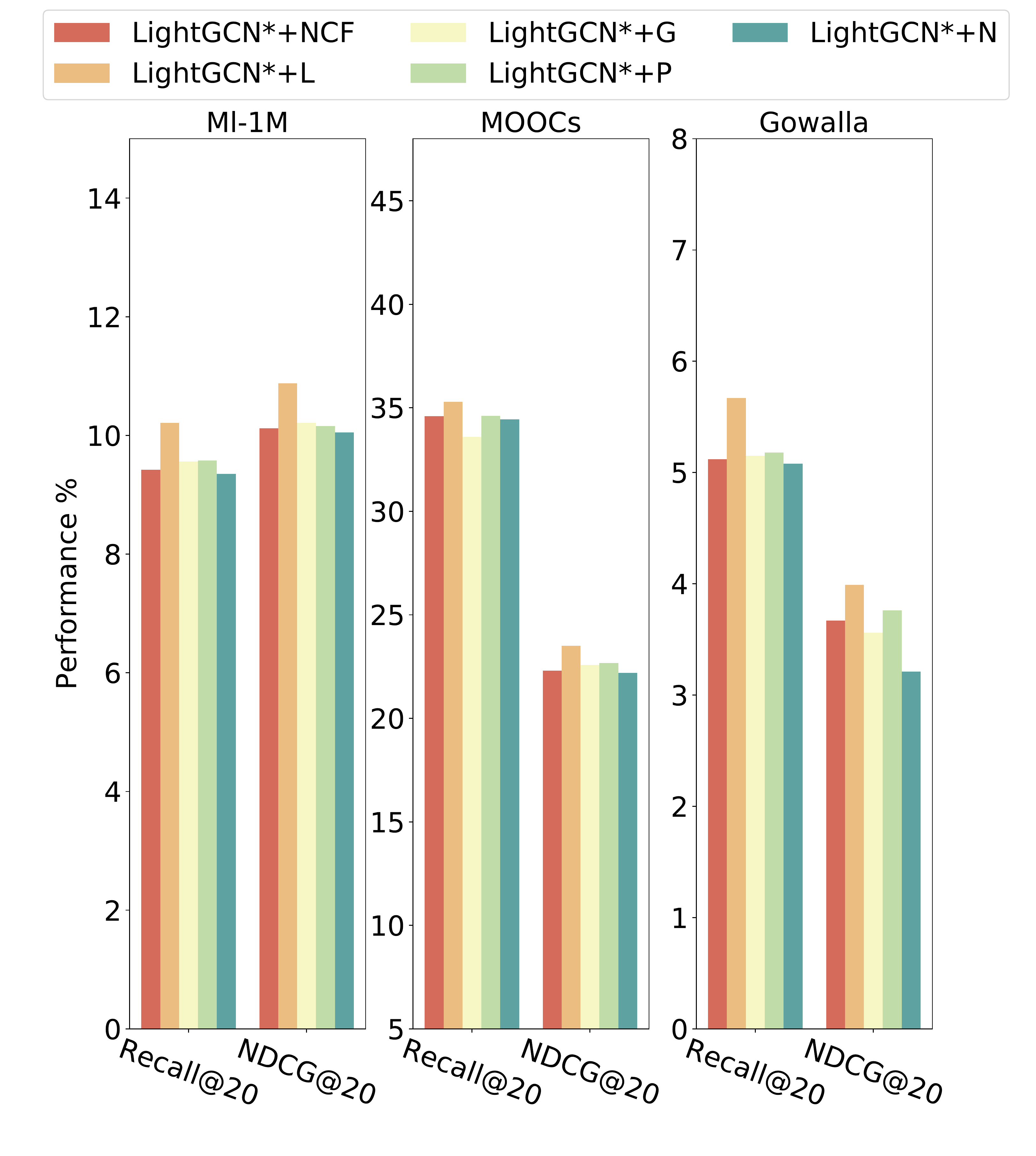}
		}
	
	\hspace{-0.3cm}
		
		\subfigure[\scriptsize  Transformer Encoder
		]{\label{subfig:ground_truth_transformer}
			\includegraphics[width=0.46\textwidth]{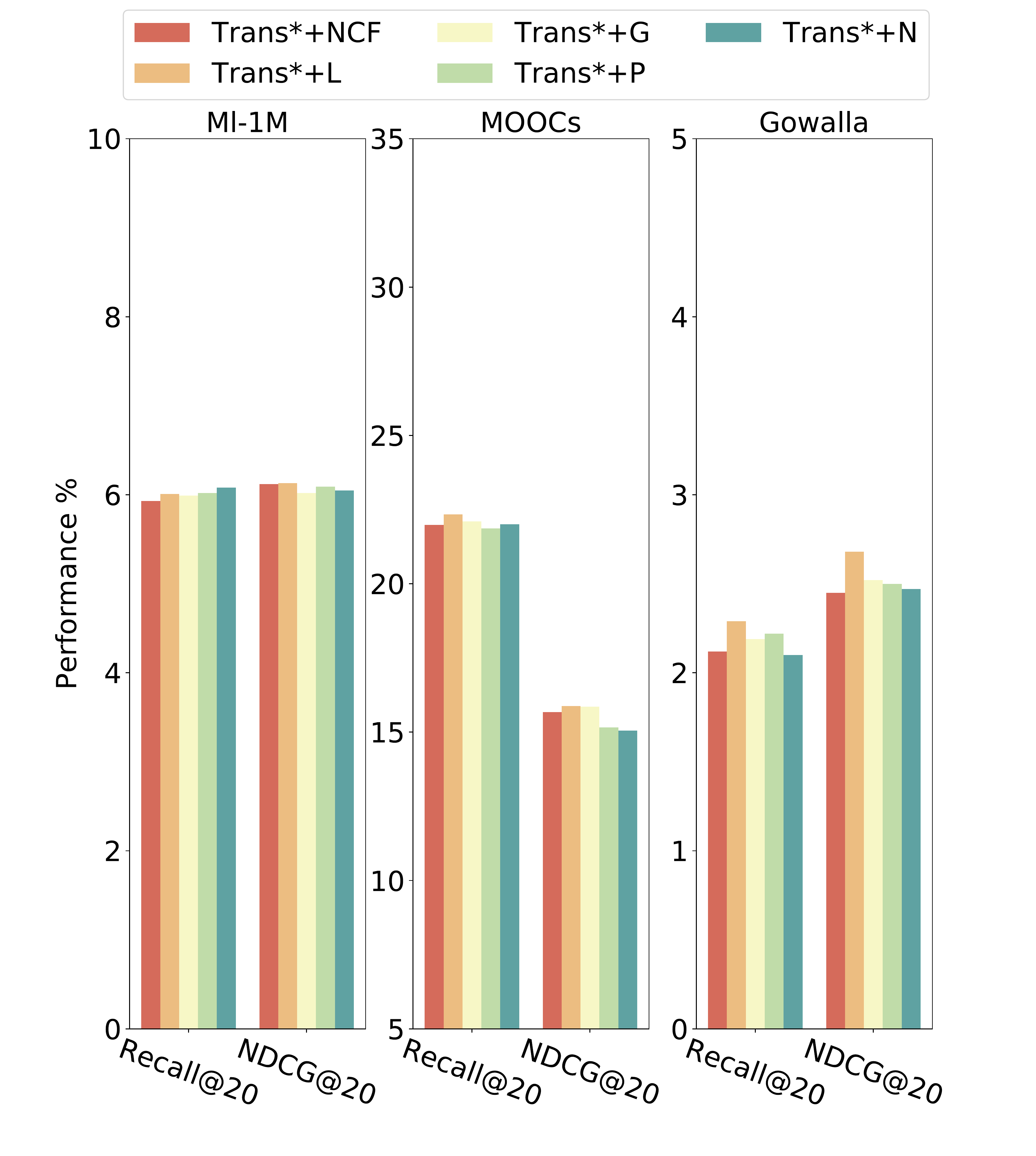}
		}
		
	}

	\caption{\label{fig:ground_truth}   \small{Sensitive analysis of ground-truth embeddings. } }
\end{figure}

As mentioned in~\secref{sec:preliminaries}, when performing embedding reconstruction with GNN and Transformer encoders, we choose NCF to learn the ground-truth embeddings. However, one may consider whether \nRC's performance is sensitive to the ground-truth embedding. To this end, we use the baseline GNN models to learn the ground-truth embeddings, and only perform embedding reconstruction task with LightGCN or Transformer.
For NCF, we concatenate embeddings produced by both the MLP and GMF modules as ground-truth embeddings. While for PinSage, GCMC, NGCF and LightGCN, we  combine the embeddings obtained at each layer to form the ground-truth
matrix, i.e., $\textbf{E} =  \textbf{E}^{(0)}$   + $\cdots$+  $\textbf{E}^{(L)} $, where  $\textbf{E}^l \in \mathcal{R}^{(|\mathcal{U}|+|\mathcal{I}|) \times d}$ is the concatenated user-item embedding matrix at $l$-th convolution step.
 Fig.~\ref{subfig:ground-truth_gnns} and Fig.~\ref{subfig:ground_truth_transformer} shows the recommendation performance using LightGCN and Transformer encoder, respectively. Suffix
+${\rm NCF}$,  +${\rm L}$, +${\rm G}$, +${\rm P}$ and +${\rm N}$ denote that the ground-truth embeddings are obtained  by NCF, LightGCN, GCMC, PinSage and NGCF, respectively. 
We find that:  
\begin{itemize}[ leftmargin=10pt ]

\item All the models that equipped with different ground-truth embeddings achieve almost the same performance. This indicates our model is not sensitive to the ground-truth embeddings, as the NCF and vanilla GNN models are good enough to learn high-quality user or item embeddings from the abundant interactions.
\item Besides, when LightGCN is used to obtain
the ground-truth embeddings (denoted as ${\rm LightGCN}^{*}$+${\rm L}$ and ${\rm Trans}^{*}$+${\rm L}$), we can obtain marginal performance gain compared with other variant models.
\end{itemize}

\begin{figure}
	\centering

	\mbox{ 
		
		\subfigure[\scriptsize  LightGCN
		]{\label{subfig:cl_gnn}
			\includegraphics[width=0.46 \textwidth]{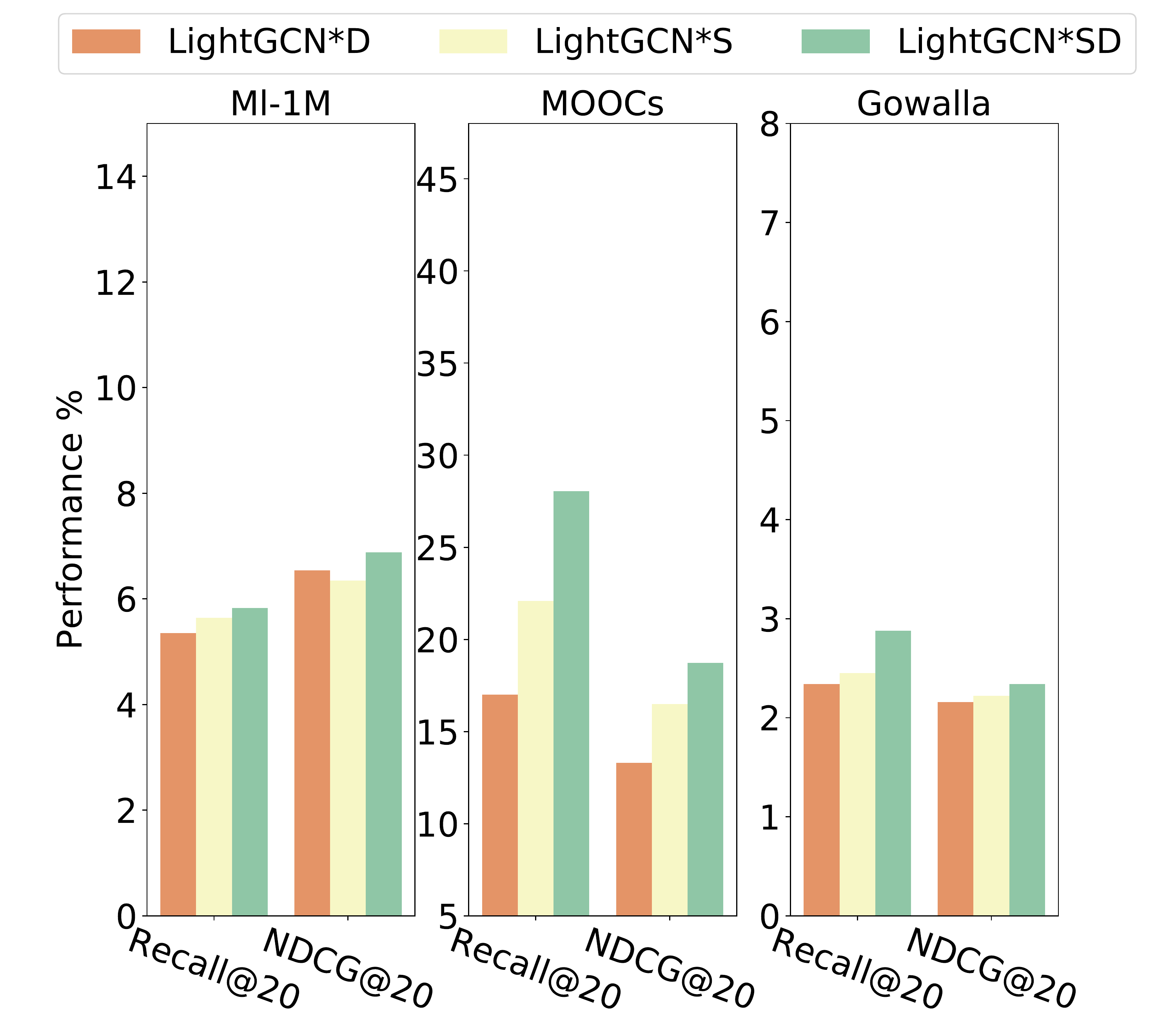}
		}
		
		\hspace{-0.3cm}
		
		\subfigure[\scriptsize  Transformer Encoder
		]{\label{subfig:cl_transformer}
			\includegraphics[width=0.46\textwidth]{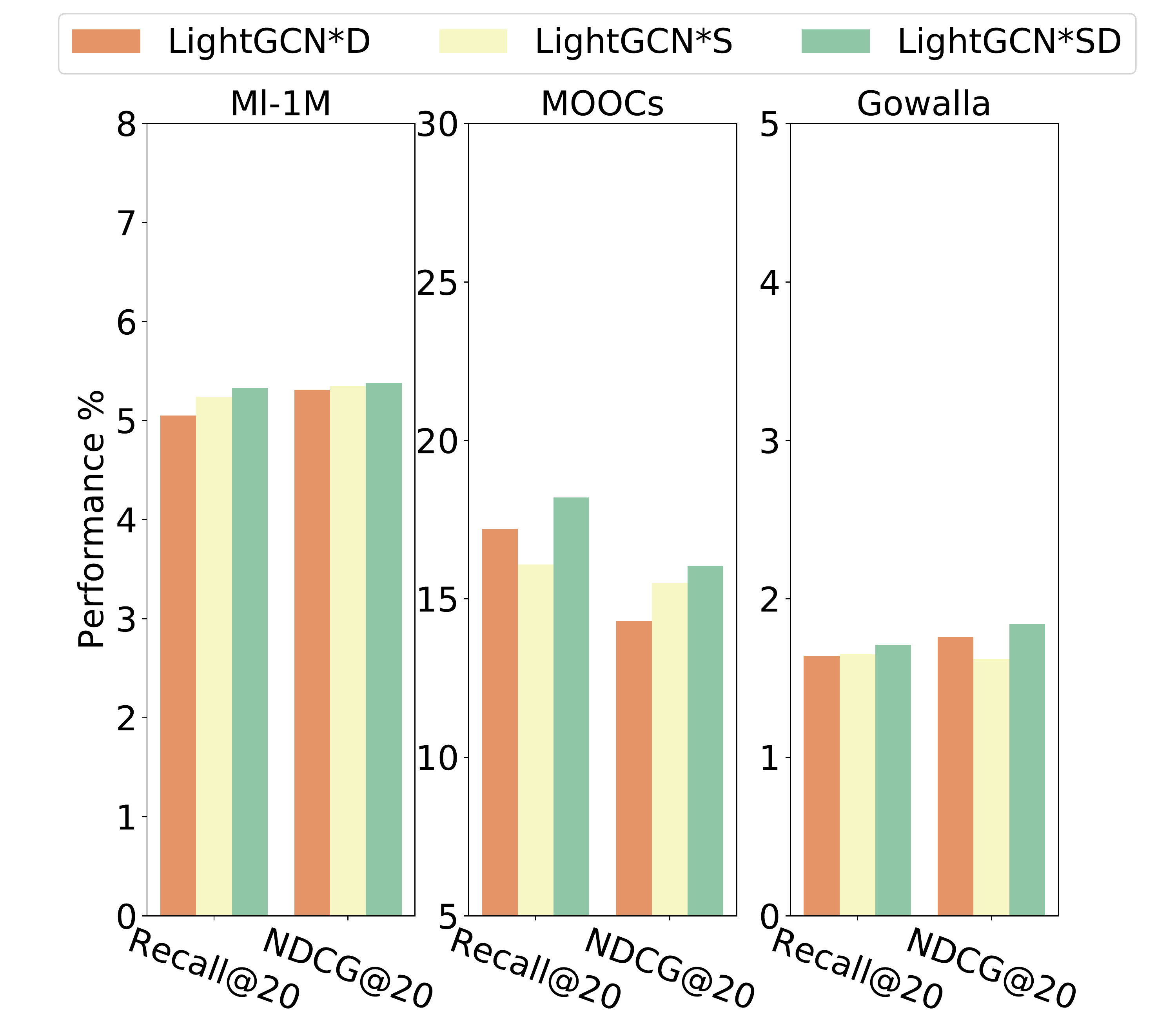}
		}
		
	}

	\caption{\label{fig:cl}  \small{Recommendation performance under individual or compositional data augmentations. $c$\%= 20\%, $L$=4.}}
\end{figure}

\subsubsection{Effect of Data Augmentation}
To understand the effects of individual or compositional data augmentations in the contrastive learning (CL) task using GNNs or Transformer encoder, we investigate the performance of CL pretext tasks in \snRC when applying augmentations individually or in pairs. We report the performance in Fig.~\ref{fig:cl}. Notation  ${\rm LightGCN}^{*}{\rm D}$, ${\rm LightGCN}^{*}{\rm S}$ and ${\rm LightGCN}^{*}{\rm SD}$ mean we apply LightGCN into the CL task and use deletion, substitution and their combinations, respectively; Notation  ${\rm Trans}^{*}{\rm D}$, ${\rm Trans}^{*}{\rm S}$ and ${\rm Trans}^{*}{\rm SD}$ denote we use Transformer encoder into the CL task and use  deletion, substitution and their combinations, respectively. We find
that:
\begin{itemize}[ leftmargin=10pt ]
 \item Composing augmentations can lead to better recommendation performance than performing individual data augmentation. 
 \item Aligning Fig.~\ref{fig:cl} and Table~\ref{tb:recommendation}, we find combining substitution and deletion using GNNs has competitive performance than SGL. The reason is that, same as SGL, our proposed contrastive data augmentation can also change the graph structure, and thus inter-correlations of nodes can be captured.
\end{itemize}

\begin{figure}[t]
	\centering

	\mbox{
		\subfigure[\scriptsize User embedding inference
		]{\label{subfig:up_ml_layer_recall}
			\includegraphics[width=0.30\textwidth]{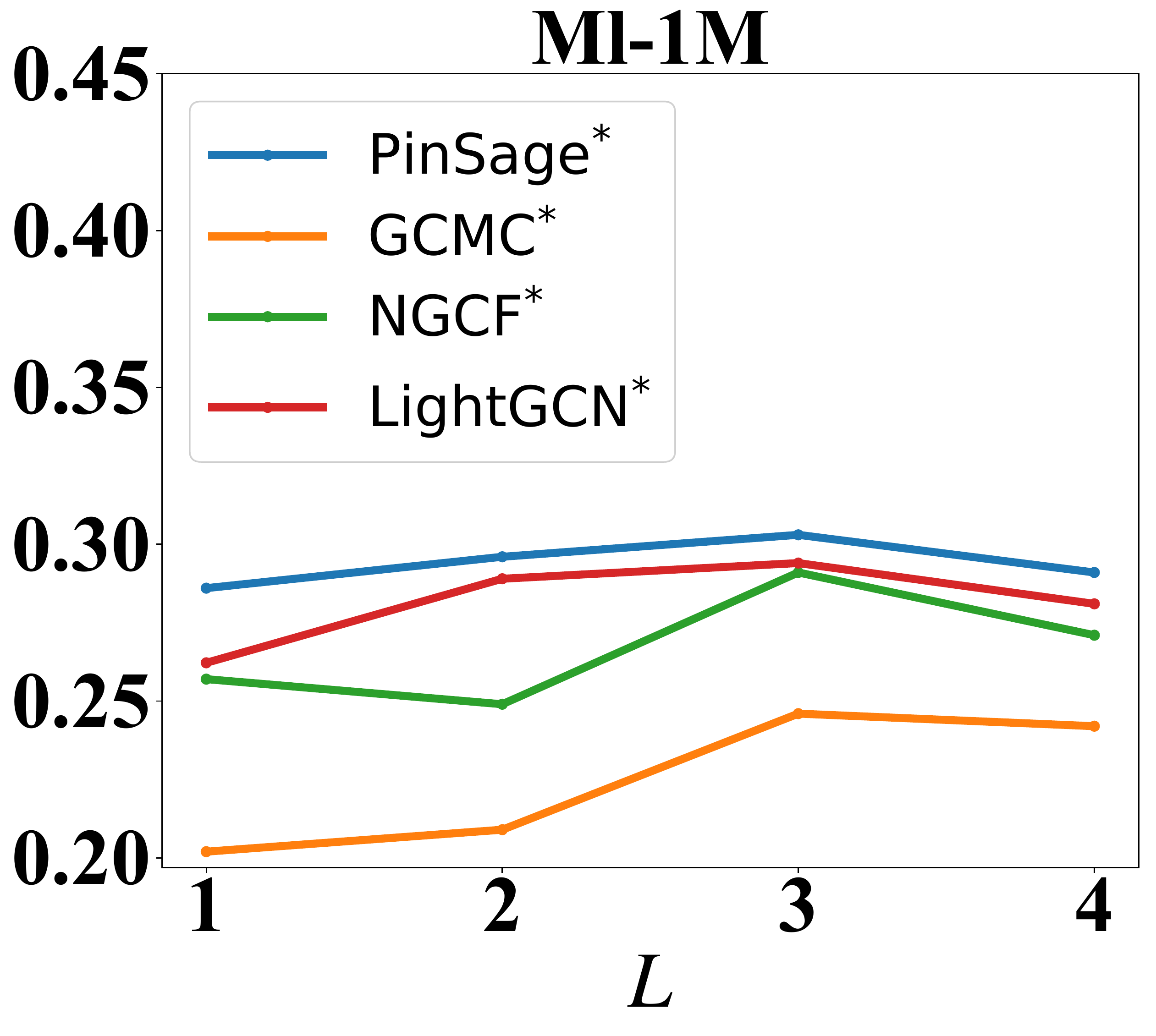}
		}
		
		\subfigure[\scriptsize  Item embedding inference
		]{\label{subfig:_upml_layer_ndcg}
			\includegraphics[width=0.30\textwidth]{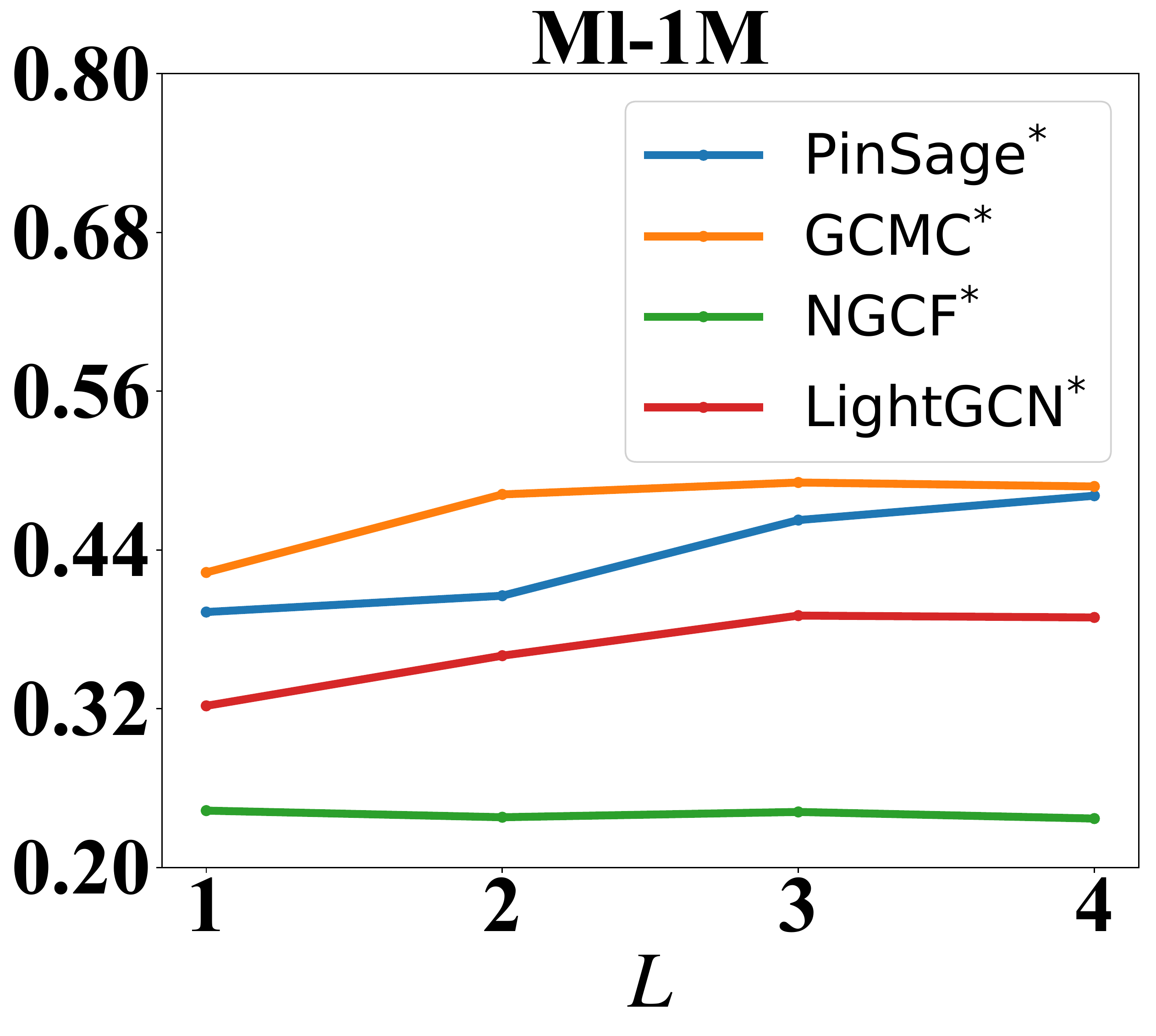}
		}
	}

	\mbox{ 
		
		\subfigure[\scriptsize  Recommendation Recall@20
		]{\label{subfig:ml_layer_recall}
			\includegraphics[width=0.30\textwidth]{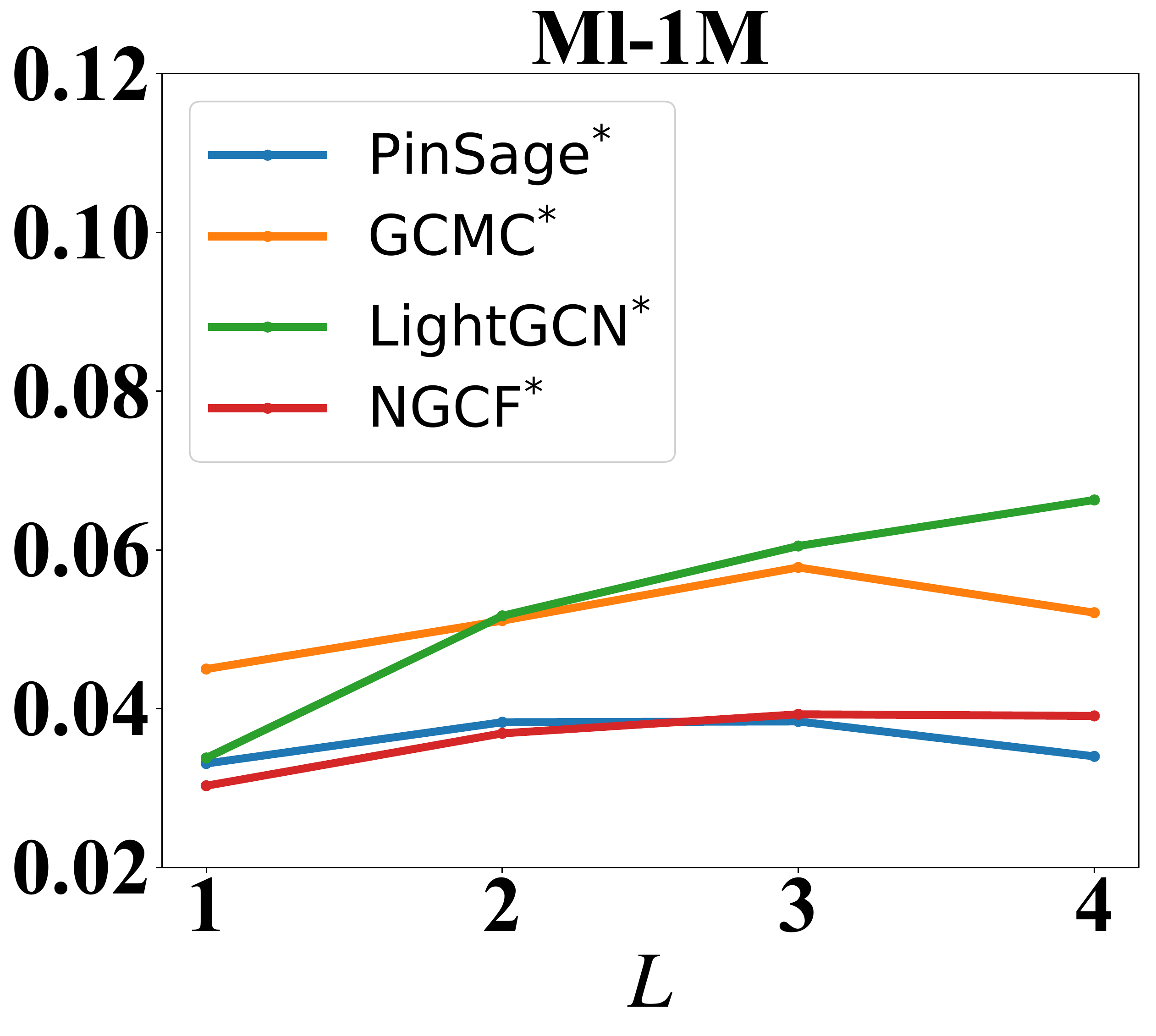}
		}
		
		\subfigure[\scriptsize  Recommendation NDCG@20
		]{\label{subfig:ml_layer_ndcg}
			\includegraphics[width=0.30\textwidth]{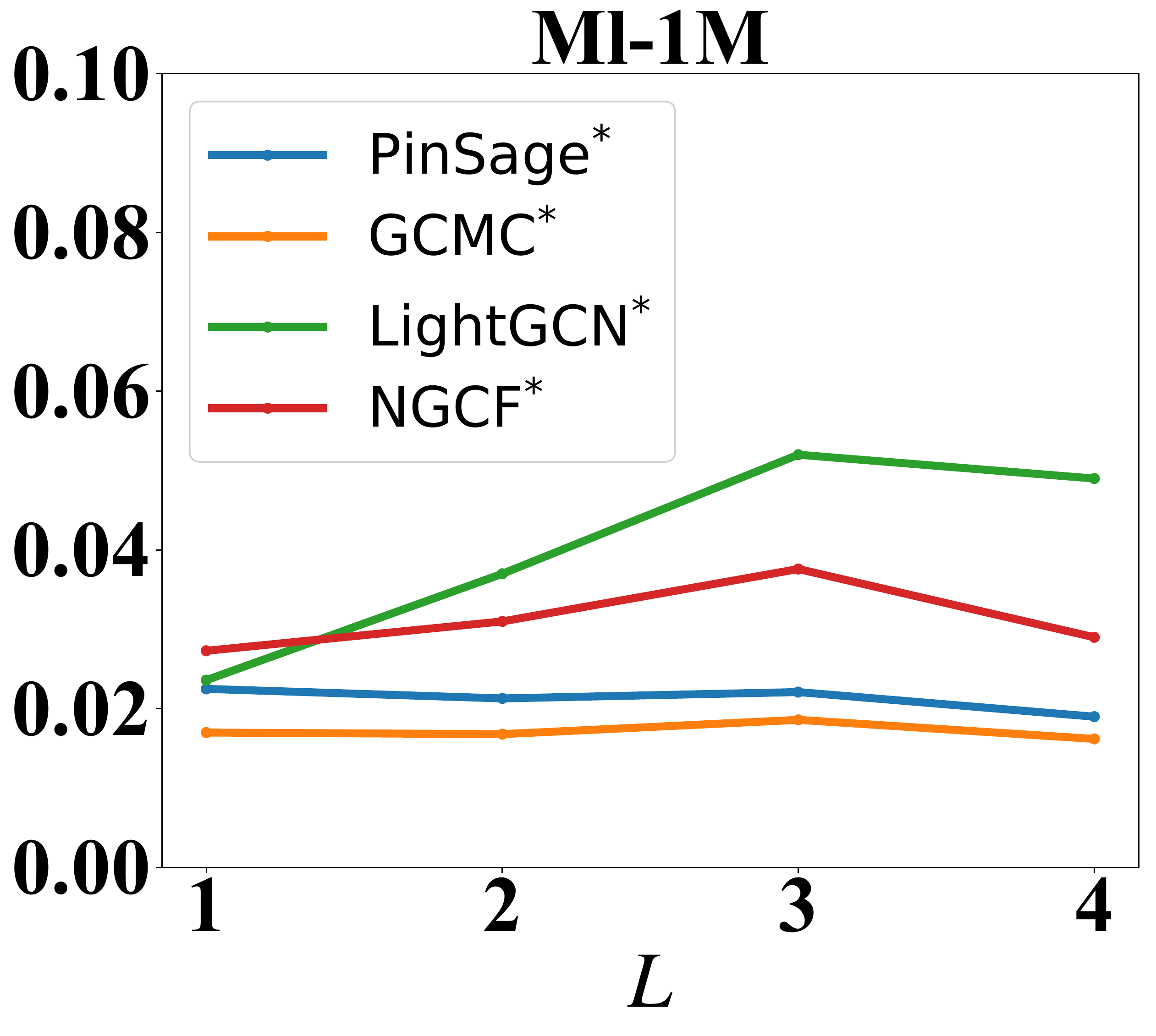}
		}
	}

	\caption{\label{fig:layer_analysis}   \small{Embedding inference and recommendation performance under different  layer $L$.}
	}
\end{figure}

\begin{figure}[t]
	\centering

	\mbox{
		\subfigure[\scriptsize Embedding inference
		]{\label{subfig:upstream_path}
			\includegraphics[width=0.30\textwidth]{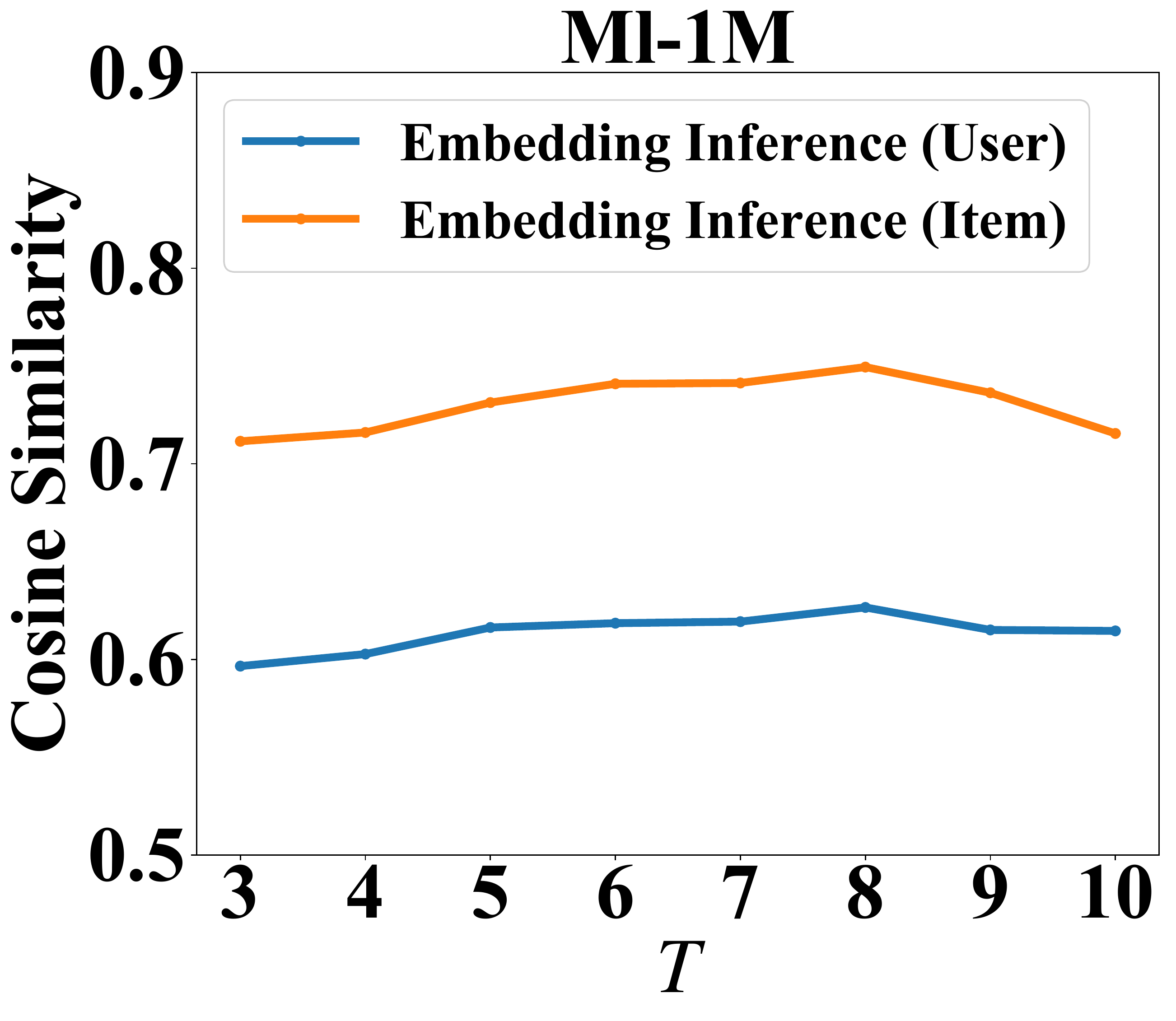}
		}
		
		\subfigure[\scriptsize  Recommendation
		]{\label{subfig:_downstream_path}
			\includegraphics[width=0.30\textwidth]{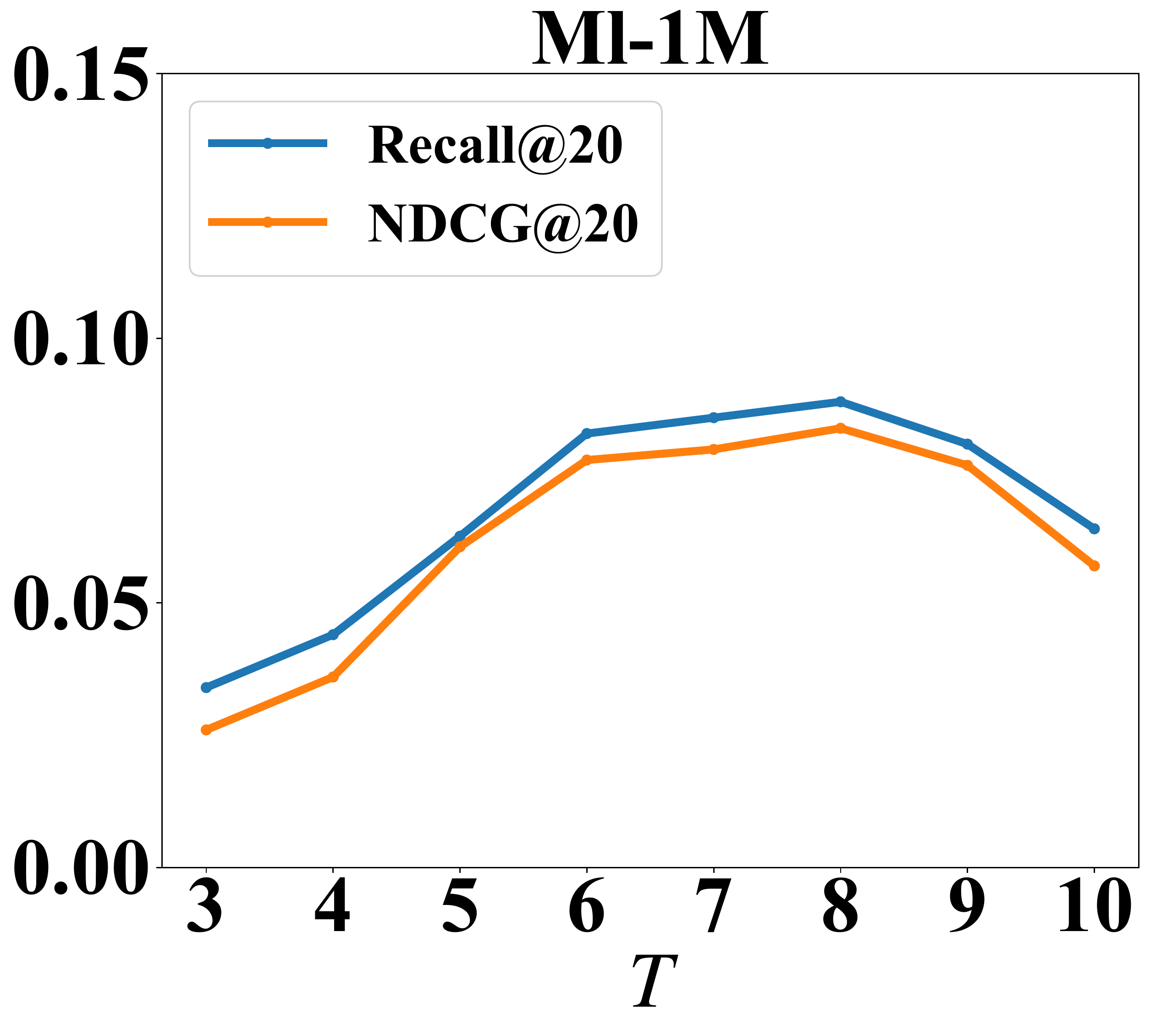}
		}
	}

	\caption{\label{fig:path_analysis}   \small{Embedding inference and recommendation performance  under different  path length $T$.}
	}
\end{figure}

\subsubsection{Effect of Hyperparameters}
\label{sub:hyper_parameters}
We move on to study different designs of the layer depth $L$,  the deletion ratio $a$\%, the substitution ratio $b$\% and the user-item path length $T$  in \nRC. Due to the space limitation, we omit the results on MOOCs and Gowalla which have a similar trend to Ml-1M. 

\begin{figure}
	\centering

	\mbox{ 
		
		\subfigure[\scriptsize  Contrastive learning with LightGCN
		]{\label{subfig:ratio_recall}
			\includegraphics[width=0.30 \textwidth]{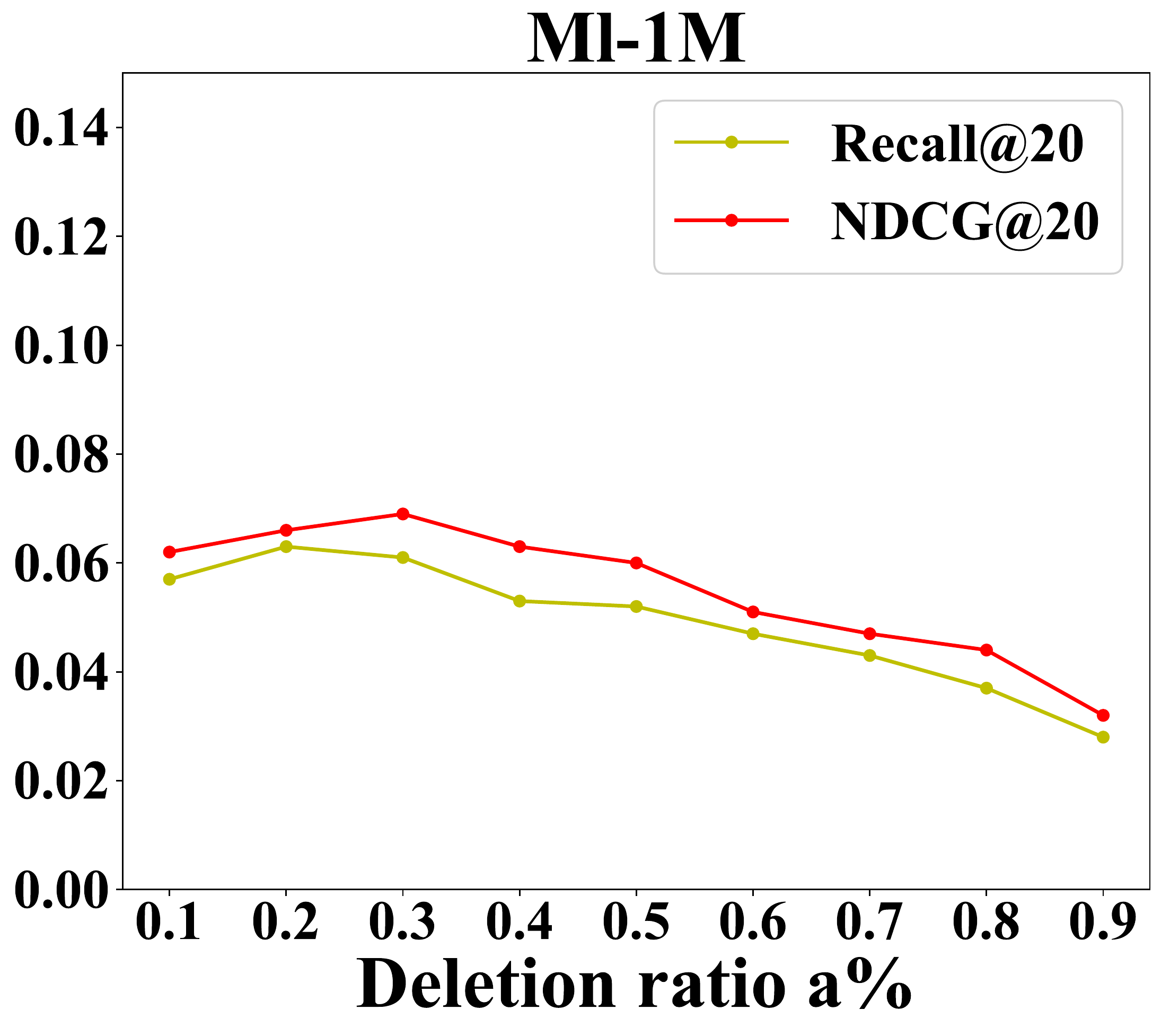}
		}
		
		\subfigure[\scriptsize  Contrastive learning with LightGCN
		]{\label{subfig:ratio_ndcg}
			\includegraphics[width=0.30\textwidth]{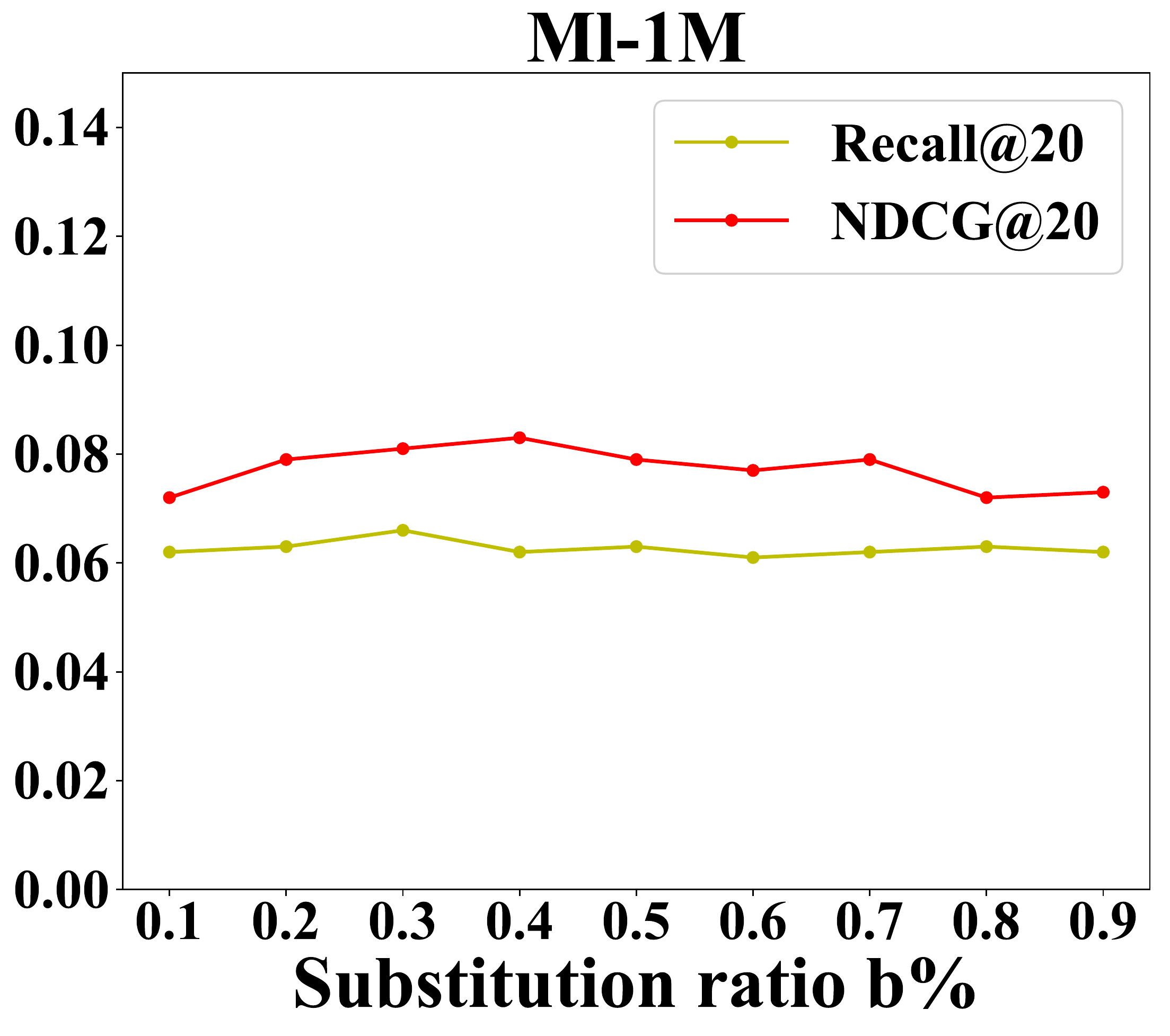}
		}
		
	}

	\mbox{ 
		
		\subfigure[\scriptsize   Contrastive learning with Transformer 
		]{\label{subfig:ml_ratio_recall}
			\includegraphics[width=0.30 \textwidth]{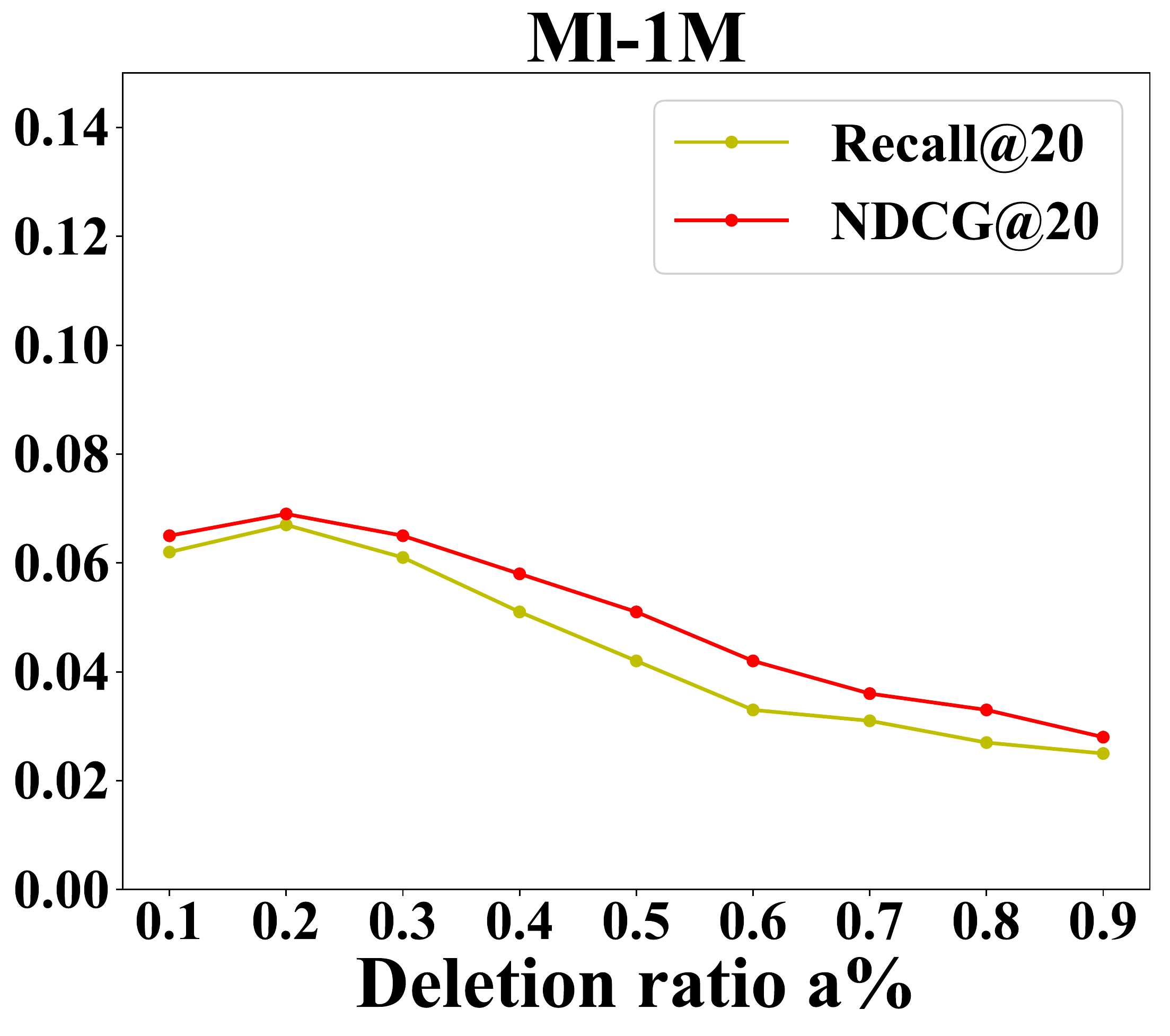}
		}
		
		\subfigure[\scriptsize   Contrastive learning with Transformer 
		]{\label{subfig:ml_ratio_ndcg}
			\includegraphics[width=0.30\textwidth]{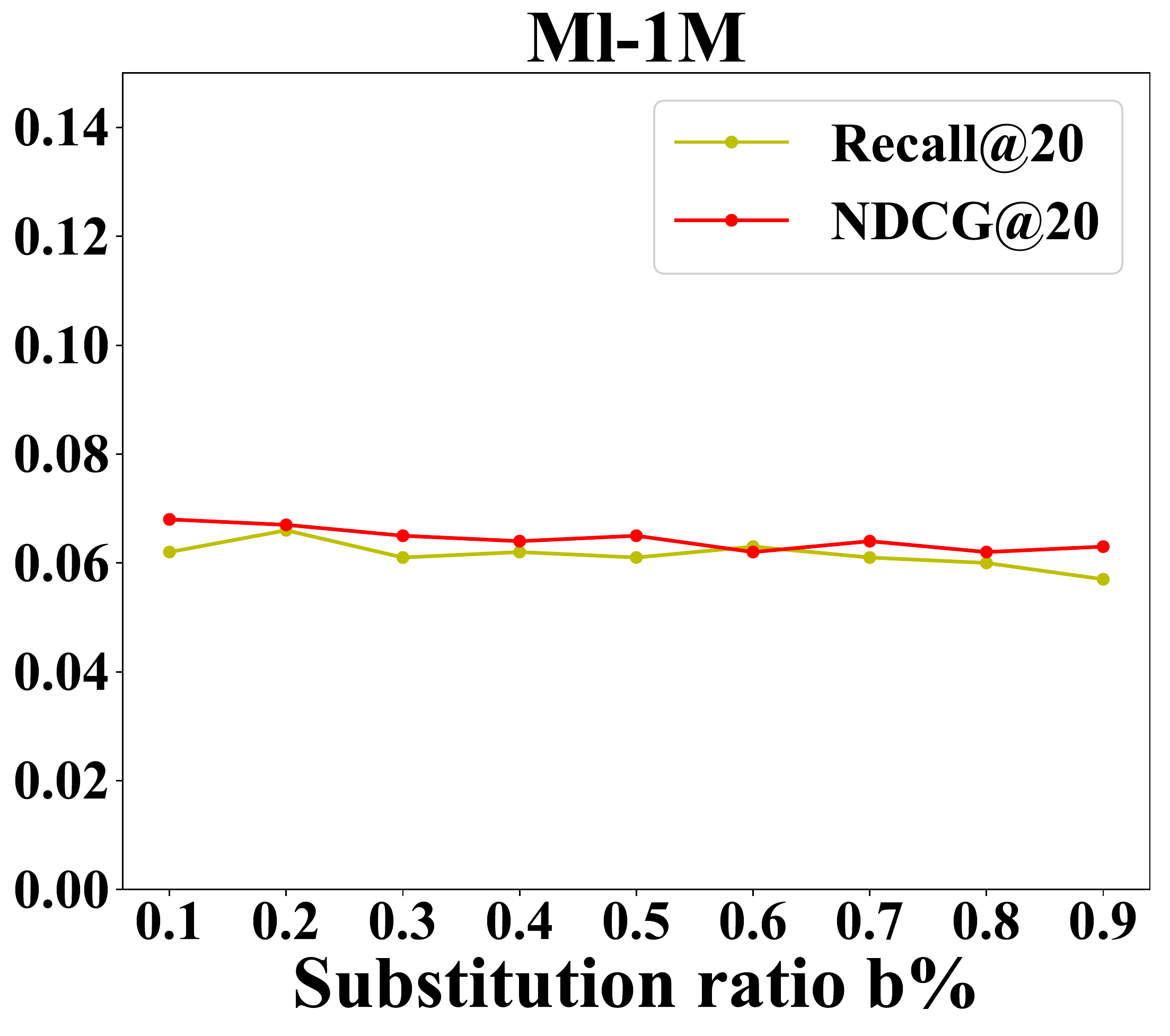}
		}
		
	}

	\caption{\label{fig:d_s_ratio}   \small{Sensitive analysis under different deletion and substitution ratios in the contrastive learning pretext tasks.}}
\end{figure}

\begin{itemize}[ leftmargin=10pt ]
	\item  We only perform embedding reconstruction with GNNs in \nRC, and report the results in  Fig.~\ref{fig:layer_analysis}. We find that the performance first increases and then drops when increasing $L$ from 1 to 4. The peak point is 3 at most cases. This indicates GNN can only capture short-range dependencies, which is  consistent with LightGCN’s~\cite{xiangnanhe_lightgcn20} finding.
	
	\item  We only perform embedding reconstruction with Transformer encoder in \nRC, and report the results in Fig.~\ref{fig:path_analysis}. We find that the performance first improves when the path length $T$ increases from 3 to 8, and then drops from 8 to 10. This indicates Transformer encoder is good at capturing long-range rather than short-range dependencies of nodes.

	\item We perform  contrastive learning with LightGCN and  contrastive learning with Transformer encoder under different deletion ratio $a$\% and substitution $b$\% independently, and report the recommendation results in Fig~\ref{fig:d_s_ratio}. Based on the results, we find that the recommendation performance is not sensitive to the substitution ratio $b$\% but is a little bit sensitive to the deletion ratio $a$\%. 
	In terms of the deletion ratio $a$\%,
	the recommendation performance first increases and then drops. The peak point is around 0.2-0.3 for most cases.    
	The reason is that when we perform the substitution operation, we randomly replace $u$ or $i$ with one of its parent’s interacted first-order neighbors, thus the replaced nodes can still reflect its parent's characteristic to some extent. However, once the deletion ratio gets larger (e.g., 0.9), it will result in a highly skewed graph structure, which can not bring much useful collaborative signals to the target node.
	
\end{itemize}

\section{Related Work}

\vpara{Self-supervised Learning with GNNs.} 
The recent advance on self-supervised learning for recommendation has received much attention~\cite{DBLP:journals/corr/abs-2203-15876,QiuHYW22,YuY000H21,0013YYWC021,0013YYSC21}, where self-supervised learning with GNNs aim to find the correlation of nodes in the graph itself, so as to alleviate the data sparsity issue in graphs~\cite{gpt_gnnhu,qiu2020gcc,velickovic2019deep,ChenYCNPL19,infograph,YCTSCWNIPS20,yuhxtlqsigir22,YuYLWH021}.
More recently, some researchers explore pre-training GNNs on user-item graphs for recommendation. For example, \RC~\cite{haopretrain21} reconstructs embeddings under the meta-learning setting. SGL~\cite{WWFHSIGIR2021} contrasts  node representation by node dropout, edge dropout and random walk data augmentation operations. PMGT~\cite{LYSLWTZArxiv2020} reconstructs graph structure and node feature using side information.
GCN-P/COM-P~\cite{MLMArxiv2021} learns the representations of entities constructed from the side information.
However, these methods suffer from  ineffective long-range dependency problem, as the GNN model can only capture  low-order correlations of  nodes. 
Besides,  the pretext tasks of these methods consider either
intra-correlations~\cite{haopretrain21,LYSLWTZArxiv2020,MLMArxiv2021}  or inter-correlations~\cite{WWFHSIGIR2021} of nodes, rather than both.
Further, the side information is not always available, making it difficult to construct the pretext tasks.
To solve the above problems, we propose \nRC, which  considers both  short- and  long-range dependencies of nodes, and both  intra- and   inter-correlations of nodes.

\vpara{Cold-start Recommendation.}
Cold-start recommendation is an intractable problem in Recommender System. Some researchers try to incorporate the side information such as knowledge graphs (KGs)~\cite{wang2019multi,wang2019kgat} or the contents of users/items~\cite{hongzhi_side,YinWWCZ17,ChenYS0GM20,YinW0LYZ19,hongzhi_tkde_20,WangYWNHC19,GharibshahZHC20} to enhance the representations of users/items. However, the contents are not always available and it is difficult to align the users/items with the entities in KGs.
Another research line is to explicitly improve the quality of cold-start users/items, which uses either  meta-learning~\cite{Finnmaml17,munkhdalaimetaneworks17,vinyalsmatching16,Snellprotocal17} or the graph neural networks (GNNs)~\cite{pinsage,wangncgf19,xiangnanhe_lightgcn20,hongzhi_tkde_20} for recommendation. However, the meta learning technique does not explicitly address the high-order neighbors of the cold-start users/items. The GNNs do not explicitly address the high-order cold-start users/items, which can not thoroughly improve the embedding quality of the cold-start users/items.

\section{Conclusion}
We propose a multi-strategy based pre-training method, \nRC,
which extends PT-GNN from the perspective of  model
architecture and  pretext tasks to improve the cold-start
recommendation performance. Specifically, in addition to the short-range dependencies of
nodes captured by the GNN encoder, we add a
transformer encoder to capture long-range dependencies.
In addition to considering the intra-correlations of
nodes by the embedding reconstruction task,
we add embedding contrastive learning task to capture
inter-correlations of nodes.
Experiments on three datasets demonstrate the effectiveness of our
proposed \snRC model against the vanilla GNN  and pre-training GNN models. 
\section*{ACKNOWLEDGMENTS}
This work is supported by National Key Research \& Develop Plan (Grant No. 2018YFB1004401), National Natural Science Foundation of China (Grant No. 62072460, 62076245, 62172424), Beijing Natural Science Foundation (Grant No. 4212022), Australian Research Council Future
Fellowship (Grant No. FT210100624) and Discovery Project
(Grant No. DP190101985).

\normalem
\bibliographystyle{ACM-Reference-Format}
\bibliography{sample-base}


\begin{thebibliography}{52}


\ifx \showCODEN    \undefined \def \showCODEN     #1{\unskip}     \fi
\ifx \showDOI      \undefined \def \showDOI       #1{#1}\fi
\ifx \showISBNx    \undefined \def \showISBNx     #1{\unskip}     \fi
\ifx \showISBNxiii \undefined \def \showISBNxiii  #1{\unskip}     \fi
\ifx \showISSN     \undefined \def \showISSN      #1{\unskip}     \fi
\ifx \showLCCN     \undefined \def \showLCCN      #1{\unskip}     \fi
\ifx \shownote     \undefined \def \shownote      #1{#1}          \fi
\ifx \showarticletitle \undefined \def \showarticletitle #1{#1}   \fi
\ifx \showURL      \undefined \def \showURL       {\relax}        \fi
\providecommand\bibfield[2]{#2}
\providecommand\bibinfo[2]{#2}
\providecommand\natexlab[1]{#1}
\providecommand\showeprint[2][]{arXiv:#2}

\bibitem[\protect\citeauthoryear{Burges}{Burges}{2010}]%
        {dynamic1}
\bibfield{author}{\bibinfo{person}{Christopher~JC Burges}.}
  \bibinfo{year}{2010}\natexlab{}.
\newblock \showarticletitle{From ranknet to lambdarank to lambdamart: An
  overview}.
\newblock \bibinfo{journal}{\emph{Learning}} \bibinfo{volume}{11},
  \bibinfo{number}{23-581} (\bibinfo{year}{2010}), \bibinfo{pages}{81}.
\newblock


\bibitem[\protect\citeauthoryear{Chen, Yin, Chen, Nguyen, Peng, and Li}{Chen
  et~al\mbox{.}}{2019}]%
        {ChenYCNPL19}
\bibfield{author}{\bibinfo{person}{Hongxu Chen}, \bibinfo{person}{Hongzhi Yin},
  \bibinfo{person}{Tong Chen}, \bibinfo{person}{Quoc Viet~Hung Nguyen},
  \bibinfo{person}{Wen{-}Chih Peng}, {and} \bibinfo{person}{Xue Li}.}
  \bibinfo{year}{2019}\natexlab{}.
\newblock \showarticletitle{Exploiting Centrality Information with Graph
  Convolutions for Network Representation Learning}. In
  \bibinfo{booktitle}{\emph{ICDE'19}}. \bibinfo{publisher}{{IEEE}},
  \bibinfo{pages}{590--601}.
\newblock


\bibitem[\protect\citeauthoryear{Chen, Yin, Sun, Chen, Gabrys, and Musial}{Chen
  et~al\mbox{.}}{2020b}]%
        {ChenYS0GM20}
\bibfield{author}{\bibinfo{person}{Hongxu Chen}, \bibinfo{person}{Hongzhi Yin},
  \bibinfo{person}{Xiangguo Sun}, \bibinfo{person}{Tong Chen},
  \bibinfo{person}{Bogdan Gabrys}, {and} \bibinfo{person}{Katarzyna Musial}.}
  \bibinfo{year}{2020}\natexlab{b}.
\newblock \showarticletitle{Multi-level Graph Convolutional Networks for
  Cross-platform Anchor Link Prediction}. In
  \bibinfo{booktitle}{\emph{SIGKDD'20}}. \bibinfo{publisher}{{ACM}},
  \bibinfo{pages}{1503--1511}.
\newblock


\bibitem[\protect\citeauthoryear{Chen, Ma, and Xiao}{Chen
  et~al\mbox{.}}{2018}]%
        {chenfastgcn18}
\bibfield{author}{\bibinfo{person}{Jie Chen}, \bibinfo{person}{Tengfei Ma},
  {and} \bibinfo{person}{Cao Xiao}.} \bibinfo{year}{2018}\natexlab{}.
\newblock \showarticletitle{FastGCN: Fast Learning with Graph Convolutional
  Networks via Importance Sampling}. In \bibinfo{booktitle}{\emph{ICLR' 18}}.
\newblock


\bibitem[\protect\citeauthoryear{Chen, Kornblith, Norouzi, and Hinton}{Chen
  et~al\mbox{.}}{2020a}]%
        {simclr}
\bibfield{author}{\bibinfo{person}{Ting Chen}, \bibinfo{person}{Simon
  Kornblith}, \bibinfo{person}{Mohammad Norouzi}, {and}
  \bibinfo{person}{Geoffrey~E. Hinton}.} \bibinfo{year}{2020}\natexlab{a}.
\newblock \showarticletitle{A Simple Framework for Contrastive Learning of
  Visual Representations}. In \bibinfo{booktitle}{\emph{ICML'20}}
  \emph{(\bibinfo{series}{Proceedings of Machine Learning Research},
  Vol.~\bibinfo{volume}{119})}. \bibinfo{publisher}{{PMLR}},
  \bibinfo{pages}{1597--1607}.
\newblock


\bibitem[\protect\citeauthoryear{Chen and Wong}{Chen and Wong}{2020}]%
        {CWKDD20}
\bibfield{author}{\bibinfo{person}{Tianwen Chen} {and}
  \bibinfo{person}{Raymond~Chi{-}Wing Wong}.} \bibinfo{year}{2020}\natexlab{}.
\newblock \showarticletitle{Handling Information Loss of Graph Neural Networks
  for Session-based Recommendation}. In \bibinfo{booktitle}{\emph{SIGKDD'20}}.
  \bibinfo{publisher}{{ACM}}, \bibinfo{pages}{1172--1180}.
\newblock


\bibitem[\protect\citeauthoryear{Finn, Abbeel, and Levine}{Finn
  et~al\mbox{.}}{2017}]%
        {Finnmaml17}
\bibfield{author}{\bibinfo{person}{Chelsea Finn}, \bibinfo{person}{Pieter
  Abbeel}, {and} \bibinfo{person}{Sergey Levine}.}
  \bibinfo{year}{2017}\natexlab{}.
\newblock \showarticletitle{Model-Agnostic Meta-Learning for Fast Adaptation of
  Deep Networks}. In \bibinfo{booktitle}{\emph{ICML'17}},
  Vol.~\bibinfo{volume}{70}. \bibinfo{pages}{1126--1135}.
\newblock


\bibitem[\protect\citeauthoryear{Gharibshah, Zhu, Hainline, and
  Conway}{Gharibshah et~al\mbox{.}}{2020}]%
        {GharibshahZHC20}
\bibfield{author}{\bibinfo{person}{Zhabiz Gharibshah},
  \bibinfo{person}{Xingquan Zhu}, \bibinfo{person}{Arthur Hainline}, {and}
  \bibinfo{person}{Michael Conway}.} \bibinfo{year}{2020}\natexlab{}.
\newblock \showarticletitle{Deep Learning for User Interest and Response
  Prediction in Online Display Advertising}.
\newblock \bibinfo{journal}{\emph{Data Sci. Eng.}} \bibinfo{volume}{5},
  \bibinfo{number}{1} (\bibinfo{year}{2020}), \bibinfo{pages}{12--26}.
\newblock


\bibitem[\protect\citeauthoryear{Gilmer, Schoenholz, Riley, Vinyals, and
  Dahl}{Gilmer et~al\mbox{.}}{2017}]%
        {GSSPPMLR17}
\bibfield{author}{\bibinfo{person}{Justin Gilmer}, \bibinfo{person}{Samuel~S.
  Schoenholz}, \bibinfo{person}{Patrick~F. Riley}, \bibinfo{person}{Oriol
  Vinyals}, {and} \bibinfo{person}{George~E. Dahl}.}
  \bibinfo{year}{2017}\natexlab{}.
\newblock \showarticletitle{Neural Message Passing for Quantum Chemistry}. In
  \bibinfo{booktitle}{\emph{ICML'17}} \emph{(\bibinfo{series}{Proceedings of
  Machine Learning Research}, Vol.~\bibinfo{volume}{70})}.
  \bibinfo{publisher}{{PMLR}}, \bibinfo{pages}{1263--1272}.
\newblock


\bibitem[\protect\citeauthoryear{Hamilton, Ying, and Leskovec}{Hamilton
  et~al\mbox{.}}{2017}]%
        {williamgraphsage17}
\bibfield{author}{\bibinfo{person}{William~L. Hamilton},
  \bibinfo{person}{Zhitao Ying}, {and} \bibinfo{person}{Jure Leskovec}.}
  \bibinfo{year}{2017}\natexlab{}.
\newblock \showarticletitle{Inductive Representation Learning on Large Graphs}.
  In \bibinfo{booktitle}{\emph{NeurlPS'17}}. \bibinfo{pages}{1024--1034}.
\newblock


\bibitem[\protect\citeauthoryear{Hao, Zhang, Yin, Li, and Chen}{Hao
  et~al\mbox{.}}{2021}]%
        {haopretrain21}
\bibfield{author}{\bibinfo{person}{Bowen Hao}, \bibinfo{person}{Jing Zhang},
  \bibinfo{person}{Hongzhi Yin}, \bibinfo{person}{Cuiping Li}, {and}
  \bibinfo{person}{Hong Chen}.} \bibinfo{year}{2021}\natexlab{}.
\newblock \showarticletitle{Pre-Training Graph Neural Networks for Cold-Start
  Users and Items Representation}. In \bibinfo{booktitle}{\emph{WSDM'21}}.
  \bibinfo{publisher}{{ACM}}, \bibinfo{pages}{265--273}.
\newblock


\bibitem[\protect\citeauthoryear{Harper and Konstan}{Harper and
  Konstan}{2016}]%
        {harper2016movielens}
\bibfield{author}{\bibinfo{person}{F.~Maxwell Harper} {and}
  \bibinfo{person}{Joseph~A. Konstan}.} \bibinfo{year}{2016}\natexlab{}.
\newblock \showarticletitle{The MovieLens Datasets: History and Context}.
\newblock \bibinfo{journal}{\emph{{ACM} Trans. Interact. Intell. Syst.}}
  (\bibinfo{year}{2016}), \bibinfo{pages}{19:1--19:19}.
\newblock


\bibitem[\protect\citeauthoryear{He, Deng, Wang, Li, Zhang, and Wang}{He
  et~al\mbox{.}}{2020}]%
        {xiangnanhe_lightgcn20}
\bibfield{author}{\bibinfo{person}{Xiangnan He}, \bibinfo{person}{Kuan Deng},
  \bibinfo{person}{Xiang Wang}, \bibinfo{person}{Yan Li},
  \bibinfo{person}{Yongdong Zhang}, {and} \bibinfo{person}{Meng Wang}.}
  \bibinfo{year}{2020}\natexlab{}.
\newblock \showarticletitle{LightGCN: Simplifying and Powering Graph
  Convolution Network for Recommendation}. In
  \bibinfo{booktitle}{\emph{SIGIR'20}}.
\newblock


\bibitem[\protect\citeauthoryear{He, Liao, Zhang, Nie, Hu, and Chua}{He
  et~al\mbox{.}}{2017}]%
        {he2017neural}
\bibfield{author}{\bibinfo{person}{Xiangnan He}, \bibinfo{person}{Lizi Liao},
  \bibinfo{person}{Hanwang Zhang}, \bibinfo{person}{Liqiang Nie},
  \bibinfo{person}{Xia Hu}, {and} \bibinfo{person}{Tat{-}Seng Chua}.}
  \bibinfo{year}{2017}\natexlab{}.
\newblock \showarticletitle{Neural Collaborative Filtering}. In
  \bibinfo{booktitle}{\emph{WWW'17}}. \bibinfo{pages}{173--182}.
\newblock


\bibitem[\protect\citeauthoryear{Hu, Dong, Wang, Chang, and Sun}{Hu
  et~al\mbox{.}}{2020}]%
        {gpt_gnnhu}
\bibfield{author}{\bibinfo{person}{Ziniu Hu}, \bibinfo{person}{Yuxiao Dong},
  \bibinfo{person}{Kuansan Wang}, \bibinfo{person}{Kai{-}Wei Chang}, {and}
  \bibinfo{person}{Yizhou Sun}.} \bibinfo{year}{2020}\natexlab{}.
\newblock \showarticletitle{GPT-GNN: Generative Pre-Training of Graph Neural
  Networks}. In \bibinfo{booktitle}{\emph{SIGKDD'20}}.
\newblock


\bibitem[\protect\citeauthoryear{Kipf and Welling}{Kipf and Welling}{2017}]%
        {Thomasgcn}
\bibfield{author}{\bibinfo{person}{Thomas~N. Kipf} {and} \bibinfo{person}{Max
  Welling}.} \bibinfo{year}{2017}\natexlab{}.
\newblock \showarticletitle{Semi-Supervised Classification with Graph
  Convolutional Networks}. In \bibinfo{booktitle}{\emph{ICLR'17}}.
\newblock


\bibitem[\protect\citeauthoryear{Liang, Charlin, McInerney, and Blei}{Liang
  et~al\mbox{.}}{2016}]%
        {gowalla}
\bibfield{author}{\bibinfo{person}{Dawen Liang}, \bibinfo{person}{Laurent
  Charlin}, \bibinfo{person}{James McInerney}, {and} \bibinfo{person}{David~M.
  Blei}.} \bibinfo{year}{2016}\natexlab{}.
\newblock \showarticletitle{Modeling User Exposure in Recommendation}. In
  \bibinfo{booktitle}{\emph{WWW'16}}. \bibinfo{publisher}{{ACM}},
  \bibinfo{pages}{951--961}.
\newblock


\bibitem[\protect\citeauthoryear{Linden, Smith, and York}{Linden
  et~al\mbox{.}}{2003}]%
        {linden2003amazon}
\bibfield{author}{\bibinfo{person}{Greg Linden}, \bibinfo{person}{Brent Smith},
  {and} \bibinfo{person}{Jeremy York}.} \bibinfo{year}{2003}\natexlab{}.
\newblock \showarticletitle{Amazon.com Recommendations: Item-to-Item
  Collaborative Filtering}.
\newblock \bibinfo{journal}{\emph{{IEEE} Internet Comput.}}
  (\bibinfo{year}{2003}), \bibinfo{pages}{76--80}.
\newblock


\bibitem[\protect\citeauthoryear{Liu, Yang, Lei, Wang, Tang, Zhang, Sun, and
  Miao}{Liu et~al\mbox{.}}{2021}]%
        {LYSLWTZArxiv2020}
\bibfield{author}{\bibinfo{person}{Yong Liu}, \bibinfo{person}{Susen Yang},
  \bibinfo{person}{Chenyi Lei}, \bibinfo{person}{Guoxin Wang},
  \bibinfo{person}{Haihong Tang}, \bibinfo{person}{Juyong Zhang},
  \bibinfo{person}{Aixin Sun}, {and} \bibinfo{person}{Chunyan Miao}.}
  \bibinfo{year}{2021}\natexlab{}.
\newblock \showarticletitle{Pre-training Graph Transformer with Multimodal Side
  Information for Recommendation}. In \bibinfo{booktitle}{\emph{MM'21}}.
  \bibinfo{publisher}{{ACM}}, \bibinfo{pages}{2853--2861}.
\newblock


\bibitem[\protect\citeauthoryear{Meng, Liu, Macdonald, and Ounis}{Meng
  et~al\mbox{.}}{2021}]%
        {MLMArxiv2021}
\bibfield{author}{\bibinfo{person}{Zaiqiao Meng}, \bibinfo{person}{Siwei Liu},
  \bibinfo{person}{Craig Macdonald}, {and} \bibinfo{person}{Iadh Ounis}.}
  \bibinfo{year}{2021}\natexlab{}.
\newblock \showarticletitle{Graph Neural Pre-training for Enhancing
  Recommendations using Side Information}.
\newblock \bibinfo{journal}{\emph{CoRR}}  \bibinfo{volume}{abs/2107.03936}
  (\bibinfo{year}{2021}).
\newblock
\urldef\tempurl%
\url{https://arxiv.org/abs/2107.03936}
\showURL{%
\tempurl}


\bibitem[\protect\citeauthoryear{Munkhdalai and Yu}{Munkhdalai and Yu}{2017}]%
        {munkhdalaimetaneworks17}
\bibfield{author}{\bibinfo{person}{Tsendsuren Munkhdalai} {and}
  \bibinfo{person}{Hong Yu}.} \bibinfo{year}{2017}\natexlab{}.
\newblock \showarticletitle{Meta Networks}. In
  \bibinfo{booktitle}{\emph{ICML'17}} \emph{(\bibinfo{series}{Proceedings of
  Machine Learning Research}, Vol.~\bibinfo{volume}{70})},
  \bibfield{editor}{\bibinfo{person}{Doina Precup} {and}
  \bibinfo{person}{Yee~Whye Teh}} (Eds.). \bibinfo{pages}{2554--2563}.
\newblock


\bibitem[\protect\citeauthoryear{Perozzi, Al{-}Rfou, and Skiena}{Perozzi
  et~al\mbox{.}}{2014}]%
        {deepwalk}
\bibfield{author}{\bibinfo{person}{Bryan Perozzi}, \bibinfo{person}{Rami
  Al{-}Rfou}, {and} \bibinfo{person}{Steven Skiena}.}
  \bibinfo{year}{2014}\natexlab{}.
\newblock \showarticletitle{DeepWalk: online learning of social
  representations}. In \bibinfo{booktitle}{\emph{SIGKDD'14}}.
  \bibinfo{publisher}{{ACM}}, \bibinfo{pages}{701--710}.
\newblock


\bibitem[\protect\citeauthoryear{Qiu, Chen, Dong, Zhang, Yang, Ding, Wang, and
  Tang}{Qiu et~al\mbox{.}}{2020}]%
        {qiu2020gcc}
\bibfield{author}{\bibinfo{person}{Jiezhong Qiu}, \bibinfo{person}{Qibin Chen},
  \bibinfo{person}{Yuxiao Dong}, \bibinfo{person}{Jing Zhang},
  \bibinfo{person}{Hongxia Yang}, \bibinfo{person}{Ming Ding},
  \bibinfo{person}{Kuansan Wang}, {and} \bibinfo{person}{Jie Tang}.}
  \bibinfo{year}{2020}\natexlab{}.
\newblock \showarticletitle{GCC: Graph Contrastive Coding for Graph Neural
  Network Pre-Training}. In \bibinfo{booktitle}{\emph{SIGKDD'20}}.
\newblock


\bibitem[\protect\citeauthoryear{Qiu, Huang, Yin, and Wang}{Qiu
  et~al\mbox{.}}{2022}]%
        {QiuHYW22}
\bibfield{author}{\bibinfo{person}{Ruihong Qiu}, \bibinfo{person}{Zi Huang},
  \bibinfo{person}{Hongzhi Yin}, {and} \bibinfo{person}{Zijian Wang}.}
  \bibinfo{year}{2022}\natexlab{}.
\newblock \showarticletitle{Contrastive Learning for Representation
  Degeneration Problem in Sequential Recommendation}. In
  \bibinfo{booktitle}{\emph{WSDM'22}}. \bibinfo{publisher}{{ACM}},
  \bibinfo{pages}{813--823}.
\newblock


\bibitem[\protect\citeauthoryear{Snell, Swersky, and Zemel}{Snell
  et~al\mbox{.}}{2017}]%
        {Snellprotocal17}
\bibfield{author}{\bibinfo{person}{Jake Snell}, \bibinfo{person}{Kevin
  Swersky}, {and} \bibinfo{person}{Richard~S. Zemel}.}
  \bibinfo{year}{2017}\natexlab{}.
\newblock \showarticletitle{Prototypical Networks for Few-shot Learning}. In
  \bibinfo{booktitle}{\emph{NeurlPS'17}}. \bibinfo{pages}{4077--4087}.
\newblock


\bibitem[\protect\citeauthoryear{Sun, Hoffmann, Verma, and Tang}{Sun
  et~al\mbox{.}}{2020}]%
        {infograph}
\bibfield{author}{\bibinfo{person}{Fan{-}Yun Sun}, \bibinfo{person}{Jordan
  Hoffmann}, \bibinfo{person}{Vikas Verma}, {and} \bibinfo{person}{Jian Tang}.}
  \bibinfo{year}{2020}\natexlab{}.
\newblock \showarticletitle{InfoGraph: Unsupervised and Semi-supervised
  Graph-Level Representation Learning via Mutual Information Maximization}. In
  \bibinfo{booktitle}{\emph{ICLR'20}}.
\newblock


\bibitem[\protect\citeauthoryear{Sutton, McAllester, Singh, and Mansour}{Sutton
  et~al\mbox{.}}{1999}]%
        {policygradient}
\bibfield{author}{\bibinfo{person}{Richard~S. Sutton},
  \bibinfo{person}{David~A. McAllester}, \bibinfo{person}{Satinder~P. Singh},
  {and} \bibinfo{person}{Yishay Mansour}.} \bibinfo{year}{1999}\natexlab{}.
\newblock \showarticletitle{Policy Gradient Methods for Reinforcement Learning
  with Function Approximation}. In \bibinfo{booktitle}{\emph{NeurIPS'99}}.
  \bibinfo{pages}{1057--1063}.
\newblock


\bibitem[\protect\citeauthoryear{van~den Berg, Kipf, and Welling}{van~den Berg
  et~al\mbox{.}}{2018}]%
        {BRDTKDD2018}
\bibfield{author}{\bibinfo{person}{Rianne van~den Berg},
  \bibinfo{person}{Thomas~N. Kipf}, {and} \bibinfo{person}{Max Welling}.}
  \bibinfo{year}{2018}\natexlab{}.
\newblock \showarticletitle{Graph Convolutional Matrix Completion}. In
  \bibinfo{booktitle}{\emph{SIGKDD'18}}. \bibinfo{publisher}{{ACM}}.
\newblock


\bibitem[\protect\citeauthoryear{Vaswani, Shazeer, Parmar, Uszkoreit, Jones,
  Gomez, Kaiser, and Polosukhin}{Vaswani et~al\mbox{.}}{2017}]%
        {self_attention17}
\bibfield{author}{\bibinfo{person}{Ashish Vaswani}, \bibinfo{person}{Noam
  Shazeer}, \bibinfo{person}{Niki Parmar}, \bibinfo{person}{Jakob Uszkoreit},
  \bibinfo{person}{Llion Jones}, \bibinfo{person}{Aidan~N. Gomez},
  \bibinfo{person}{Lukasz Kaiser}, {and} \bibinfo{person}{Illia Polosukhin}.}
  \bibinfo{year}{2017}\natexlab{}.
\newblock \showarticletitle{Attention is All you Need}. In
  \bibinfo{booktitle}{\emph{NeurlPS'17}}. \bibinfo{pages}{5998--6008}.
\newblock


\bibitem[\protect\citeauthoryear{Velickovic, Fedus, Hamilton, Li{\`o}, Bengio,
  and Hjelm}{Velickovic et~al\mbox{.}}{2019}]%
        {velickovic2019deep}
\bibfield{author}{\bibinfo{person}{Petar Velickovic}, \bibinfo{person}{William
  Fedus}, \bibinfo{person}{William~L Hamilton}, \bibinfo{person}{Pietro
  Li{\`o}}, \bibinfo{person}{Yoshua Bengio}, {and} \bibinfo{person}{R~Devon
  Hjelm}.} \bibinfo{year}{2019}\natexlab{}.
\newblock \showarticletitle{Deep Graph Infomax}. In
  \bibinfo{booktitle}{\emph{ICLR'19}}.
\newblock


\bibitem[\protect\citeauthoryear{Vinyals, Blundell, Lillicrap, Kavukcuoglu, and
  Wierstra}{Vinyals et~al\mbox{.}}{2016}]%
        {vinyalsmatching16}
\bibfield{author}{\bibinfo{person}{Oriol Vinyals}, \bibinfo{person}{Charles
  Blundell}, \bibinfo{person}{Tim Lillicrap}, \bibinfo{person}{Koray
  Kavukcuoglu}, {and} \bibinfo{person}{Daan Wierstra}.}
  \bibinfo{year}{2016}\natexlab{}.
\newblock \showarticletitle{Matching Networks for One Shot Learning}. In
  \bibinfo{booktitle}{\emph{NeurlPS'16}}. \bibinfo{pages}{3630--3638}.
\newblock


\bibitem[\protect\citeauthoryear{Wang, Zhang, Zhao, Li, Xie, and Guo}{Wang
  et~al\mbox{.}}{2019d}]%
        {wang2019multi}
\bibfield{author}{\bibinfo{person}{Hongwei Wang}, \bibinfo{person}{Fuzheng
  Zhang}, \bibinfo{person}{Miao Zhao}, \bibinfo{person}{Wenjie Li},
  \bibinfo{person}{Xing Xie}, {and} \bibinfo{person}{Minyi Guo}.}
  \bibinfo{year}{2019}\natexlab{d}.
\newblock \showarticletitle{Multi-Task Feature Learning for Knowledge Graph
  Enhanced Recommendation}. In \bibinfo{booktitle}{\emph{WWW' 19}}.
  \bibinfo{pages}{2000--2010}.
\newblock


\bibitem[\protect\citeauthoryear{Wang, Yin, Wang, Nguyen, Huang, and Cui}{Wang
  et~al\mbox{.}}{2019c}]%
        {WangYWNHC19}
\bibfield{author}{\bibinfo{person}{Qinyong Wang}, \bibinfo{person}{Hongzhi
  Yin}, \bibinfo{person}{Hao Wang}, \bibinfo{person}{Quoc Viet~Hung Nguyen},
  \bibinfo{person}{Zi Huang}, {and} \bibinfo{person}{Lizhen Cui}.}
  \bibinfo{year}{2019}\natexlab{c}.
\newblock \showarticletitle{Enhancing Collaborative Filtering with Generative
  Augmentation}. In \bibinfo{booktitle}{\emph{SIGKDD'19}}.
  \bibinfo{publisher}{{ACM}}, \bibinfo{pages}{548--556}.
\newblock


\bibitem[\protect\citeauthoryear{Wang, He, Cao, Liu, and Chua}{Wang
  et~al\mbox{.}}{2019a}]%
        {wang2019kgat}
\bibfield{author}{\bibinfo{person}{Xiang Wang}, \bibinfo{person}{Xiangnan He},
  \bibinfo{person}{Yixin Cao}, \bibinfo{person}{Meng Liu}, {and}
  \bibinfo{person}{Tat{-}Seng Chua}.} \bibinfo{year}{2019}\natexlab{a}.
\newblock \showarticletitle{{KGAT:} Knowledge Graph Attention Network for
  Recommendation}. In \bibinfo{booktitle}{\emph{SIGKDD'19}}.
  \bibinfo{pages}{950--958}.
\newblock


\bibitem[\protect\citeauthoryear{Wang, He, Wang, Feng, and Chua}{Wang
  et~al\mbox{.}}{2019b}]%
        {wangncgf19}
\bibfield{author}{\bibinfo{person}{Xiang Wang}, \bibinfo{person}{Xiangnan He},
  \bibinfo{person}{Meng Wang}, \bibinfo{person}{Fuli Feng}, {and}
  \bibinfo{person}{Tat{-}Seng Chua}.} \bibinfo{year}{2019}\natexlab{b}.
\newblock \showarticletitle{Neural Graph Collaborative Filtering}. In
  \bibinfo{booktitle}{\emph{SIGIR'19}}. \bibinfo{pages}{165--174}.
\newblock


\bibitem[\protect\citeauthoryear{Williams}{Williams}{1992}]%
        {williams_policy}
\bibfield{author}{\bibinfo{person}{Ronald~J. Williams}.}
  \bibinfo{year}{1992}\natexlab{}.
\newblock \showarticletitle{Simple Statistical Gradient-Following Algorithms
  for Connectionist Reinforcement Learning}.
\newblock \bibinfo{journal}{\emph{Mach. Learn.}}  \bibinfo{volume}{8},
  \bibinfo{pages}{229--256}.
\newblock


\bibitem[\protect\citeauthoryear{Wu, Wang, Feng, He, Chen, Lian, and Xie}{Wu
  et~al\mbox{.}}{2021}]%
        {WWFHSIGIR2021}
\bibfield{author}{\bibinfo{person}{Jiancan Wu}, \bibinfo{person}{Xiang Wang},
  \bibinfo{person}{Fuli Feng}, \bibinfo{person}{Xiangnan He},
  \bibinfo{person}{Liang Chen}, \bibinfo{person}{Jianxun Lian}, {and}
  \bibinfo{person}{Xing Xie}.} \bibinfo{year}{2021}\natexlab{}.
\newblock \showarticletitle{Self-supervised Graph Learning for Recommendation}.
  In \bibinfo{booktitle}{\emph{SIGIR'21}}. \bibinfo{publisher}{{ACM}},
  \bibinfo{pages}{726--735}.
\newblock


\bibitem[\protect\citeauthoryear{Wu, Xiong, Yu, and Lin}{Wu
  et~al\mbox{.}}{2018}]%
        {WXYSCVPR18}
\bibfield{author}{\bibinfo{person}{Zhirong Wu}, \bibinfo{person}{Yuanjun
  Xiong}, \bibinfo{person}{Stella~X. Yu}, {and} \bibinfo{person}{Dahua Lin}.}
  \bibinfo{year}{2018}\natexlab{}.
\newblock \showarticletitle{Unsupervised Feature Learning via Non-Parametric
  Instance Discrimination}. In \bibinfo{booktitle}{\emph{CVPR'18}}.
  \bibinfo{publisher}{Computer Vision Foundation / {IEEE} Computer Society},
  \bibinfo{pages}{3733--3742}.
\newblock


\bibitem[\protect\citeauthoryear{Xia, Yin, Yu, Shao, and Cui}{Xia
  et~al\mbox{.}}{2021a}]%
        {0013YYSC21}
\bibfield{author}{\bibinfo{person}{Xin Xia}, \bibinfo{person}{Hongzhi Yin},
  \bibinfo{person}{Junliang Yu}, \bibinfo{person}{Yingxia Shao}, {and}
  \bibinfo{person}{Lizhen Cui}.} \bibinfo{year}{2021}\natexlab{a}.
\newblock \showarticletitle{Self-Supervised Graph Co-Training for Session-based
  Recommendation}. In \bibinfo{booktitle}{\emph{CIKM'21}}.
  \bibinfo{publisher}{{ACM}}, \bibinfo{pages}{2180--2190}.
\newblock


\bibitem[\protect\citeauthoryear{Xia, Yin, Yu, Wang, Cui, and Zhang}{Xia
  et~al\mbox{.}}{2021b}]%
        {0013YYWC021}
\bibfield{author}{\bibinfo{person}{Xin Xia}, \bibinfo{person}{Hongzhi Yin},
  \bibinfo{person}{Junliang Yu}, \bibinfo{person}{Qinyong Wang},
  \bibinfo{person}{Lizhen Cui}, {and} \bibinfo{person}{Xiangliang Zhang}.}
  \bibinfo{year}{2021}\natexlab{b}.
\newblock \showarticletitle{Self-Supervised Hypergraph Convolutional Networks
  for Session-based Recommendation}. In \bibinfo{booktitle}{\emph{AAAI'21}}.
  \bibinfo{publisher}{{AAAI} Press}, \bibinfo{pages}{4503--4511}.
\newblock


\bibitem[\protect\citeauthoryear{Yin, Cui, Sun, Hu, and Chen}{Yin
  et~al\mbox{.}}{2014}]%
        {hongzhi_side}
\bibfield{author}{\bibinfo{person}{Hongzhi Yin}, \bibinfo{person}{Bin Cui},
  \bibinfo{person}{Yizhou Sun}, \bibinfo{person}{Zhiting Hu}, {and}
  \bibinfo{person}{Ling Chen}.} \bibinfo{year}{2014}\natexlab{}.
\newblock \showarticletitle{{LCARS:} {A} Spatial Item Recommender System}.
\newblock \bibinfo{journal}{\emph{TOIS'14}} \bibinfo{volume}{32},
  \bibinfo{number}{3} (\bibinfo{year}{2014}), \bibinfo{pages}{11:1--11:37}.
\newblock


\bibitem[\protect\citeauthoryear{Yin, Wang, Zheng, Li, Yang, and Zhou}{Yin
  et~al\mbox{.}}{2019}]%
        {YinW0LYZ19}
\bibfield{author}{\bibinfo{person}{Hongzhi Yin}, \bibinfo{person}{Qinyong
  Wang}, \bibinfo{person}{Kai Zheng}, \bibinfo{person}{Zhixu Li},
  \bibinfo{person}{Jiali Yang}, {and} \bibinfo{person}{Xiaofang Zhou}.}
  \bibinfo{year}{2019}\natexlab{}.
\newblock \showarticletitle{Social Influence-Based Group Representation
  Learning for Group Recommendation}. In \bibinfo{booktitle}{\emph{ICDE'19}}.
  \bibinfo{publisher}{{IEEE}}, \bibinfo{pages}{566--577}.
\newblock


\bibitem[\protect\citeauthoryear{Yin, Wang, Zheng, Li, and Zhou}{Yin
  et~al\mbox{.}}{2020}]%
        {hongzhi_tkde_20}
\bibfield{author}{\bibinfo{person}{Hongzhi Yin}, \bibinfo{person}{Qinyong
  Wang}, \bibinfo{person}{Kai Zheng}, \bibinfo{person}{Zhixu Li}, {and}
  \bibinfo{person}{Xiaofang Zhou}.} \bibinfo{year}{2020}\natexlab{}.
\newblock \showarticletitle{Overcoming Data Sparsity in Group Recommendation}.
\newblock \bibinfo{journal}{\emph{{IEEE} Trans. Knowl. Data Eng.}}
  (\bibinfo{year}{2020}).
\newblock


\bibitem[\protect\citeauthoryear{Yin, Wang, Wang, Chen, and Zhou}{Yin
  et~al\mbox{.}}{2017}]%
        {YinWWCZ17}
\bibfield{author}{\bibinfo{person}{Hongzhi Yin}, \bibinfo{person}{Weiqing
  Wang}, \bibinfo{person}{Hao Wang}, \bibinfo{person}{Ling Chen}, {and}
  \bibinfo{person}{Xiaofang Zhou}.} \bibinfo{year}{2017}\natexlab{}.
\newblock \showarticletitle{Spatial-Aware Hierarchical Collaborative Deep
  Learning for {POI} Recommendation}.
\newblock \bibinfo{journal}{\emph{{IEEE} Trans. Knowl. Data Eng.}}
  \bibinfo{volume}{19}, \bibinfo{number}{11} (\bibinfo{year}{2017}),
  \bibinfo{pages}{2537--2551}.
\newblock


\bibitem[\protect\citeauthoryear{Ying, He, Chen, Eksombatchai, Hamilton, and
  Leskovec}{Ying et~al\mbox{.}}{2018}]%
        {pinsage}
\bibfield{author}{\bibinfo{person}{Rex Ying}, \bibinfo{person}{Ruining He},
  \bibinfo{person}{Kaifeng Chen}, \bibinfo{person}{Pong Eksombatchai},
  \bibinfo{person}{William~L. Hamilton}, {and} \bibinfo{person}{Jure
  Leskovec}.} \bibinfo{year}{2018}\natexlab{}.
\newblock \showarticletitle{Graph Convolutional Neural Networks for Web-Scale
  Recommender Systems}. In \bibinfo{booktitle}{\emph{SIGKDD"18}}.
  \bibinfo{pages}{974--983}.
\newblock


\bibitem[\protect\citeauthoryear{You, Chen, Sui, Chen, Wang, and Shen}{You
  et~al\mbox{.}}{2020}]%
        {YCTSCWNIPS20}
\bibfield{author}{\bibinfo{person}{Yuning You}, \bibinfo{person}{Tianlong
  Chen}, \bibinfo{person}{Yongduo Sui}, \bibinfo{person}{Ting Chen},
  \bibinfo{person}{Zhangyang Wang}, {and} \bibinfo{person}{Yang Shen}.}
  \bibinfo{year}{2020}\natexlab{}.
\newblock \showarticletitle{Graph Contrastive Learning with Augmentations}. In
  \bibinfo{booktitle}{\emph{NeurIPS'20}}.
\newblock


\bibitem[\protect\citeauthoryear{Yu, Yin, Gao, Xia, Zhang, and Hung}{Yu
  et~al\mbox{.}}{2021a}]%
        {YuY000H21}
\bibfield{author}{\bibinfo{person}{Junliang Yu}, \bibinfo{person}{Hongzhi Yin},
  \bibinfo{person}{Min Gao}, \bibinfo{person}{Xin Xia},
  \bibinfo{person}{Xiangliang Zhang}, {and} \bibinfo{person}{Nguyen Quoc~Viet
  Hung}.} \bibinfo{year}{2021}\natexlab{a}.
\newblock \showarticletitle{Socially-Aware Self-Supervised Tri-Training for
  Recommendation}. In \bibinfo{booktitle}{\emph{SIGKDD'21}}.
  \bibinfo{publisher}{{ACM}}, \bibinfo{pages}{2084--2092}.
\newblock


\bibitem[\protect\citeauthoryear{Yu, Yin, Li, Wang, Hung, and Zhang}{Yu
  et~al\mbox{.}}{2021b}]%
        {YuYLWH021}
\bibfield{author}{\bibinfo{person}{Junliang Yu}, \bibinfo{person}{Hongzhi Yin},
  \bibinfo{person}{Jundong Li}, \bibinfo{person}{Qinyong Wang},
  \bibinfo{person}{Nguyen Quoc~Viet Hung}, {and} \bibinfo{person}{Xiangliang
  Zhang}.} \bibinfo{year}{2021}\natexlab{b}.
\newblock \showarticletitle{Self-Supervised Multi-Channel Hypergraph
  Convolutional Network for Social Recommendation}. In
  \bibinfo{booktitle}{\emph{WWW'21}}. \bibinfo{publisher}{{ACM} / {IW3C2}},
  \bibinfo{pages}{413--424}.
\newblock


\bibitem[\protect\citeauthoryear{Yu, Yin, Xia, Chen, Li, and Huang}{Yu
  et~al\mbox{.}}{2022a}]%
        {DBLP:journals/corr/abs-2203-15876}
\bibfield{author}{\bibinfo{person}{Junliang Yu}, \bibinfo{person}{Hongzhi Yin},
  \bibinfo{person}{Xin Xia}, \bibinfo{person}{Tong Chen},
  \bibinfo{person}{Jundong Li}, {and} \bibinfo{person}{Zi Huang}.}
  \bibinfo{year}{2022}\natexlab{a}.
\newblock \showarticletitle{Self-Supervised Learning for Recommender Systems:
  {A} Survey}.
\newblock \bibinfo{journal}{\emph{CoRR}}  \bibinfo{volume}{abs/2203.15876}
  (\bibinfo{year}{2022}).
\newblock
\urldef\tempurl%
\url{https://doi.org/10.48550/arXiv.2203.15876}
\showDOI{\tempurl}


\bibitem[\protect\citeauthoryear{Yu, Yin, Xia, Tong~Chen, and Nguyen}{Yu
  et~al\mbox{.}}{2022b}]%
        {yuhxtlqsigir22}
\bibfield{author}{\bibinfo{person}{Junliang Yu}, \bibinfo{person}{Hongzhi Yin},
  \bibinfo{person}{Xin Xia}, \bibinfo{person}{Lizhen~Cui Tong~Chen}, {and}
  \bibinfo{person}{Quoc Viet~Hung Nguyen}.} \bibinfo{year}{2022}\natexlab{b}.
\newblock \showarticletitle{Are Graph Augmentations Necessary? Simple Graph
  Contrastive Learning for Recommendation}. In
  \bibinfo{booktitle}{\emph{SIGIR'22}}.
\newblock


\bibitem[\protect\citeauthoryear{Zhang, Hao, Chen, Li, Chen, and Sun}{Zhang
  et~al\mbox{.}}{2019}]%
        {zhanghrl19}
\bibfield{author}{\bibinfo{person}{Jing Zhang}, \bibinfo{person}{Bowen Hao},
  \bibinfo{person}{Bo Chen}, \bibinfo{person}{Cuiping Li},
  \bibinfo{person}{Hong Chen}, {and} \bibinfo{person}{Jimeng Sun}.}
  \bibinfo{year}{2019}\natexlab{}.
\newblock \showarticletitle{Hierarchical Reinforcement Learning for Course
  Recommendation in MOOCs}. In \bibinfo{booktitle}{\emph{AAAI'19}}.
  \bibinfo{pages}{435--442}.
\newblock


\bibitem[\protect\citeauthoryear{Zhang, Chen, Wang, and Yu}{Zhang
  et~al\mbox{.}}{2013}]%
        {dynamic2}
\bibfield{author}{\bibinfo{person}{Weinan Zhang}, \bibinfo{person}{Tianqi
  Chen}, \bibinfo{person}{Jun Wang}, {and} \bibinfo{person}{Yong Yu}.}
  \bibinfo{year}{2013}\natexlab{}.
\newblock \showarticletitle{Optimizing top-n collaborative filtering via
  dynamic negative item sampling}. In \bibinfo{booktitle}{\emph{SIGIR'13}}.
  \bibinfo{publisher}{{ACM}}, \bibinfo{pages}{785--788}.
\newblock


\end{thebibliography}

\end{document}